\documentclass[11pt]{article}
\pdfoutput=1

\usepackage{jheppub}
\usepackage{multirow}
\usepackage{url,comment}
\usepackage{times}
\usepackage{latexsym}
\usepackage{graphicx, graphics, hyperref, amsmath, amssymb, slashed, xcolor, bbm,amsthm, array}
\usepackage{subfigure}

\usepackage{rotating}
\usepackage{afterpage}

\newcommand{\nc}{\newcommand}

\nc{\beq}{\begin{equation}}
\nc{\eeq}{\end{equation}}
\nc{\barray}{\begin{eqnarray}}
\nc{\earray}{\end{eqnarray}}
\nc{\barrayn}{\begin{eqnarray*}}
\nc{\earrayn}{\end{eqnarray*}}
\nc{\bcenter}{\begin{center}}
\nc{\ecenter}{\end{center}}
\nc{\mc}{\mathcal}
\nc{\er}[1]{(\ref{eq:#1})}
\nc{\onehalf}{\frac{1}{2}} 
\nc{\partialbar}{\bar{\partial}}
\nc{\psit}{\widetilde{\psi}}
\nc{\Tr}{\mbox{Tr}}
\nc{\hc}{\mbox{H.c.}}
\nc{\ev}{\;\mathrm{eV}}
\nc{\mev}{\;\mathrm{MeV}}
\nc{\gev}{\;\mathrm{GeV}}
\nc{\tev}{\;\mathrm{TeV}}

\def\chii0{\chi_i^0}
\def\chij0{\chi_j^0}

\newcommand{\gsim}{\lower.7ex\hbox{$\;\stackrel{\textstyle>}{\sim}\;$}}
\newcommand{\lsim}{\lower.7ex\hbox{$\;\stackrel{\textstyle<}{\sim}\;$}}
\nc{\ttbar}{t\bar t}
\def\ifb{{\ \rm fb}^{-1}}

\newcommand{\fref}[1]{Fig.~\ref{f.#1}}
\newcommand{\eref}[1]{Eq.~(\ref{e.#1})}

\newcommand{\sref}[1]{Section~\ref{s.#1}}
\newcommand{\ssref}[1]{Section~\ref{ss.#1}}

\newcommand{\cref}[1]{Chapter~\ref{c.#1}}
\newcommand{\tref}[1]{Table~\ref{t.#1}}

\def\Br{\mathrm{Br}}

\graphicspath{{plots/}}

\title{Discovering Uncolored Naturalness\\ in Exotic Higgs Decays}

\author{David Curtin,}
\author{Christopher B. Verhaaren}

\affiliation{Maryland Center for Fundamental Physics, Department of Physics, University of Maryland, College Park, MD 20742-411\\}

\emailAdd{dcurtin1@umd.edu}
\emailAdd{cver@umd.edu}

\abstract{
Solutions to the hierarchy problem usually require top partners. In standard SUSY or composite Higgs theories, the partners carry SM color and are becoming increasingly constrained by LHC searches. However, theories like Folded SUSY (FS), Twin Higgs (TH) and Quirky Little Higgs (QLH) introduce uncolored top partners, which can be SM singlets or carry electroweak charge. Their small production cross section left doubt as to whether the LHC can effectively probe such scenarios. Typically, these partners are charged under their own mirror color gauge group. In FS and QLH, the absence of light mirror matter allows glueballs to form at the bottom of the mirror spectrum. This is also the case in some TH realizations. The Higgs can decay to these mirror glueballs, with the glueballs decaying into SM particles with potentially observable lifetimes. We undertake the first detailed study of this glueball signature and quantitatively demonstrate the discovery potential of uncolored naturalness via exotic Higgs decays at the LHC and a potential future 100 TeV collider. Our findings indicate that mirror glueballs are the smoking gun signature of natural FS and QLH type theories, in analogy to tree-level Higgs coupling shifts for the TH. We show that glueball masses in the $\sim$ 10-60 GeV mass range are theoretically preferred. Careful treatment of lifetime, mirror-hadronization and nonperturbative uncertainties is required to perform meaningful collider studies. We outline several new search strategies for exotic Higgs decays of the form $h\to XX \to 4f$ at the LHC, with $X$ having lifetimes in the $10 \mu m$ to $km$ range. We find that FS stops can be probed with masses up to 600 (1100) GeV at the LHC with 300 (3000) $\ifb$ of data, and TH top partners could be accessible with masses up to 900 (1500) GeV. This makes exotic Higgs decays the prime discovery channel for uncolored naturalness at the LHC.
}

\begin{document}

\maketitle

\section{Introduction}\label{s.intro}

The Standard Model (SM) is beautifully completed by the discovery of the Higgs boson at the LHC~\cite{Aad:2012tfa,Chatrchyan:2012ufa}. A vigorous program to explore the new particle's properties is already underway, and all available high-energy data is, so far, in perfect agreement with SM predictions. This experimental triumph has made the hierarchy problem all the more vexing. Stabilizing the electroweak scale requires new radiative Higgs mass contributions around a TeV. The Higgs has been found, but the associated beyond-SM (BSM) physics has not revealed itself to date.

There are two known symmetry mechanisms for stabilizing the electroweak scale.\footnote{For an interesting alternative approach, see \cite{Graham:2015cka}.} Supersymmetry~\cite{Martin:1997ns} introduces a partner whose spin differs by a half-unit but has identical gauge charge for each SM particle, with stops canceling the top quadratic correction to the Higgs mass. In the Composite Higgs framework \cite{Bellazzini:2014yua} or its 5D-duals, the Higgs is realized as a pseudo Nambu-Goldstone boson (pNGB) transforming under an approximate shift symmetry. Additional protection mechanisms like collective breaking in Little Higgs theories \cite{ArkaniHamed:2001nc, ArkaniHamed:2002qx, ArkaniHamed:2002qy, Schmaltz:2004de} and holographic Higgs models \cite{Contino:2003ve,Agashe:2004rs, Contino:2006qr}  control the shape of the Higgs potential, which result in fermionic top partner states regulating the quadratic Higgs mass divergence. 
Since the symmetry that protects the Higgs commutes with gauge symmetry, both scalar and fermionic partners have SM color charge and can be copiously produced at the LHC if they are lighter than a TeV. Theories with colored top partners are therefore becoming increasingly constrained by LHC null results (see e.g. \cite{Chatrchyan:2013xna,CMS:2014wsa,Aad:2014bva,Aad:2014kra}).

This might lead one to think that naturalness, as a guide to new physics, is in a bit of a tight spot, but this naive view is too pessimistic.
For example, various $R$-parity violating scenarios or non-minimal models like Stealth SUSY \cite{Fan:2011yu} can remain natural while protecting the Higgs mass. 
Alternatively, the colored partners could simply sit on top of SM backgrounds in kinematic blind spots \cite{Martin:2007gf} (though those areas are coming under increased scrutiny \cite{Martin:2008sv,LeCompte:2011fh,Belanger:2012mk,Rolbiecki:2012gn,Curtin:2014zua,Kim:2014eva,Czakon:2014fka,CMS:2014exa, Rolbiecki:2015lsa, An:2015uwa}). 
Finally, it remains possible that colored top partners are simply `around the corner' and will show up early in run 2 of the LHC. 
It is clearly too early to abandon standard SUSY or composite Higgs theories. However, it is worthwhile to consider alternatives in an attempt to consider the full theory space of natural solutions to the hierarchy problem.

Recently, there has been an upswell of interest in theories with \emph{uncolored} top partners. In these models the symmetry protecting the Higgs mass does not commute with color, leaving the weak scale to be stabilized by SM singlets or states with only electroweak (EW) quantum numbers. An essential component of these models is an $SU(3)_B$ gauge group in the BSM sector to match the color $SU(3)_A$ of the SM. The first and prototypical examples of such theories are the Twin Higgs \cite{Chacko:2005pe} and Folded SUSY \cite{Burdman:2006tz} models.

In the original Mirror Twin Higgs model, the entire SM is duplicated in a hidden mirror sector. The full Higgs sector respects an approximate global symmetry, and the SM-like Higgs is a pNGB of its spontaneous breaking. Gauge and Yukawa couplings explicitly break the global symmetry to a discreet $\mathbb{Z}_2$, which is nevertheless sufficient to make quadratic corrections to the Higgs mass from quark and gauge loops respect the original global symmetry, protecting the pNGB Higgs mass at one-loop. Most significant to phenomenology, all twin particles are singlets under the SM gauge groups, the top partners are fermions, and the couplings of the Higgs to other SM fields are reduced by the usual pNGB factor.

In Folded SUSY the mass of the Higgs is protected by an accidental low-energy SUSY limit, with the SM fields and their superpartners  charged under different $SU(3)$ gauge groups. Because superpartners of the SM fermions couple to the SUSY Higgs fields in the usual way they must carry EW charges. Thus, in this model the top partners are EW charged scalars.

These models, and related constructions like Quirky Little Higgs \cite{Cai:2008au}, include a \emph{mirror sector}\footnote{We shall use the terms `mirror sector' and `mirror particles' to refer to the new sectors and particles in all  uncolored top partner theories.}  with its own strong force under which the top partners are charged. This leads to either mirror baryons or, in the absence of light $SU(3)_\text{B}$-charged matter, glueballs (see \cite{Morningstar:1999rf, Juknevich:2009ji, Juknevich:2009gg}) at the bottom of the mirror sector spectrum, connecting  Hidden Valley phenomenology \cite{Strassler:2006im,Strassler:2006ri,Strassler:2006qa,Han:2007ae,Strassler:2008bv,Strassler:2008fv} to naturalness. Another commonality in all these scenarios is that they only solve the \emph{little hierarchy problem} at one-loop level, necessitating a more complete theory at $\mathcal{O}(5 - 10 \tev)$.

There has been much interest in generalizing these models and understanding their underlying mechanisms \cite{Craig:2014aea, Craig:2014roa}, as well as exploring more UV-complete implementations within a supersymmetric \cite{Craig:2013fga}, Randall-Sundrum \cite{Geller:2014kta}, composite Higgs \cite{Batra:2008jy,Barbieri:2015lqa, Low:2015nqa}, or deconstructed \cite{Craig:2014fka} setup. The mirror sectors can have important cosmological consequences, such as providing a dark matter candidate (see \cite{Garcia:2015loa,Craig:2015xla, Garcia:2015toa,Farina:2015uea} for recent studies within the Twin Higgs framework), or leading to detectable gravitational waves \cite{Schwaller:2015tja}. There may even be connections to the SM neutrino sector \cite{Batell:2015aha}. Future high-energy colliders, dark matter experiments, and cosmological observations might explore this rich new world, but what can the LHC teach us now? 

A few phenomenologial studies of uncolored naturalness exist \cite{Burdman:2008ek, Burdman:2014zta}, but these were not focused on the vital third generation partners. Recently, the authors of~\cite{Craig:2015pha} pointed out the exciting connection between exotic Higgs decays and uncolored top partner models. In this paper we will explore that direction further, and place it in the broader context of what experimental signals are ``required'' by uncolored naturalness.

In order to understand the experimental signatures of naturalness, \emph{broadly defined}, it is informative to classify existing (or possibly existing) theories according to the physical properties of the top partners, namely SM gauge charge and spin (which corresponds to the kind of symmetry protecting the Higgs). It is these physical properties that give rise to a set of experimental signatures which, as we will argue, hold the promise of discovering every theory of uncolored naturalness.

\begin{table}
\begin{center}
\includegraphics[width=12cm]{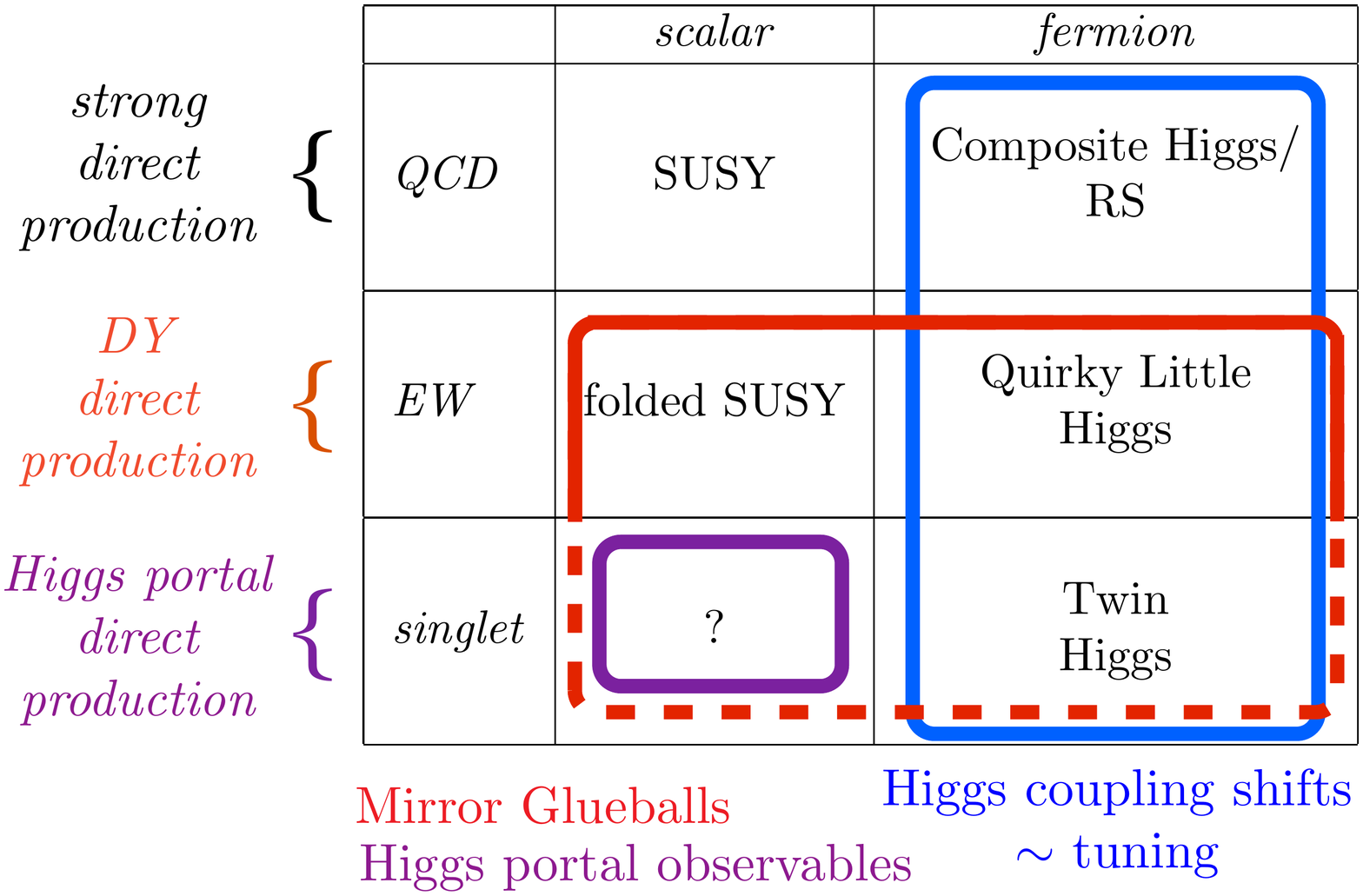}
\end{center}
\vspace{-1.8cm}
\caption{
The ``theory space'' of solutions to the hierarchy problem with top partners, organized by SM gauge charge and spin, with a representative model example in each field. The gauge charge dictates the direct top partner production mode, which makes the LHC suitable for discovery of colored top partners. For uncolored top partners, mirror glueballs are highly favored for EW-charged mirror sectors, and possible for singlet top partners. Higgs coupling shifts of same order as tuning are present in all known fermionic top partner theories. Together, these two signatures allow discovery of all known uncolored top partner theories. A hypothetical ``singlet-stop'' theory is indicated with a question mark, and would have to be discovered by either probing the UV completion or, for partner masses of a few 100 GeV, with Higgs portal observables (see text).
}
\label{t.modeltable}
\end{table}

\tref{modeltable} shows the grid of top partner SM charges and spins, with a specific model in each field acting as a representative of its class of top partner theories. The top partner production channel is given by the gauge charge. For theories of uncolored naturalness, the `smoking-gun' type signature that would lead to the discovery of that particular model class is indicated via the colored boxes. 

All known theories with fermionic top partners assume the Higgs is a pNGB. This leads to unavoidable \emph{Higgs Coupling Shifts} relative to the SM. They are generated at tree-level and are of the same order as the tuning in the theory.
The LHC will only be sensitive to $\mathcal{O}(10\%)$ deviations, but future lepton colliders like the ILC, TLEP or CEPC will constrain these coupling at the sub-percent level~\cite{Burdman:2014zta, Dawson:2013bba}. Therefore, natural fermionic-top-partner solutions to the hierarchy problem should produce measurable deviations. While diagnosing the details of the theory might be challenging, possibly requiring access to the UV completion with a 100 TeV collider, these couplings serve as a smoking-gun for the discovery of e.g. Twin-Higgs type theories. 
(Higgs coupling deviations are, of course, also generated at loop-level if the mirror sector has any SM charge. However, the small size of these deviations make them an unlikely discovery channel for partner masses above a few 100 GeV~\cite{Burdman:2014zta}.)

A more declarative signature is \emph{mirror glueballs}.
Our understanding of confining pure gauge theories has advanced significantly since the first uncolored naturalness theories were proposed \cite{Morningstar:1999rf, Juknevich:2009ji, Juknevich:2009gg}. A Minimal (``Fraternal'') Twin Higgs setup without light first and second generation partners~\cite{Craig:2015pha} could have glueballs at the bottom of the mirror sector spectrum. These glueballs couple to the visible-sector Higgs through a top partner loop, leading to glueball production in exotic Higgs decays, with subsequent glueball decay to two SM fermions via an off-shell Higgs. 
While the quantitative phenomenological details of this signature were not fully explored, it is clear that the corresponding decay lifetimes can be in the observable range, leading to striking LHC signatures.  
 
As exciting as this experimental signature is, it is not a requirement for generic Twin-Higgs type models---the SM-singlet sector could easily have relatively light quarks, making for a hadron spectrum more like that of the visible sector.
 On the other hand, mirror glueballs, and their associated signals, are a \emph{requirement} for uncolored naturalness theories with EW-charged mirror sectors, like Folded SUSY or Quirky Little Higgs. This is due to LEP limits forbidding BSM particles with EW charge lighter than about 100 GeV \cite{Beringer:1900zz}. If the structure of the mirror sector is based on our own, it cannot contain very light strongly interacting matter, resulting in glueballs at the bottom of the mirror-QCD spectrum. Crucially, this makes mirror glueball signals the smoking-gun discovery signal for Folded-SUSY type theories.

It is interesting to think about the empty square in \tref{modeltable}. So far, no explicit theory with SM-singlet scalar top partners has been proposed. If such a theory existed, and there were no other SM-charged states required near the weak scale, discovery could be quite difficult. In a Folded-SUSY like spectrum with weak-scale soft masses we might again expect the existence of mirror glueballs, with their accompanying experimental signatures. If, however, the mirror sector contains light matter or mirror-QCD was broken, discovery would have to proceed through \emph{Higgs-portal observables}: invisible direct top partner production $h^* \to \tilde t \tilde t$ \cite{Curtin:2014jma, Craig:2014lda}, Higgs cubic coupling shifts \cite{Curtin:2014jma, Curtin:2015bka} at a 100 TeV collider, or sub-percent $\sigma_{Zh}$ shifts at a lepton collider \cite{Craig:2013xia}. In each case, the currently understood sensitivity extends only to singlet stop masses of about 300 GeV for the possible future machines under consideration (depending on the number and coupling structure of the partners). If the partners are heavier, we must rely on probing the UV completion to discover these hypothetical models.  

In constructing this picture we assume the existence of a mirror QCD under which the mirror sector is charged. This ensures similar running of the visible and hidden sector Yukawa couplings to protect the one-loop cancellation (see \cite{Craig:2015pha} for a recent discussion), but depending on the UV completion scale this is not technically essential.\footnote{There are uncolored top partner theories without mirror color \cite{Poland:2008ev}, but in those models the scale of the UV-completion, defined broadly as the scale where additional states appear, is only a few TeV in the fully natural case.}  
We also focus on signatures that are due to the $3^\mathrm{rd}$ generation partners, because of their direct link to the little hierarchy problem. Other signatures (like electroweak precision tests, or direct production of the first two generations and subsequent quirky annihilation~\cite{Burdman:2008ek}) are certainly possible. Finally, it is likely possible to engineer theories that avoid the indicated smoking-gun signatures. Nevertheless, this generic expectation gives an instructive overview of the experimental potential for probing uncolored naturalness. 
	
Given the importance of mirror glueball signatures, we here set about studying their phenomenology in detail. We find that, for representative mirror sectors giving rise to glueball signatures, the lightest state is favored to have a mass in the $\sim 10 - 60 \gev$ range, which permits production via exotic Higgs decays. There are still important uncertainties in our understanding of pure glue dynamics, most importantly possible mixing effects between glueballs and the Higgs and details of hadronization. We outline how to effectively account for these unknowns in a collider study, and demonstrate that concrete sensitivity predictions can still be made.

In the course of conducting our collider analyses we make use of efficiency tables for the reconstruction of displaced vertices supplied by the ATLAS studies \cite{Aad:2015asa, Aad:2015uaa}. We hope that this simple method for estimating signal yield can serve as a template for future theory studies of scenarios involving long-lived particles.\footnote{The data driven techniques employed by \cite{Halyo:2013yfa} may also aid in further studies.}

We estimate the sensitivity of the LHC to discover these mirror glueballs, and find sensitivity to~$\sim$~TeV top partners (scalar or fermion) across the entire mass range with $3000\ifb$ of luminosity. This assumes present-day detector capabilities, which is likely to be conservative. 
New searches would be required to achieve this coverage, but our results provide strong  motivation to implement the required experimental analyses, some of which require the reconstruction of displaced vertices $50 \mu$m from the interaction point.\footnote{The issue of how to trigger on exotic Higgs decays is a pressing one (see e.g.~\cite{Curtin:2013fra, Craig:2015pha}), but we show that standard trigger strategies give significant reach to uncolored naturalness models.}

We also estimate the reach of a 100 TeV collider by scaling the same searches to higher energy. Due to likely superior triggering and reconstruction capabilities compared to the LHC, these estimates are very pessimistic. Even so, they demonstrate the impressive reach achievable at such a machine.

There is potential for exciting complementarity between experimental signatures of uncolored naturalness. 
Top partner direct pair production and annihilation could not only produce another detectable glueball signal, but also allow hidden sector masses and couplings to be determined, testing the solution to the little hierarchy problem. 
Higgs coupling measurements would independently hint at the mass of fermionic top partners. 
Finally, the mirror sector's connections to cosmology might also be probed: the existence of glueballs implies an absence of light quark flavors, which gives rise to a strong first-order chiral phase transition in the early universe. In that case there may be detectable gravitational-wave signals \cite{Schwaller:2015tja}. 
Correlating these cosmological and LHC signals would serve as a powerful diagnostic of the mirror sector dynamics.

This paper is structured as follows. \sref{models} reviews the prototypical models of colorless naturalness and establish notation for subsequent analyses. We then describe, in \sref{glueballs}, the expected spectrum and properties of the mirror glueballs associated with these models. In \sref{higgsdecays} we find the expected experimental reach for these models from exotic decays of the Higgs into mirror glueballs. We then give a brief preliminary discussion of the signals generated by the direct production of uncolored top partners in \sref{toppartners} and conclude in \sref{conclusion}.

\section{Models of Uncolored Naturalness}\label{s.models}

In this section we briefly review the salient features of three representative theories of uncolored naturalness: Folded SUSY, Twin Higgs, and Quirky Little Higgs. Each of these solves the little hierarchy problem by protecting the Higgs mass from large corrections at one-loop, up to some cutoff $\Lambda\sim \mathcal{O}(5-10) \tev$. All of them require a more complete theory at higher scales, with supersymmetry or compositeness addressing the full Hierarchy Problem beyond the cutoff of the low-energy description. 

Mirror glueballs arise automatically in Folded SUSY and Quirky Little Higgs theories. Certain `minimal' incarnations of the Twin Higgs setup can feature them as well. This motivates our study of exotic Higgs decay for probing uncolored naturalness.

\subsection{Folded SUSY\label{ss.fsusydef}}

In typical supersymmetric extensions of the SM every known particle has partner particles (or superpartners) associated with it. These partners differ from their SM counterpart by an half unit of spin, but otherwise carry the same quantum numbers. In particular, the gauge structure of the theories commutes with supersymmetry so the particles and their superpartners carry identical gauge charges.

In Folded SUSY \cite{Burdman:2006tz} theories the superpartners are not charged under SM color, but carry a different $SU(3)_\text{B}$ charge. This can be arranged in the context of a 5D theory with the extra dimension of radius $R$ compactified over a $S_1/\mathbb{Z}_1$ orbifold with branes at the orbifold fixed points.\footnote{For an alternative, and 4D, construction see \cite{Craig:2014fka}.} The gauge structure is $SU(3)_\text{A}\times SU(3)_\text{B}\times SU(2)_\text{L}\times U(1)_\text{Y}$, along with a $\mathbb{Z}_2$ symmetry that exchanges the vector superfields of the two $SU(3)$s. 
The 5D $\mathcal{N}=1$ SUSY corresponds to $\mathcal{N} = 2$ in 4D, which is broken by boundary conditions on each brane to a different $\mathcal{N}=1$ through the Scherk-Schwarz mechanism~\cite{Scherk:1978ta}. This breaks $\mathcal{N} = 2 \to \mathcal{N} = 0$ globally. 

The boundary conditions of the fields are chosen such that the light quark states identified with SM are charged under $SU(3)_\text{A}$ while the squarks that cancel the quadratic Higgs mass corrections of the quarks are charged under $SU(3)_\text{B}$. The boundary conditions of the gluinos ensure that they are not part of the low energy theory. The (s)lepton and EWino spectrum is the same as in the MSSM, with soft masses given by the scale of the extra dimension in the minimal model.

Therefore, the low lying spectrum is made up of SM field zero modes and the scalar zero modes of the minimal supersymmetric standard model (MSSM).\footnote{Note that there is no tuning associated with splitting the folded squarks from the gauginos. In the extra dimensional construction the folded squarks have zero mass tree level; the scale of their soft masses is given by gauge (and Higgs for the third generation) interactions, see \cite{Burdman:2006tz} for the explicit soft masses. Thus, the squarks are generally lighter than the gauginos by the gauge coupling squared divided by a loop factor.} The MSSM scalars are charged under the mirror $SU(3)_\text{B}$, but the $SU(2)_\text{L} \times U(1)_\text{Y}$ gauge group is shared.  In this or similar constructions, the electroweak charges of the superpartners are unavoidable, because the cancellation of Higgs mass quadratic divergence proceeds via the usual SUSY mechanism, so the stops charged under $SU(3)_\text{B}$  must couple to the SM-like Higgs identically to stops in the usual MSSM.

The mirror-stops are not charged under SM color, which drastically reduces their production cross section at hadron colliders compared to colored stops in standard SUSY theories. However, because they still carry EW charge, LEP limits require that they not be lighter than $\sim 100 \gev$. The lightest states in the mirror sector are therefore the $SU(3)_\text{B}$ glueballs, which can be produced in decays of the SM-like Higgs and lead to the discovery signature discussed in this work. 

Direct production of the mirror squarks in Drell-Yan-like processes and subsequent annihilation through quirky dynamics may yield additional discovery signatures \cite{Burdman:2008ek}, including $W\gamma$ resonances at the LHC~\cite{Burdman:2014zta}. Possible glueball signals from these processes are briefly discussed in \sref{toppartners}.

\subsection{Twin Higgs\label{ss.twinhiggsdef}}

The Twin Higgs model \cite{Chacko:2005pe} posits a copy of the SM, called the Twin (or mirror) sector, along with a discrete $\mathbb{Z}_2$ symmetry that exchanges the two. Additionally, the total Higgs sector, including both SM and Twin fields, is approximately invariant under a a global symmetry, either $SU(4)\times U(1)$ or $SO(8)$. This symmetry is spontaneously broken when one of the fields gets a VEV $f/\sqrt{2}$. The SM Higgs is then identified as one of the pNGBs of the broken symmetry.

This global symmetry is explicitly broken by the gauging of the Higgs doublet in the SM and Twin sectors and by Yukawa couplings, making the mass of the Higgs susceptible to quadratic corrections that depend on gague or fermion loops. However, the $\mathbb{Z}_2$ symmetry guarantees that the SM and Twin contributions cancel in each case, thus the mass of the Higgs is protected from large 1-loop contributions.

In general the VEVs of the Higgs field $v_\text{A}/\sqrt{2}$ and its Twin $v_\text{B}/\sqrt{2}$ satisfy $f^2=v_\text{A}^2+v_\text{B}^2$. We characterize how much of the total VEV is in each sector by defining
\begin{equation}
\begin{array}{cc}
\displaystyle v_\text{A}\equiv f\sin\left(\frac{v}{f} \right)=f\sin\vartheta, & \displaystyle v_\text{B}\equiv f\cos\left(\frac{v}{f} \right)=f\cos\vartheta,
\end{array}\label{e.twinvevs}
\end{equation}
where $v_\text{A}$= 246 GeV. Of the seven pNGBs produced when the symmetry is broken, six are eaten by the vector bosons of the SM and Twin SU(2)s. The remaining pNGB is identified with the SM Higgs. If the $\mathbb{Z}_2$ symmetry is exact then $\vartheta =\pi/4$ and the Higgs boson is an equal mixture of SM and Twin sectors. The couplings $g$ of the SM particles to the Higgs are $g=g_{\text{SM}}\cos\vartheta$, making such are large value of $\vartheta$ excluded by experiment.

A soft breaking of the $\mathbb{Z}_2$, however, allows the Higgs mass to be protected while setting $\vartheta\ll 1$ and thus making the observed Higgs a mostly SM object. This same difference in the VEVs raises the masses of the Twin fermions. The masses of a SM fermion $F$ and its twin $F_T$ are given by
\begin{equation}
\label{e.cotvscaling}
\begin{array}{cc}
\displaystyle m_F=y_F\frac{f\sin\vartheta}{\sqrt{2}}, & \displaystyle m_{F_T}=y_F\frac{f\cos\vartheta}{\sqrt{2}},
\end{array}
\end{equation}
respectively. Or, expressed another way
\begin{equation}
m_{F_T}=m_F\cot\vartheta\approx m_F\cdot \frac{f}{v},\label{e.twinbmasses}
\end{equation}
where the final relation is taken in the $v\ll f$ limit. While we do not show it here, the masses of the SM and Twin gauge bosons are related by the same factor.

Breaking the $\mathbb{Z}_2$ by just the right amount to give small $v$ constitutes a tuning of the model. 
Twin top loops give the dominant contribution to the Higgs mass parameter:
\begin{equation}
\delta m^2=\frac{3y_t^2m_T^2}{4\pi^2}\ln\left(\frac{\Lambda}{m_T} \right),
\end{equation}
with $\Lambda$ the cutoff of the model. For $\Lambda$ of a few TeV this implies that the tuning goes like
\begin{equation}
\left|\frac{\delta m^2}{m^2} \right|^{-1} \sim \mathcal{O}\left(\frac{v^2}{f^2}\right).
\end{equation}

While the first incarnation of this model assumed a full doubling of the SM, this is not the minimal model required by naturalness. The so-called Fraternal Twin Higgs \cite{Craig:2015pha} is such a minimal model and only includes the third generation in the mirror sector. In such a model, the lightest mirror states may be glueballs of $SU(3)_\text{B}$, depending on the mass of the mirror bottom quark.

\subsection{Quirky Little Higgs\label{ss.quirkylittlehiggsdef}}

The Quirky Little Higgs model shares features with both Twin Higgs and Folded SUSY. As in Twin Higgs, the mass of the Higgs is protected by its being a pNGB of a spontaneously broken symmetry. This leads to similar reductions of the Higgs couplings relative to the SM by the factor $\cos\vartheta$ and thence to the same induced coupling of the Higgs to mirror gluons.

However, the entire SM is not copied in this model. There is instead a 5D construction, with extra dimension called $y$, similar to Folded SUSY's including the $S_1/\mathbb{Z}_1$ orbifold and branes at $y=0$ and $y=\pi R$. The bulk gauge structure is $SU(6)\times SU(3)_\text{W}\times U(1 )_\text{x}$ which (before any fields get VEVs) is also preserved at $y=\pi R$. Boundary conditions on the $y=0$ brane break the gauge symmetry to $SU(3)_\text{A}\times SU(3)_\text{B}\times SU(2)_\text{L}\times U(1)_\text{Y}$. This includes the SM electroweak and color gague groups along with a new $SU(3)_\text{B}$.

The $SU(3)_\text{W}$ gauge group is also broken by a field $\Phi$, which lives on the $y=0$ brane, getting a VEV. This spontaneous breaking of the symmetry leads to 5 pNGBs, 4 of which are identified with the SM Higgs doublet. The matter content of the theory is composed of fields that transform as fundamentals of both $SU(6)$ and $SU(3)_\text{W}$. This effectively doubles the matter content of the usual Little Higgs model, one set charged under SM color and the other under $SU(3)_\text{B}$. Orbifold boundary conditions are chosen such that the low energy spectrum of the third generation consists of the SM top quark $t$ and the top quirk $T$.

As in Folded SUSY, the quadratic corrections to the Higgs mass of the SM fermions are canceled by partners charged under a different $SU(3)$, but carry the same electroweak charges. This means that they can be produced directly at colliders and give rise to phenomenology similar to Folded SUSY. The most significant difference is that the top partners are fermions, which also leads to tree-level Higgs coupling deviations as for the Twin Higgs.

\section{Mirror Glueballs\label{s.glueballs}}

As outlined above, the mirror sector of many theories of uncolored naturalness is pure gauge $SU(3)_\text{B}$ at low energies. In this section, we briefly review the resulting mirror glueball spectrum and derive the range of glueball masses favored by renormalization group (RG) evolution for a range of mirror sectors. We find that mirror glueballs, if they exist, are likely to have masses in the $\sim$ 10 - 60 GeV range. We also show the form of the effective mirror-glue coupling to the visible SM-like Higgs through a top partner loop, and discuss the resulting mirror glueball and Higgs decays.

\subsection{Spectrum}
\label{ss.spectrum}

The low-energy spectrum of a pure $SU(3)$ gauge theory was computed on the lattice by \cite{Morningstar:1999rf, Chen:2005mg}. 
There are 12 stable (in the absence of other interactions) $J^{PC}$ eigenstates, shown in \fref{glueballspectrum}. 
Masses are given as multiples of $m_0$, the mass of the $0^{++}$ scalar glueball state at the bottom of the spectrum. In terms of the familiar $\overline{\mathrm{MS}}$ QCD confinement scale, $m_0 \approx 7 \Lambda_{\text{QCD}}$ to a precision of about 5\%. The other glueball masses as a multiple of $m_0$ are known to $\sim$ few $\%$ or better.
Above $\sim $ $2$ - $3$ $m_0$ there is a continuum of glueball states that decay down to the 12 stable states shown in \fref{glueballspectrum}. Hadronization will be discussed in \ssref{exotichiggsdecays}.

The mass of the mirror glueballs is entirely determined by the running of the strong coupling constant in the B sector. Given the matter content of a mirror sector we can compute the running of the mirror strong coupling $\alpha_s^\text{B}(\mu)$ using the standard one-loop beta function.
Define $\mu^\text{B}_\mathrm{pole}$ as the scale satisfying ${\alpha_s^\text{B}(\mu^\text{B}_\mathrm{pole})}^{-1} = 0$, and similarly $\mu^\text{A}_\mathrm{pole}$ for the visible sector. The A-sector beta function is also computed at one-loop and matched to the measured value of $\alpha_s(m_Z)$. The mirror glueball mass can then be obtained using the result of \cite{Chen:2005mg}:
\begin{equation}
\label{e.m01}
m_0 \ =  \  a_0 \ \cdot \ r_0^{-1} \ \  \ \ \ , \ \ \ \ \ \ a_0 = 4.16 \pm 0.12 \ \ 
\end{equation}
where $r_0^{-1}$ is the hadronic scale, with $(r_0^\mathrm{SM})^{-1} = 410 \pm 20 \mev$. Rather than computing $r_0$ in the mirror sector directly from the running gauge coupling, which would require a more sophisticated treatment than one-loop RGEs, we estimate it by a simple rescaling of the one-loop Landau poles. This gives
\begin{equation}
 \label{e.m0calculation}
 m_0 \ = \ a_0 \  \cdot \ (r_0^\mathrm{SM})^{-1}  \frac{\mu_\mathrm{pole}^\text{B}}{\mu_\mathrm{pole}^\text{A}} \ \  .
\end{equation}
We can then compute the well-motivated range of $m_0$ in several representative examples of uncolored naturalness theories, showing that glueball masses below half the Higgs mass are strongly theoretically favored. Unless otherwise noted, both $a_0$ and $r_0^\mathrm{SM}$ are taken at their central value for these estimates.

\begin{figure}
\begin{center}
\includegraphics[width=10cm]{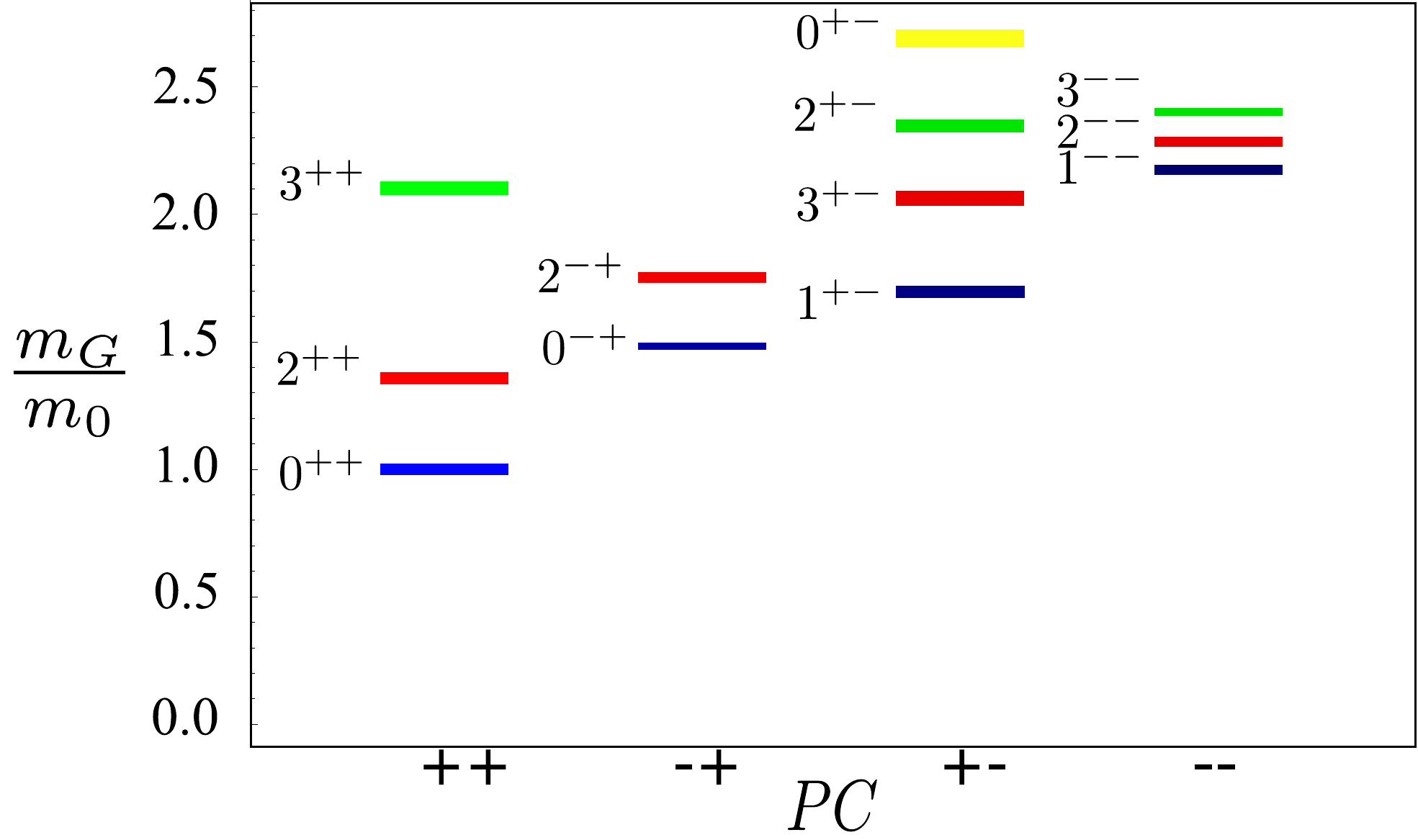}
\end{center}
\caption{
Spectrum of glueballs in pure $SU(3)$ theory \cite{Morningstar:1999rf}, arranged by $J^{PC}$ quantum numbers. Plot taken from \cite{Juknevich:2009gg}. Masses given in units of $m_{0^{++}} = m_0 \approx 7 \Lambda_\mathrm{QCD}$.
}
\label{f.glueballspectrum}
\end{figure}

\subsubsection*{Folded SUSY}

In Folded SUSY without soft masses or Yukawa terms, the KK-states of A-quarks (B-squarks) have masses $\{0, 1/R, 2/R, \ldots\}$, while the A-squarks (B-quarks) have masses of $\{1/(2R), 3/(2R), \ldots\}$. Both sectors have identical gauge-KK-towers, with no zero-mode gauginos. At each threshold $n/(2R)$ the A- and B-states have different spin but identical gauge quantum numbers and multiplicities. Their contributions to the $\alpha_s^{\text{A,B}}$ beta-functions are identical, so the two $SU(3)_{\text{A,B}}$ strong interactions have identical couplings.\footnote{Small differences are introduced at two-loop \cite{Machacek:1983tz}, but this should not affect our estimate of $m_0$ in a significant way.} 

The introduction of soft masses and Yukawa terms results in very small shifts to the KK-towers, assuming $m_\mathrm{KK}^2 \gg m_\mathrm{soft}^2, m_\mathrm{Yukawa}^2$. The most significant effect is the lifting of zero modes. Assuming the largest B-squark soft mass to be larger than the top mass in the A-sector, the two strong couplings track each other from some UV-completion scale $\mu = \Lambda_{\text{UV}}$ down to $\mu =m_{\mathbb{Z}_2}$, which we designate as the scale (near the largest B-squark soft mass) where the $\mathbb{Z}_2$ symmetry between the two strong couplings is broken. 

Without knowing the soft mass spectrum of the theory it is impossible to predict the mirror glueball mass $m_0$ precisely. 
However, it is possible to highlight the range $m_0$ can take. 
Heavier mirror sector soft masses lead to heavier glueballs, since less light matter causes the mirror-QCD to confine more quickly as we approach the infrared (IR). Therefore, we find the most probable range of glueball masses by considering opposite extremes of the possible particle masses. 

Without loss of generality, $\tilde t_1^\text{B}$ can be designated as the lightest B-squark zero mode. Its mass, $m_{\tilde t_1}$, sets the bottom of the mirror sector matter spectrum. For a given $m_{\mathbb{Z}_2}$ where $\alpha_s^\text{A}(m_{\mathbb{Z}_2}) = \alpha_s^\text{B}(m_{\mathbb{Z}_2})$, we then compute $\alpha_s^\text{B}(\mu)$ at one-loop order for two scenarios: one where all the B-squarks except $\tilde t_1$ have mass $m_{\mathbb{Z}_2}$, and one where all B-squarks are degenerate with $\tilde t_1$. The resulting minimum and maximum values of $m_0$ for different $m_{\mathbb{Z}_2}$ are shown in \fref{MinMaxm0massfoldedsusy}. For values of $m_{\mathbb{Z}_2}$ up to 20 TeV, which is very high considering the Higgs mass is only protected at one-loop, the glueball mass ranges from $\sim 12 - 55 \gev$.  While these extremes do not represent the most natural realizations of the framework, they span the lightest to heaviest glueball possibilities, so that the more motivated models lie within these boundaries. For instance, if instead of keeping only the lightest stop light we keep the entire third generation at the light stop mass (a ``Natural SUSY"-like construction) then the upper glueball mass bound is lowered slightly to 50 GeV.

\begin{figure}
\begin{center}
\includegraphics[width=16cm]{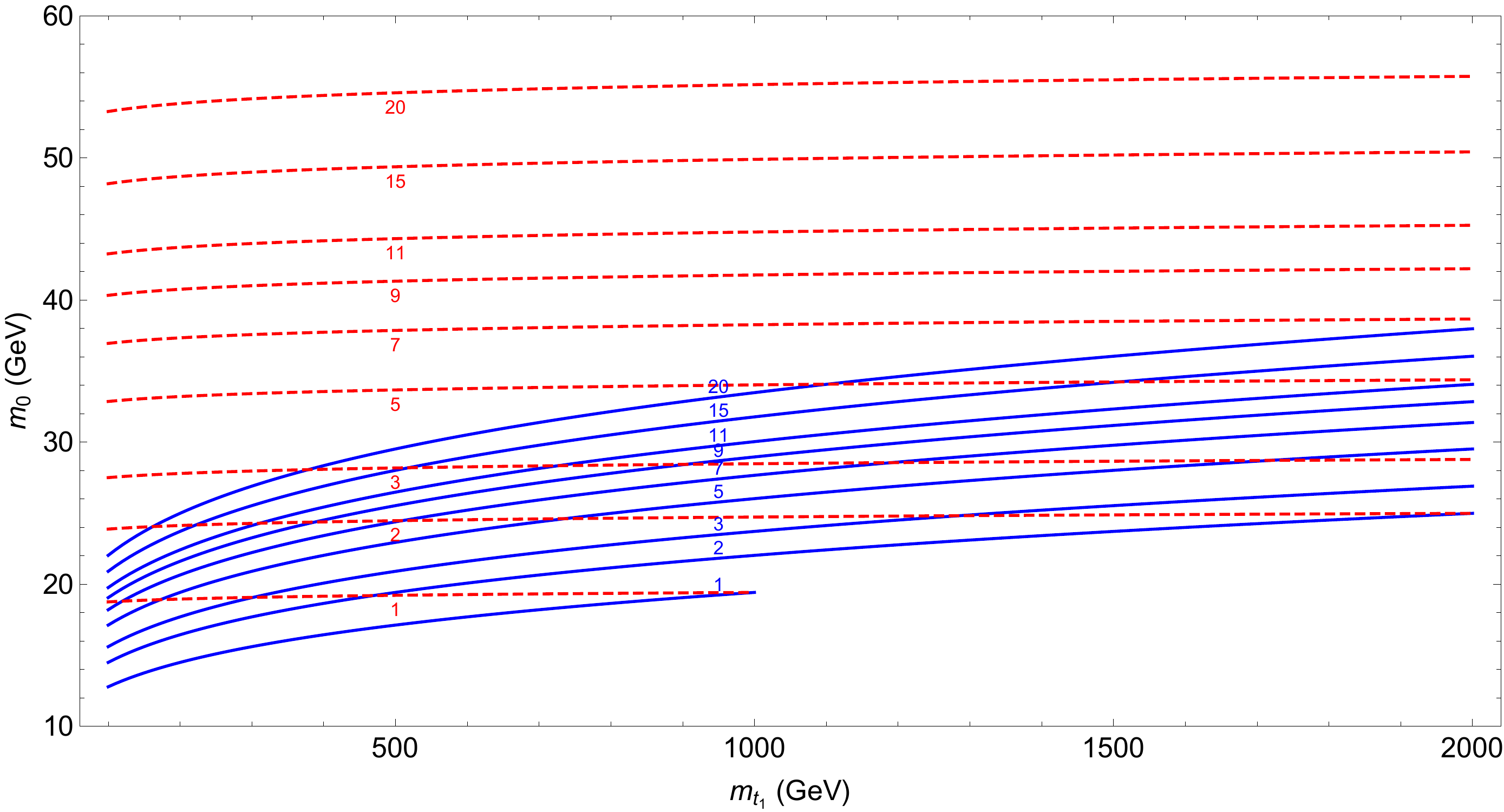}
\end{center}
\caption{
The minimum (blue) and maximum (red dashed) glueball mass $m_0 \approx 7 \Lambda_\mathrm{QCD'}$ as a function of the lightest B-squark mass in Folded SUSY, taken without loss of generality to be $\tilde t_1^\mathrm{B}$. Different contours correspond to $m_{\mathbb{Z}_2}$ varying from 1 to 20 TeV. For the minimum glueball mass, all squarks were taken to have mass $m_{t_1}$. For the maximum glueball mass, all squarks except $\tilde t_1$ were taken to have mass $m_{\mathbb{Z}_2}$. 
}
\label{f.MinMaxm0massfoldedsusy}
\end{figure}

\subsubsection*{Twin Higgs}

In the original Mirror Twin Higgs model, the entire SM fermion spectrum is duplicated in the B sector. In that case there are no glueballs in the mirror sector, because the $u_\text{B}, d_\text{B}, s_\text{B}$ and possibly also the $c_\text{B}$ quarks would be lighter than $m_0$.

Departing from the exact mirror symmetry assumption, a large variety of hidden sector spectra are possible. This makes it impossible to predict without additional information whether mirror glueballs are realized in the Twin Higgs framework, and at what masses. Even so, we can demonstrate that glueballs below half the Higgs mass are a plausible and well-motivated possibility.

\begin{figure}
\begin{center}
\hspace*{-10mm}
\includegraphics[width=16cm]{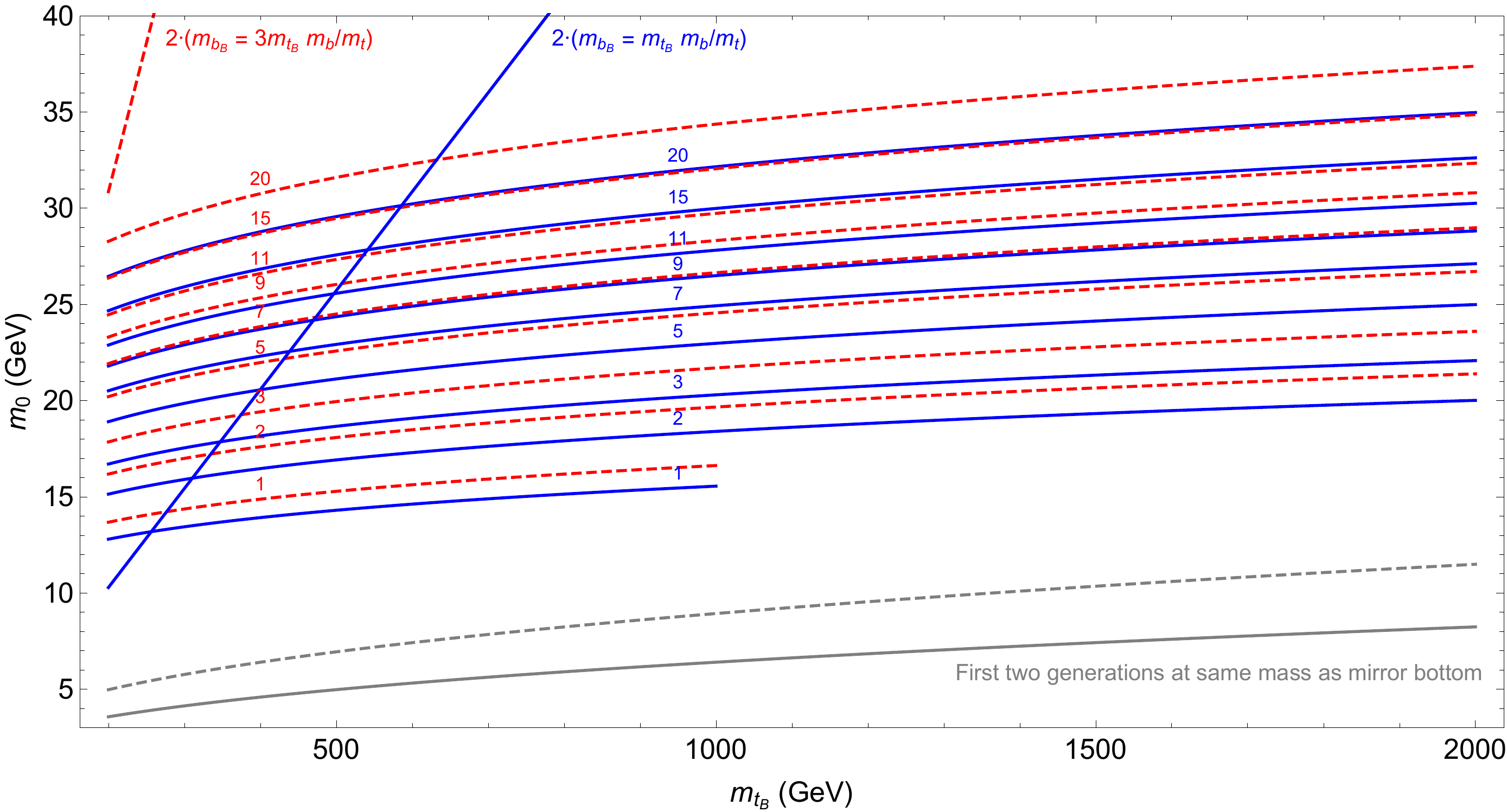}
\end{center}\vspace{-5mm}
\caption{
Glueball masses in Twin Higgs models where the mirror symmetry is broken for the first two quark generations and optionally also the bottom quark, as a function of $m_{\mathbb{Z}_2}$ (contour labels) and the mirror top mass $m_{t_\text{B}}$.
Blue (dashed red): for $m_{Q_{\text{B}_{1,2}}} = m_{\mathbb{Z}_2}$ and $m_{b_\text{B}} = r  \cdot m_{t_\text{B}} \frac{m_b}{m_t}$, with $r = 1$ (3). 
Gray (dashed gray): for $m_{Q_{\text{B}_{1,2}}} = m_{b_\text{B}} = r  \cdot m_{t_\text{B}} \frac{m_b}{m_t}$, with $r = 1$ (3). (In this case, there is no dependency on $m_{\mathbb{Z}_2}$ since all mirror states are light.)
Note that glueball states only exist if they are lighter than approximately twice the lightest hidden sector quark (straight lines), otherwise the hidden sector QCD spectrum consists of quarkonia states. 
}
\label{f.m0massMinimalTwinHiggs}
\end{figure}

For example, the Fraternal Twin Higgs model~\cite{Craig:2015pha} contains only third-generation $b_\text{B}, t_\text{B}$ quarks in the mirror sector, which is sufficient to preserve the Twin Higgs mechanism. Assuming again the two QCD forces to unify $\alpha_s^\text{A}(m_{\mathbb{Z}_2}) = \alpha_s^\text{B}(m_{\mathbb{Z}_2})$, we can then calculate the glueball mass as a function of of $m_{t_\text{B}}$ and the scale $m_{\mathbb{Z}_2}$ as for Folded SUSY above. This is shown as the blue contours in \fref{m0massMinimalTwinHiggs}~(top), and motivates glueballs in the $\sim 12 - 35 \gev$ range.

The glueball mass has to be below approximately twice the mirror bottom mass $m_{b_\text{B}} = m_{t_\text{B}} \frac{m_b}{m_t}$ for glueball states to form.\footnote{See \cite{Craig:2015pha} for a careful discussion of the relative importance of mirror bottomonium and glueballs in the Fraternal Twin Higgs model. We avoid these complications here and focus on the regime where glueball states dominate the low-energy hidden sector.} Therefore, in the above scenario, there are no glueballs for mirror tops lighter than about 400 GeV. However, Twin Higgs top partners of such low masses will lead to Higgs coupling deviations greater than 20\%, which will be effectively probed by LHC run 2 \cite{Burdman:2014zta}. 

It is possible to break the mirror symmetry even further and allow the mirror $b$-quark to depart from the $\mathbb{Z}_2$ prediction by a modest amount. The red dashed contours in \fref{m0massMinimalTwinHiggs} (Top) show the glueball mass if the mirror bottom mass is enhanced by a factor of 3. The effect on the glueball mass is minor, but glueballs can now exist for mirror tops as light as 200 GeV. Similarly, the mirror $b$-quark can be lighter than expected, which would decrease the parameter space with glueballs at the bottom of the spectrum.

Another possible scenario is a non-maximally broken mirror symmetry for the first two quark generations~---~rather than completely removing them from the spectrum (i.e. pushed to $m_{\mathbb{Z}_2}$), they could merely be significantly heavier than expected by the $\cot \vartheta$ scaling of the B sector masses, \eref{twinbmasses}. The glueball mass for this scenario, where $m_{Q_b} = r  \cdot m_{t_\text{B}} \frac{m_b}{m_t}$ with $r = 1$ or $3$, is shown as gray lines in \fref{m0massMinimalTwinHiggs}. This leads to glueballs with a mass of a few $- 10$ GeV, with no dependence on $m_{\mathbb{Z}_2}$.

Finally, it is possible that the $\mathbb{Z}_2$ symmetry is only approximate at $m_{\mathbb{Z}_2}$ (possibly due to threshold effects at even higher scale). A 10\% difference between $g_3^A$ and $g_3^B$ can change the glueball masses by about an order of magnitude in either direction compared to the above predictions~\cite{Craig:2015pha}. That being said, if the deviations are the typical size of threshold corrections at $m_{\mathbb{Z}_2}$, then the range of possible glueball masses is similar to that obtained in Folded SUSY. 

In summary, while the Twin Higgs framework makes it difficult to predict the eixstence or mass of mirror glueballs, there are many scenarios, including the Fraternal Twin Higgs, where glueballs in the $\sim 10 - 60 \gev$ mass range arise. This justifies our close examination of  glueballs in this window.

\subsubsection*{Quirky Little Higgs}

The mirror sector spectrum of the Quirky Little Higgs \cite{Burdman:2008ek} framework is similar in feel to Folded SUSY, containing a fermionic partner for each SM quark. All but the top partner, however, are given bulk masses to remove them from the low energy spectrum. This is phenomenlologically motivated by LEP limits and  the   EW charge of the B sector partners. This results in approximately the same range of preferred $m_0$ values as the Folded SUSY setup described above.

\subsection{Mirror gluon coupling to SM-like Higgs Boson}
\label{ss.yoverM}

The visible SM-like Higgs couples to mirror gluons through a top partner loop, in exact analogy to its coupling to visible gluons through a top loop. Assuming the top partner to be significantly heavier than $m_h/2$, this interaction can be described by an effective dimension-6 operator:
\begin{eqnarray}
\nonumber \delta \mathcal{L}^{(6)} &=& \frac{\alpha_s^\text{B}}{3 \pi} \left[ \frac{y^2}{M^2} \right] \  |H|^2 \ {G}^{(\text{B})}_{\mu \nu}  {{G}^{(\text{B})}}^{\mu\nu}
\\
\label{e.L6}
&=& 
 \frac{\alpha_s^\text{B}}{3 \pi} \left[ \frac{y^2}{M^2} \right] v \ \ h \ 
 {G}^{(\text{B})}_{\mu \nu}  {{G}^{(\text{B})}}^{\mu\nu}
  + \ldots 
 \end{eqnarray}
where $H$ is the SM-like Higgs doublet, $G^\text{(B)}_{\mu\nu}$ is the mirror gluon field strength, and the second line arises from the substitution $H \to (0, (v+h)/\sqrt{2})^T$. We adopt the notation of \cite{Juknevich:2009gg} and use $[y^2/M^2]$ as a coefficient that can be independently set for each theory, as we discuss below. 

Note that dimension-8 operators allowing two gluons to couple to two visible or mirror sector electroweak gauge bosons can also be generated, depending on the top partner quantum numbers. This can lead to additional decay channels opening up, but for top partners above $\sim 100 \gev$ they do not significantly contribute to the $0^{++}$ decay width~\cite{Juknevich:2009gg}, which will be our primary focus.

\subsubsection*{Folded SUSY}
The B-sector squark zero modes in Folded SUSY have a standard MSSM spectrum, with stop mass matrix
\begin{equation}
M_{\tilde t \tilde t}^2 = 
\left(
\begin{array}{cc}
\hat M_{\tilde t_L}^2 + m_t^2 & m_t X_t\\
m_t X_t & \hat M_{\tilde t_R^2} + m_t^2
\end{array}
\right),
\end{equation}
where 
\begin{equation}
\hat M_{\tilde t_L}^2 = m_{Q_3}^2 + \frac{1}{6} \cos{2 \beta} (1+2 \cos 2\theta_w) m_Z^2 
\ \ , \ \ \ \ 
\hat M_{\tilde t_R}^2 = m_{U_3}^2 + \frac{2}{3} \cos{2 \beta} \sin^2 \theta_w m_Z^2 
\end{equation}
are dominated by the Higgs independent soft masses. On the other hands, the terms depending on
\begin{equation}
m_t = \frac{1}{\sqrt{2}} \lambda_t \sin \beta v 
\ \ , \ \ \ \ \ \ 
X_t = A_t - \mu \cot \beta
\end{equation}
arise from the B-stops' interaction with the Higgs field. The leading contributions to the $hGG$ operator  in \eref{L6} are therefore (see e.g. \cite{Carmi:2012yp})
\begin{eqnarray}
\frac{y^2}{M^2} = \frac{1}{16} \frac{1}{v^2} \left[
 \frac{m_t^2}{m_{\tilde t_1}^2} +  \frac{m_t^2}{m_{\tilde t_2}^2}
 -  \frac{m_t^2 X_t^2}{m_{\tilde t_1}^2 m_{\tilde t_2}^2}\right]. \ \ \ \ \ \mbox{(Folded SUSY)}
\end{eqnarray}
A useful benchmark point is to assume $X_t = 0$ and $m_{\tilde t_1} = m_{\tilde t_2} = m_{\tilde t}$. In that case, 
\begin{equation}
\label{e.yoverMFoldedSUSY}
\frac{y^2}{M^2} = 
\frac{1}{8 v^2} \ \frac{m_t^2}{m_{\tilde t}^2}
\end{equation}
We will use the parameter $m_{\tilde t}$ to represent both stop masses in Folded SUSY.

\subsubsection*{Twin Higgs}

In the Twin Higgs model there are two Higgs doublets coupling to gluons in their sector: 
\begin{equation}
\delta \mathcal{L}^{(6)} = \frac{\alpha^\text{A}_s}{24 \pi} \frac{{y^\text{A}_t}^2}{m_{t}^2} |H_\text{A}|^2 G^\text{(A)}_{\mu\nu} G^{\text{(A)}\mu\nu} +\frac{\alpha_s^\text{B}}{24 \pi} \frac{{y_t^\text{B}}^2}{m_{T}^2} |H_\text{B}|^2 G^\text{(B)}_{\mu\nu} G^{\text{(B)}\mu\nu}.
\end{equation}
Cancellation of the quadratically divergent contributions to the light Higgs mass from the A- and B-top quarks requires $y_t^\text{A} = y_t^\text{B}$, which we assume from now on. The $SU(2)$ doublet pNGB $h$ field can be described in a non-linear sigma model (see e.g. \cite{Burdman:2014zta}):
\begin{eqnarray}
\nonumber |H_\text{B}|^2 &=& \frac{f^2}{2} - |H_\text{A}|^2 \\
 |H_\text{A}|^2 &=& \frac{f^2}{2}\sin^2\left(\frac{v+h}{f} \right)=\frac{v_\text{A}^2}{2}+hv_\text{A}\cos\vartheta+\ldots
\end{eqnarray}
where we have used the definitions in \eref{twinvevs}. The relevant dimension-6 operator coupling the visible SM-like Higgs to both visible and mirror gluons is therefore
\begin{equation}
\delta \mathcal{L}^{(6)} = |H_\text{A}|^2  \left[ \frac{\alpha^\text{A}_s}{24 \pi} \frac{{y_t}^2}{m_{t}^2}G^\text{(A)}_{\mu\nu} G^{\text{(A)}\mu\nu} -\frac{\alpha_s^\text{B}}{24 \pi} \frac{{y_t}^2}{m_{T}^2}G^\text{(B)}_{\mu\nu} G^{\text{(B)}\mu\nu}\right]
\end{equation}
The $[y/M]$ coefficient in \eref{L6} can be read off from the second term:
\begin{equation}
\label{e.yoverMTH}
\frac{y^2}{M^2} = \frac{1}{8} \frac{y_t^2}{m_T^2}\cos\vartheta = \frac{1}{4 v_{\text{A}}^2} \  \frac{m_t^2}{m_T^2}\cos\vartheta. \ \ \ \ \ \ \ \ \mbox{(Twin Higgs)}
\end{equation}

\subsubsection*{Quirky Little Higgs}

The nonlinear low energy parameterization of the Quirky Little Higgs model is nearly identical to the Twin Higgs. While there is not a complete copy of the SM, the B-sector contains a scalar $SU(2)_\text{L}$ singlet $\phi$ whose VEV is related to the EW VEV by $v_\text{EW}^2+v_{\phi}^2=f^2$, just like Twin Higgs. The induced coupling of $\phi$ to mirror gluons is then related to the coupling of the Higgs to mirror gluons as above. In short, $y/M$ is identical for the Twin Higgs and Quirkly Little Higgs models.

\subsection{Mirror Glueball Lifetime}
\label{ss.glueballlifetime}

The dimension-6 operator \eref{L6} allows glueballs to decay to SM particles through an off-shell Higgs. The corresponding decay widths were computed in \cite{Juknevich:2009gg}. For the lightest glueball decaying to two SM particles:
\begin{equation}
\label{e.glueballwidth}
\Gamma(0^{++} \to \xi \xi)  \ \ = \ \  
\left( \frac{1}{12 \pi^2} \left[\frac{y^2}{M^2} \right]\frac{v }{m_h^2 - m_0^2} \right)^2 
\
\left(4 \pi \alpha_s^\text{B} \mathbf{F_{0^{++}}^S}\right)^2
\ \ 
\Gamma^\mathrm{SM}_{h\to \xi \xi}(m_0^2),
\end{equation}
where $\mathbf{F_{0^{++}}^S}=\langle 0|\text{Tr}\ G_{\mu\nu}^\text{(B)}G^{\text{(B)}\mu\nu}|0^{++}\rangle$ is the annihilation matrix element of the glueball through the scalar operator composed of gluon field strengths, and $\Gamma^\mathrm{SM}_{h\to \xi \xi}(m_0^2)$ is the partial decay width of a SM-like Higgs with mass $m_0$, computed to high precision using HDECAY 6.42 \cite{Djouadi:1997yw}.  Mirror glueballs therefore have the same SM branching ratios as a SM-like Higgs of the same mass. 

The hadronic matrix element can be extracted from lattice studies \cite{Chen:2005mg, Meyer:2008tr}. We use the more recent result by \cite{Meyer:2008tr}:
\begin{equation}
4 \pi \alpha_s^\text{B} \mathbf{F_{0^{++}}^S} = f_0 \ \cdot \ r_0^{-3}
 \ \  \ \ \ , \ \ \ \ \  f_0 = 167 \pm 16 \ \ .
\end{equation}
The main observable of mirror-QCD is the glueball mass, so we express the matrix element in terms of $m_0$, see \eref{m01}:
\begin{equation}
4 \pi \alpha_s^\text{B} \mathbf{F_{0^{++}}^S} =  \left(\frac{f_0}{a_0^3}\right) \ \cdot \ m_0^3  \ \ \approx  \ \ (2.3) \ m_0^3 \ .
\end{equation}
Since $\Gamma^\mathrm{SM}_{h\to \xi \xi}(m_0^2) \sim m_0$, this gives the familiar scaling 
$\Gamma_{0^{++}} \sim m_0^7/(M^4 m_h^2)$. In this work we take $m_0$ as an input in our collider study. For this strategy, the main uncertainty in the total width is given by the uncertainty of the dimensionless number $f_0^2/a_0^6$, which is about $25\%$. 

The resulting decay length is shown, as a function of $m_0$ and top partner mass in Folded SUSY \eref{yoverMFoldedSUSY} and Twin Higgs \eref{yoverMTH} theories, in \fref{glueballlifetimegenericestimate}. The 25\% lifetime uncertainty on the contours is indicated with blue bands.

Clearly, discovering very light glueballs would be challenging. However, the situation is more promising for the preferred $12 - 60 \gev$ regime, with decay lengths ranging from microns  to kilometers. 

The heavier glueball states have lifetimes that are several orders of magnitude longer than $0^{++}$. Since that state already decays on macroscopic scales, we will focus exclusively on detecting $0^{++}$ decays as a probe of uncolored naturalness.

\begin{figure}
\begin{center}
\includegraphics[width=9cm]{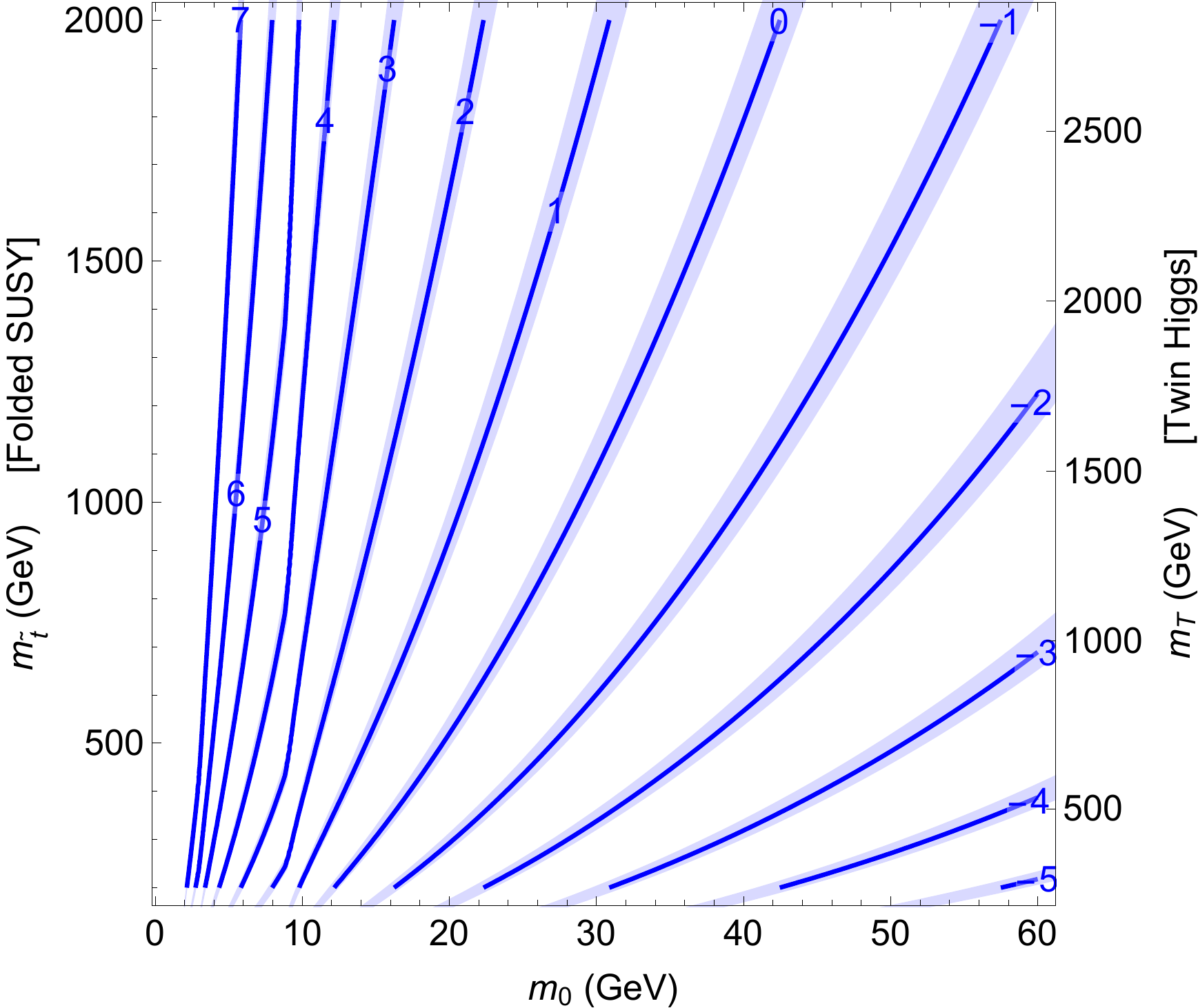}
\end{center}
\caption{Contours show  $\log_{10} c \tau/m$, where $c\tau$ is the mean decay length of the lightest glueball state $0^{++}$. Computed with \eref{glueballwidth} in Folded SUSY \eref{yoverMFoldedSUSY} and Twin Higgs \eref{yoverMTH} theories. The blue bands correspond to the shift of the contours resulting from the $25\%$ uncertainty in the total $0^{++}$ width.
}
\label{f.glueballlifetimegenericestimate}
\end{figure}

\subsection{Exotic Higgs Decays}
\label{ss.exotichiggsdecays}

For $m_0 \ll m_h/2$, the \emph{inclusive} exotic branching ratio of the Higgs to mirror-glue can be obtained from the SM branching ratio to gluons via a simple rescaling:
\begin{equation}
\label{e.brhinclusive}
\Br(h\to g_\text{B} g_\text{B}) \ \  \approx \ \  \Br(h\to g g)_\mathrm{SM} \ \cdot \ 
\left( \frac{\alpha_s^\text{B}(m_h)}{\alpha^\text{A}_s(m_h)} \ 4 v^2 \ \left[\frac{y^2}{M^2}\right] \right)^2
\end{equation}
where $\Br(h\to g g)_\mathrm{SM} \approx 8.6\%$. 

The coupling ratio $\alpha_s^\text{B}(m_h)/\alpha_s^\text{A}(m_h)$ depends on the mirror sector spectrum between $m_0$ and $m_h$. Ignoring threshold effects below $m_h$, it can be estimated by solving \eref{m0calculation} for $\mu_\mathrm{pole}^\text{B}$ and evolving to $\mu = m_h$:
\begin{equation}
\alpha_s^\text{B}(m_h)^{-1} = \frac{b}{2\pi} \log\frac{m_h}{\mu_\mathrm{pole}^\text{B}} \ \ \ , \ \ \  \ \ \ \mbox{where} \  \ \ \ \  \ \ \mu_\mathrm{pole}^\text{B} = \mu_\mathrm{pole}^\text{A} \ \cdot \ \frac{m_0}{a_0 (r_0^\mathrm{SM})^{-1}}.
\end{equation}
The minimal assumption is $b = 11$, corresponding to no mirror sector matter below $m_h$. This is almost required by LEP limits for Folded SUSY and Quirky Little Higgs. The resulting coupling ratio $\left(\alpha_s^\text{B}(m_h)/\alpha_s^\text{A}(m_h)\right)^2$ is shown as the green band in \fref{BrhtoglueballfoldedSUSY}~(left), ranging from about 1 to 2.5 for $m_0$ from 10 to 60 GeV. For a likely Fraternal Twin Higgs scenario, with a single mirror bottom below $m_h$ (assumed for illustrative purposes to be close to $m_0$ in mass), the ratio is only about 10\% higher due to the negative contribution to $b$, as indicated by the purple band. If much more matter is present there can be significant enhancement, as shown by the red band for all mirror quarks being close in mass to $m_0$ except the mirror top. However, as illustrated by \fref{m0massMinimalTwinHiggs} (bottom), in Twin Higgs scenarios this is only compatible with glueball masses below 25 GeV.

\begin{figure}
\begin{center}
\begin{tabular}{cc}
\includegraphics[height=4.5cm]{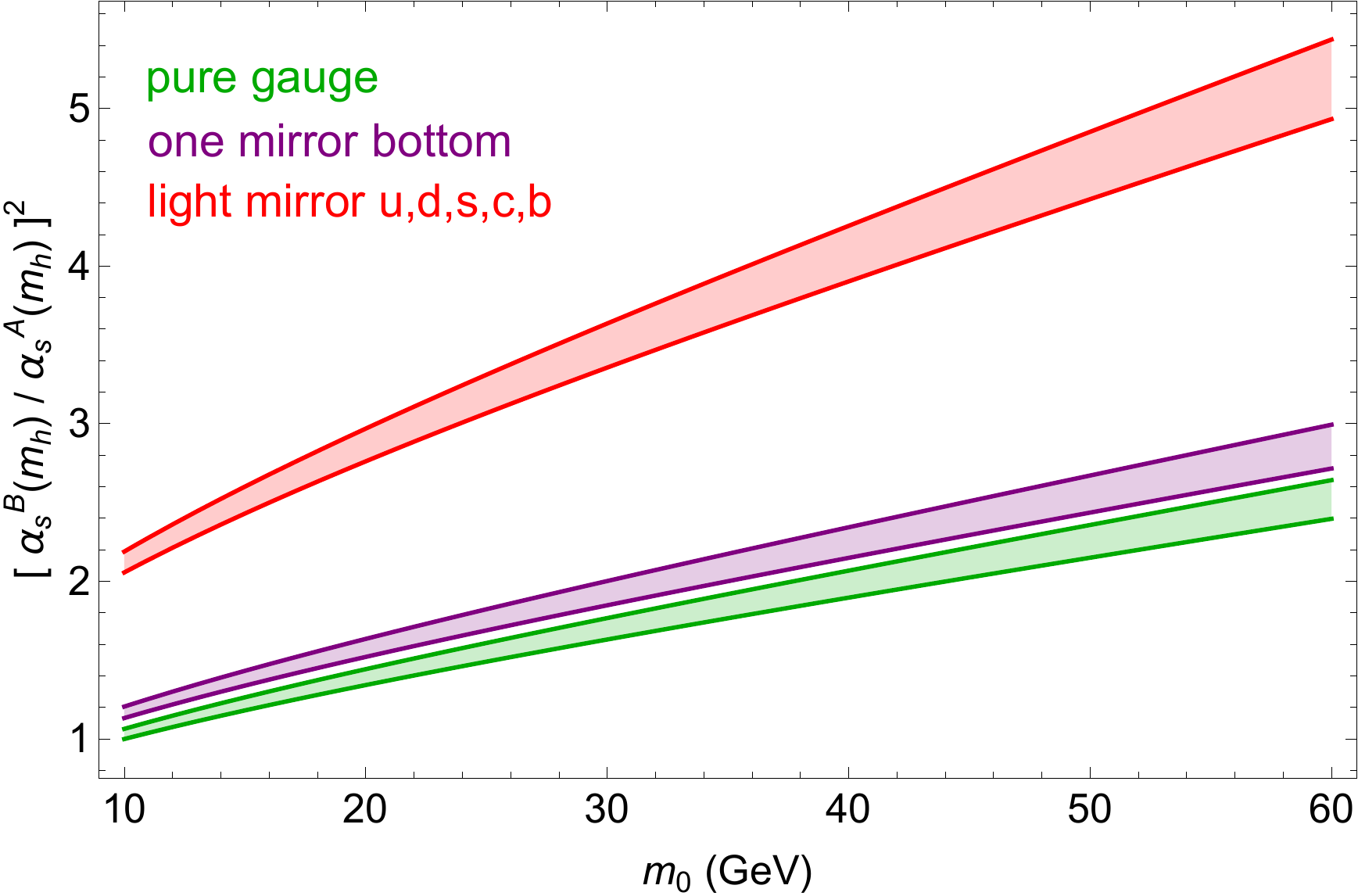}
&
\includegraphics[height=4.5cm]{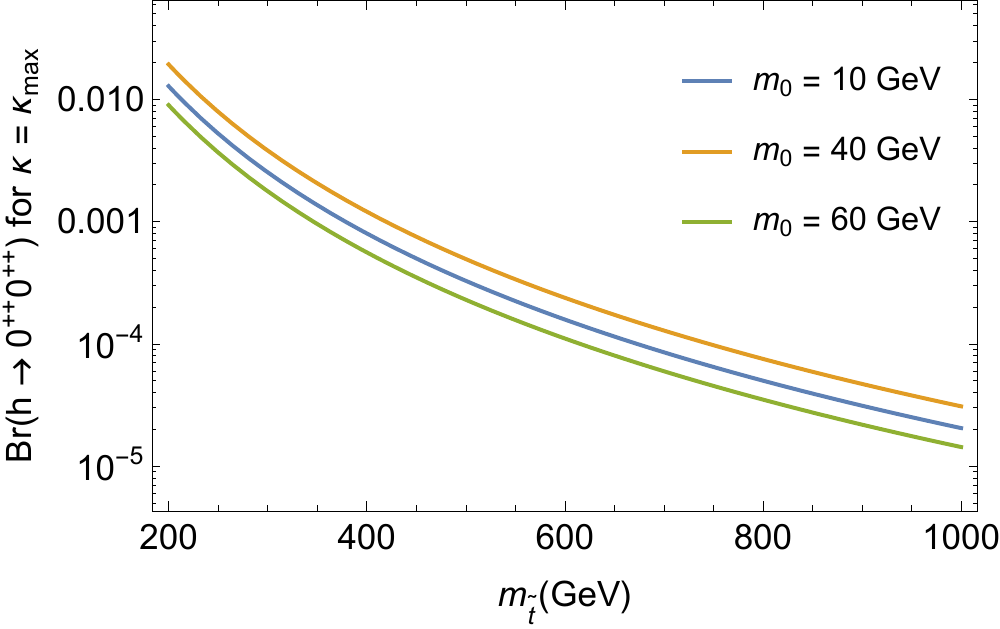}
\end{tabular}
\end{center}
\caption{
\emph{Left:} The overall $\left(\alpha_s^\text{B}(m_h)/\alpha_s^\text{A}(m_h)\right)^2$ factor in \eref{brhinclusive}, using one-loop RGE extrapolation from $m_0$, assuming either pure gauge (green, $b = 11$), one mirror bottom (purple, $b = 31/3$) or five light mirror quarks (red, $b = 23/3$). The pure gauge and one mirror bottom case closely resemble typical Folded SUSY and Fraternal Twin Higgs scenarios, respectively. The width of the band represents the range obtained by letting $a_0$ and $r_0^\mathrm{SM}$ vary independently within their uncertainties. 
\emph{Right:} Estimate of $\mathrm{Br}(h\to 0^{++} 0^{++})$ for $\kappa = \kappa_\mathrm{max} = 1$ from \eref{brhglueballs} for Folded SUSY \eref{yoverMFoldedSUSY}. 
}
\label{f.BrhtoglueballfoldedSUSY}
\end{figure}

\subsection{Estimating $0^{++}$ production}
\label{ss.exclusiveproduction}

Owing to the vastly different lifetimes of the glueball states, we need to estimate the \emph{exclusive} production rate of $0^{++}$ from exotic Higgs decays, since it will likely be the only glueball state that decays observably (though there can be exceptions). This requires detailed knowledge of pure-glue hadronization, which is not available. However, progress can be made by parameterizing our ignorance, as well as being pessimistic about signal rates for the purpose of a conservative sensitivity analysis.

First, we assume $0^{++}$ glueballs are produced in symmetric two-body Higgs decays only. 
For very light glueballs ($m_0 \ll m_h/2$) this might seem to be a poor approximation, since mirror hadronization likely leads to final states with more than two glueballs. Nevertheless, the two-body assumption is suitable for a conservative signal estimate in displaced vertex searches. Compared to a  realistic modeling of mirror hadronization, which would be challenging to do reliably, it underestimates glueball multiplicity and overestimates the $p_T$ of the resulting glueballs. The former trivially reduces the derived signal, but so does the latter, since the increased boost makes it more likely for the glueballs to escape the detector in this low-mass long-lifetime regime (see \fref{glueballlifetimegenericestimate}). 
We can then bootstrap an estimate for the exclusive Higgs branching fraction:
\begin{equation}
\label{e.brhglueballs}
\Br(h\to 0^{++} 0^{++}) \ \  \approx \ \  \Br(h\to g g)_\mathrm{SM} \ \cdot \ 
\left( \frac{\alpha_s^\text{B}(m_h)}{\alpha^\text{A}_s(m_h)} \ 4 v^2 \ \left[\frac{y^2}{M^2}\right] \right)^2
 \ \cdot \ \sqrt{1 - \frac{4 m_0^2}{m_h^2}} \ \cdot \  \kappa (m_0)
\end{equation}
For our benchmark models of Folded SUSY and Fraternal Twin Higgs we can use the lower green curve in \fref{BrhtoglueballfoldedSUSY}~(left) as a conservative estimate of $\alpha_s^B(m_h)/\alpha_s^A(m_h)$. The third phase-space factor in \eref{brhglueballs} ensures the branching ratio approaches zero at the kinematic threshold. Finally, $\kappa(m_0)$ is a nuisance parameter which encapsulates our ignorance about glueball hadronization, as well as non-perturbative mixing effects between excited ${0^{++}}^*$ states and the Higgs.  \fref{BrhtoglueballfoldedSUSY}~(right) shows the branching ratio for $\kappa = 1$.

For a given $[y/M]$ and $m_0$, $\Br(h\to 0^{++} 0^{++}) $ is  completely fixed up to the overall factor $\kappa$. A search which is sensitive to these exotic Higgs decays will therefore set an upper bound on $\kappa$. To proceed further we need to understand this factor in more detail. 

Thermal partition functions with $T \sim \Lambda_\mathrm{QCD'}$ give one estimate of the relative abundances of the different glueballs  \cite{JuknevichPhD}. Since the glueball masses are almost an order of magnitude higher than the confinement scale, Boltzmann suppression significantly favors the lightest state $0^{++}$ relative to the other species, despite the small relative mass difference. Using these arguments, we would expect $\kappa \sim 0.5$, but this estimate is unlikely to be correct in detail. Furthermore, for glueball production in exotic Higgs decays, some or all of the heavier glueball final states are forbidden if $m_0 \gtrsim 20 \gev$, further complicating the picture. We therefore choose two benchmark functions $\kappa_\mathrm{min, max}(m_0)$ to span the range of expected possibilities and roughly take the decreasing number of available final states with increasing $m_0$ into account. 

With some exceptions (discussed below) it seems unlikely that $\kappa$ be bigger than unity. Therefore we define 
\begin{equation}
\label{e.kappamax}
\kappa_\mathrm{max} = 1
\end{equation}
as the maximally optimistic signal estimate.  A more pessimistic assumption (given that $\kappa \sim 0.5$ is expected from the above thermal arguments) is that only $\sim 10\%$ of glueballs end up in the $0^{++}$ state if all two-glueball final states are kinematically allowed. This pessimistic estimate should approach the optimistic one as the glueball mass is increased to the point where only $0^{++}$ is allowed. We therefore choose the ad-hoc ratio of phase space factors 
\begin{equation}
\label{e.kappamin}
\kappa_\mathrm{min}(m_0) = \frac{\displaystyle \sqrt{1- \frac{4 m_0^2}{m_h^2}}}{{ \displaystyle \sum_i \sqrt{1 - \frac{4 m_i^2}{m_h}}}}
\end{equation}
where $i$ runs over stable stable glueball states with $m_i < m_h/2$. (Note the absence of spin multiplicities.) 
This factor ranges from about $1/12$ for $m_0 \sim 10 \gev$ to 1 for $m_0 \gtrsim 45 \gev$, see \fref{kappa}. These two assumptions should represent optimistic and pessimistic estimates of the effects of hadronization on the $0^{++}$ signal rate. We will therefore show projections for explicit exclusions using $\kappa = \kappa_\mathrm{min},  \kappa_\mathrm{max}$  to illustrate the potential reach of the LHC and a future 100 TeV collider. 

\begin{figure}
\vspace{-3mm}
\begin{center}
\includegraphics[width=6cm]{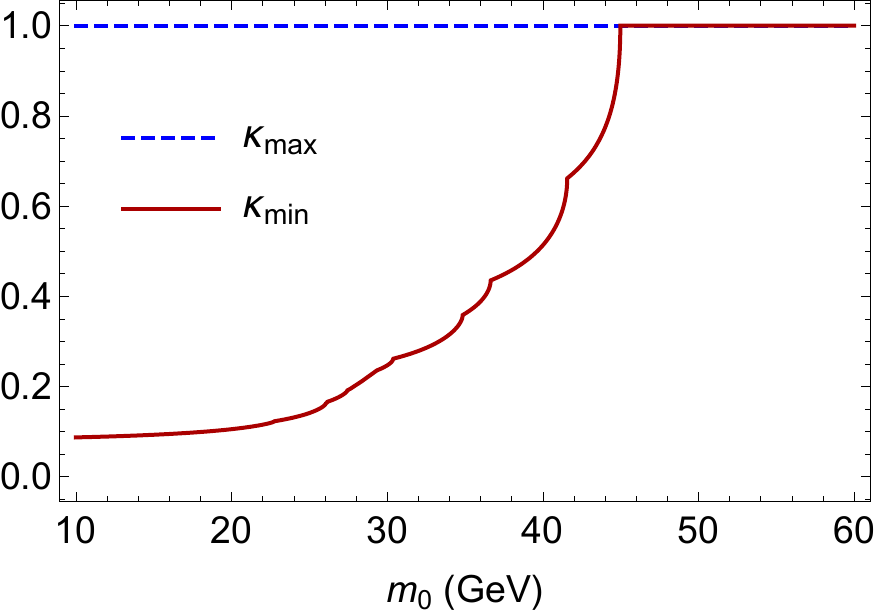}
\end{center}
\caption{
The high and low benchmark values for $\kappa(m_0)$, representing optimistic and pessimistic estimates of exclusive $h\to 0^{++} 0^{++}$ production.
}
\label{f.kappa}
\end{figure}

There are also non-perturbative effects which could, for some values of $m_0$, affect the value of $\kappa$ \cite{Craig:2015pha}. The $0^{++}$ glueball, which has the same quantum numbers as the physical Higgs boson, has a tower of excited resonances $0^{++*}_{(n)}$. There is evidence \cite{Morningstar:1999rf} for the first excited resonance $0^{++}_{(1)}$ around mass $m_{0^{++*}_{(1)}} \approx 1.5 m_0$. Going up in energy, there are likely to be a few more excited states before they get lost in the glueball continuum. These excited states could mix with the Higgs if $m_{0^{++*}_{(n)}} \approx m_h$, leading to enhancements in $\mathrm{Br}(h\to 0^{++} 0^{++})$, since the glueballs will have $\mathcal{O}(1)$ couplings amongst themselves. Similarly, if $m_h$ lies between two such resonances, $\mathrm{Br}(h\to 0^{++} 0^{++})$ could be suppressed. These non-perturbative enhancements and suppressions can be significant, though very likely smaller than a factor of 10. 

We can interpret these non-perturbative effects as \emph{possible} enhancements or suppressions of the value of $\kappa(m_0)$ for some glueball masses. To get a feeling when this could be significant, consider a toy-model of four $0^{++*}_{(n)}$ states, with $m_{0^{++*}_{(n)}}/m_0 = 1.5,\  2, \ 2.5,\  3$. In that case, 
\begin{equation}
m_0 \sim 50, 42 \gev \ \ \ \ \ \ \mbox{($\kappa$-enhancement)}
\end{equation}
would lead to enhancements in $\kappa$ due to mixing of the Higgs with $0^{++*}_{(4)}$ and $0^{++*}_{(3)}$ respectively. On the other hand, 
\begin{equation}
m_0 \sim 56, 46 \gev \ \ \ \ \ \ \mbox{($\kappa$-suppression)}
\end{equation}
could lead to $\kappa$-suppression. The above values of $m_0$ will be indicated in our limit projection plots, to indicate where $\kappa$ may be significantly different from $\kappa_{min}$ or $\kappa_\mathrm{max}$.

\section{Sensitivity of Exotic Higgs Decays}\label{s.higgsdecays}

The enormous number of Higgs bosons already produced at the LHC, and the fact that mirror glueballs are motivated to be in the $\sim 10 - 60 \gev$ mass range, make Higgs decays to mirror glueballs an excellent discovery channel for uncolored naturalness. Displaced decay searches are expected to be very sensitive to mirror glueballs.

The HL-LHC will produce about $10^8$ Higgs bosons. With $\mathrm{Br}(h\to  0^{++}0^{++}) \sim 10^{-5} - 10^{-2}$ for Folded SUSY stops in the 200 - 1000 GeV mass range (see \fref{BrhtoglueballfoldedSUSY}), the number of produced mirror glueballs could be in the thousands or millions. This should result in several detectable glueball decays even for kilometer decay lengths. 

In this section we estimate the total number of exotic Higgs decay events where, under the assumptions outlined in \ssref{exclusiveproduction}, one or more glueballs decay in various subsystems of the ATLAS detector. (The results would be qualitatively similar for CMS.)

These events form the `raw material' for displaced searches. We then apply estimated reconstruction efficiencies and trigger requirements to estimate the actual discovery potential of the LHC, as well as a hypothetical future 100 TeV collider. The HL-LHC could be sensitive to uncolored top partners with TeV scale masses. Achieving full coverage requires several new search strategies, some of which involve the reconstruction of displaced vertices within $50\mu$m of the interaction point.

\subsection{Geometrical Signal Estimates}

We start with a purely geometrical signal estimate for the three detector volumes defined in \tref{detectorgeometric}. This gives an intuition for the amount of `raw material' available for displaced-vertex searches of uncolored naturalness.  In the next section, we include triggers and reconstruction efficiencies. 

The number of expected events in which glueballs decay in these detector volumes is estimated as follows. The HAHM Madgraph model \cite{Curtin:2014cca} is used to simulate the \emph{kinematics} of Higgs bosons produced via gluon-fusion and vector-boson-fusion, with subsequent decay into two scalars $s$, which each decay dominantly into two $b$-quarks or $\tau$-leptons. This can be used to model the \emph{kinematics} of $h \to 0^{++} 0^{++} \to \bar ff \bar f^{(\prime)} f^{(\prime)}$. The hard matrix element for $h\to ss$ is different from the hard matrix element for $h\to g_\text{B} g_\text{B}$, but since the Higgs is a scalar the distribution of the final glueballs is isotropic in the Higgs rest frame, which is the case for either matrix element. Glueball decay, on the other hand, occurs long after mirror hadronization, when the glueball is a genuine scalar. This is correctly modeled by having the scalar $s$ decay to $\bar ff$. 

Matched samples with 0 or 1 extra jet are generated in Madgraph 5 and showered in Pythia 6 \cite{Alwall:2011uj, Sjostrand:2007gs}. The \emph{total signal cross section} is computed by using the Higgs working group cross sections \cite{HWG} for gluon fusion or vector boson fusion, along with \eref{brhglueballs} for the Higgs to Glueballs branching ratio for $\kappa = \kappa_\mathrm{max}$ and $\kappa_\mathrm{min}$, giving optimistic and pessimistic signal estimates under the assumption that the Higgs to glueball decays are dominantly two-body.\footnote{Assuming SM Higgs production is not exactly correct for the Twin Higgs, where the cross sections are reduced by a factor of $\cos^2 \vartheta$  due to mixing with the mirror Higgs. However, this effect is negligible (compared to other uncertainties) for top partner masses near the edges of sensitivity that we derive, so we neglect it in order to conveniently show bounds for both models in one parameter plane.} since we are interested in the top partner mass reach of different searches we neglect this effect to preserve the easy comparison of FSUSY and TH

\emph{Displaced glueball decays} are analyzed by extracting the decayed scalars $s$ (i.e. the glueballs) from each event, and using their boosted decay length $|\vec p_3|/m_0 \cdot c \tau$ and angle $\theta$ to the beam axis to compute their probability of decaying within each of the detector volumes in \tref{detectorgeometric}. This allows us to estimate the number of events with (a) at least one glueball decaying in the tracker, and (b) two glueballs decaying in the barrel HCAL or Muon System (MS). 
 
 \fref{geometricsignalestimate} shows the estimated event rates for LHC run 1, the 14 TeV LHC with $3000\ifb$ of data, and a hypothetical future 100 TeV $pp$ collider with $3000\ifb$ of data. The numbers for the 100 TeV collider should be seen as suggestive, since we use the ATLAS detector geometry to estimate signal yield, and the future detector layout will be different.

\begin{table}
\begin{center}
\begin{tabular}{| l ||  c | c | c |}
\hline
& $r$ (m) & $|z|$ (m) & $|\eta|$ \\
\hline \hline
Tracker &
(0, 1) & (0, 2.7) & $ < 2.4$
 \\
\hline
HCAL (barrel) & 
(2.25, 4.25) & (0, 4.3) & ---
\\
\hline
Muon System (barrel) & 
(5, 10) & ---  & $<1.1$
\\
\hline
\end{tabular}
\end{center}
\caption{
Extent of detector volumes for geometrical signal estimates, modeled on the ATLAS detector. 
}
\label{t.detectorgeometric}
\end{table}

\def\tempwidthone{5cm}
\begin{figure}
\begin{center}
\hspace*{-13mm}
\begin{tabular}{cccccc}

& \footnotesize $\sqrt{s} = 8 \tev$ && \footnotesize  $\sqrt{s} = 14 \tev$ && \footnotesize $\sqrt{s} = 100 \tev$\\
& \footnotesize $\mathcal{L} = 20 \ifb$ && \footnotesize $\mathcal{L} = 3000 \ifb$ && \footnotesize $\mathcal{L} = 3000 \ifb$
\\

\begin{sideways} \footnotesize \phantom{bllablablabla}  Tracker\end{sideways}
&
\includegraphics[width=\tempwidthone]{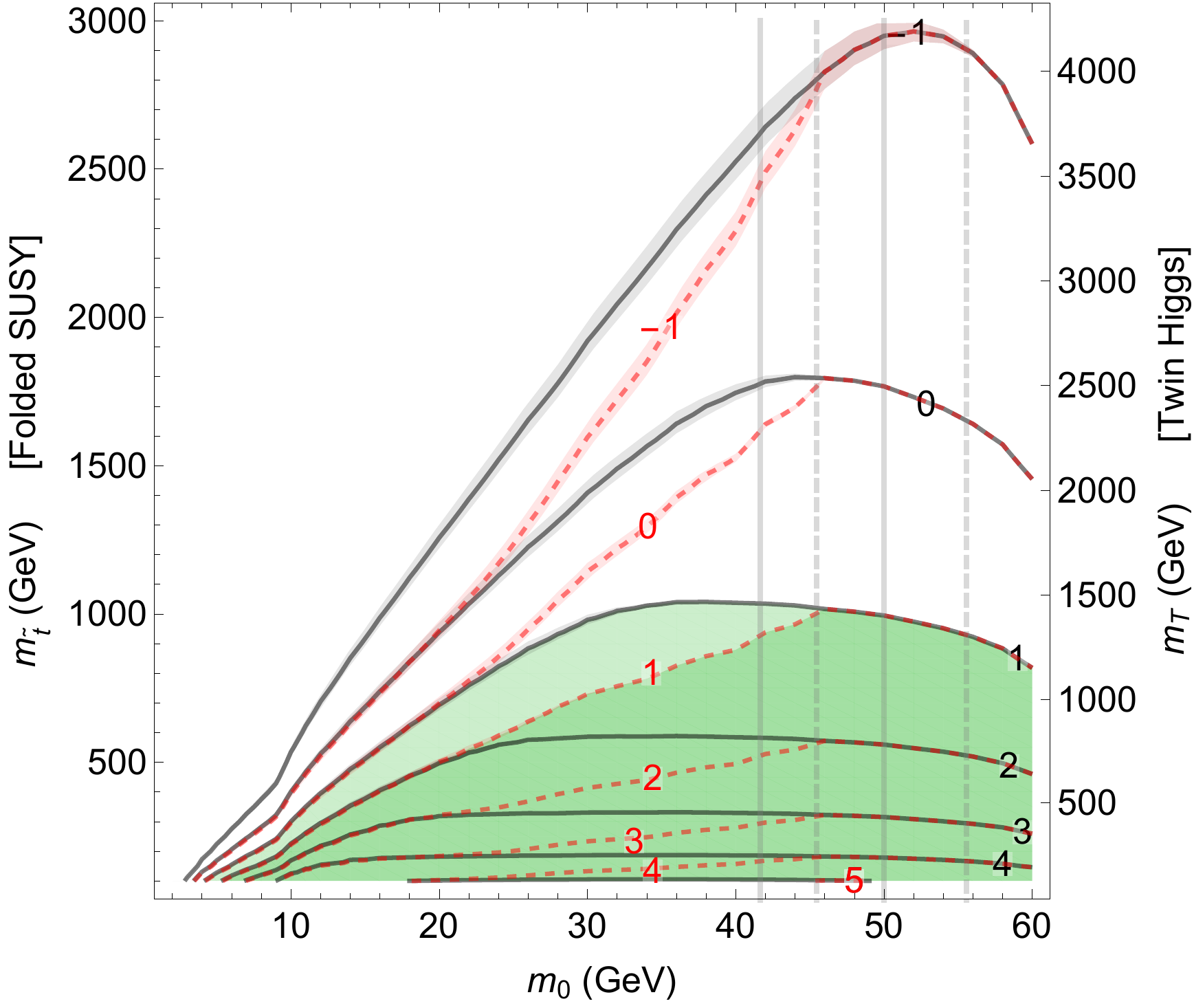}
&&
\includegraphics[width=\tempwidthone]{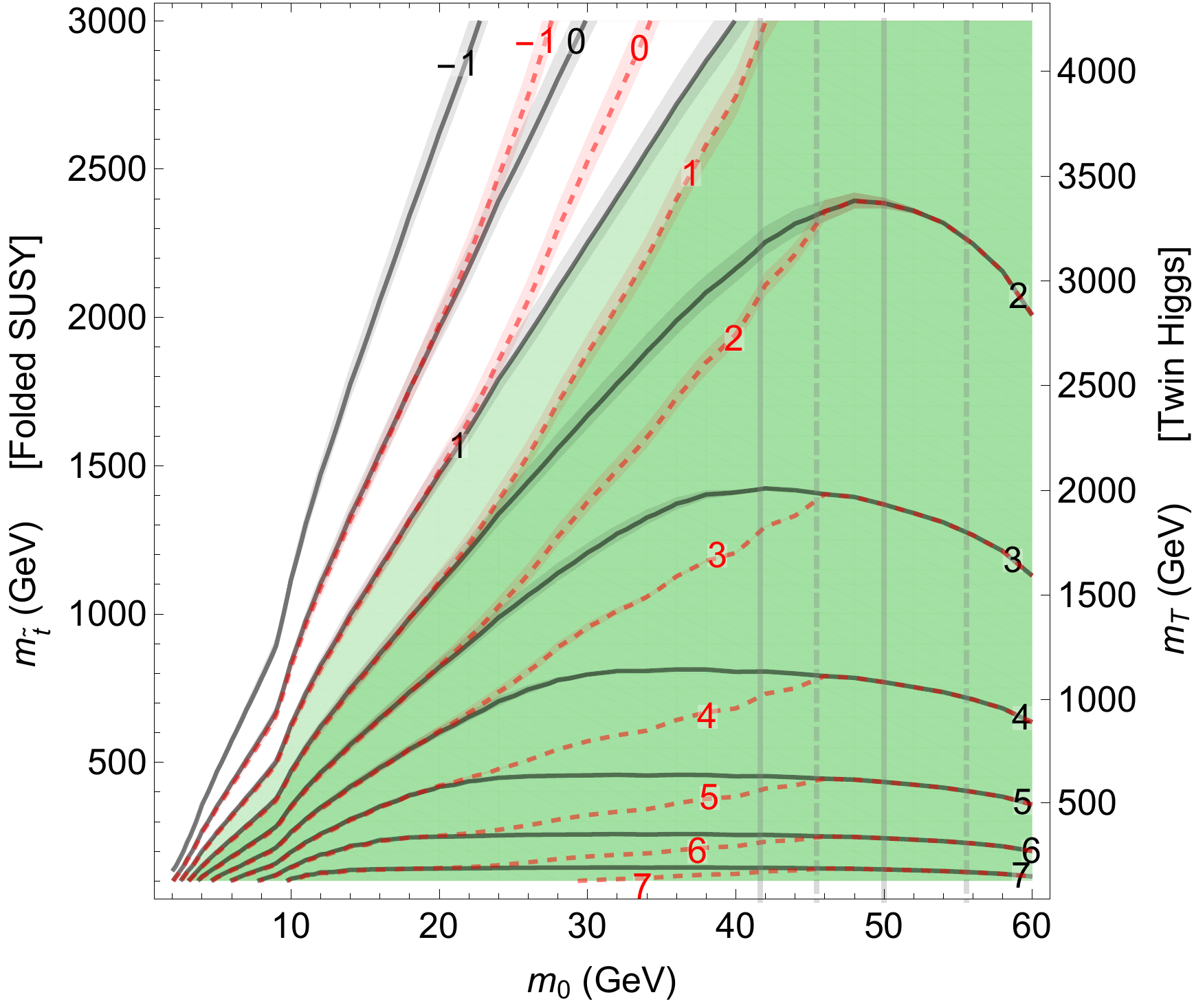}
&&
\includegraphics[width=\tempwidthone]{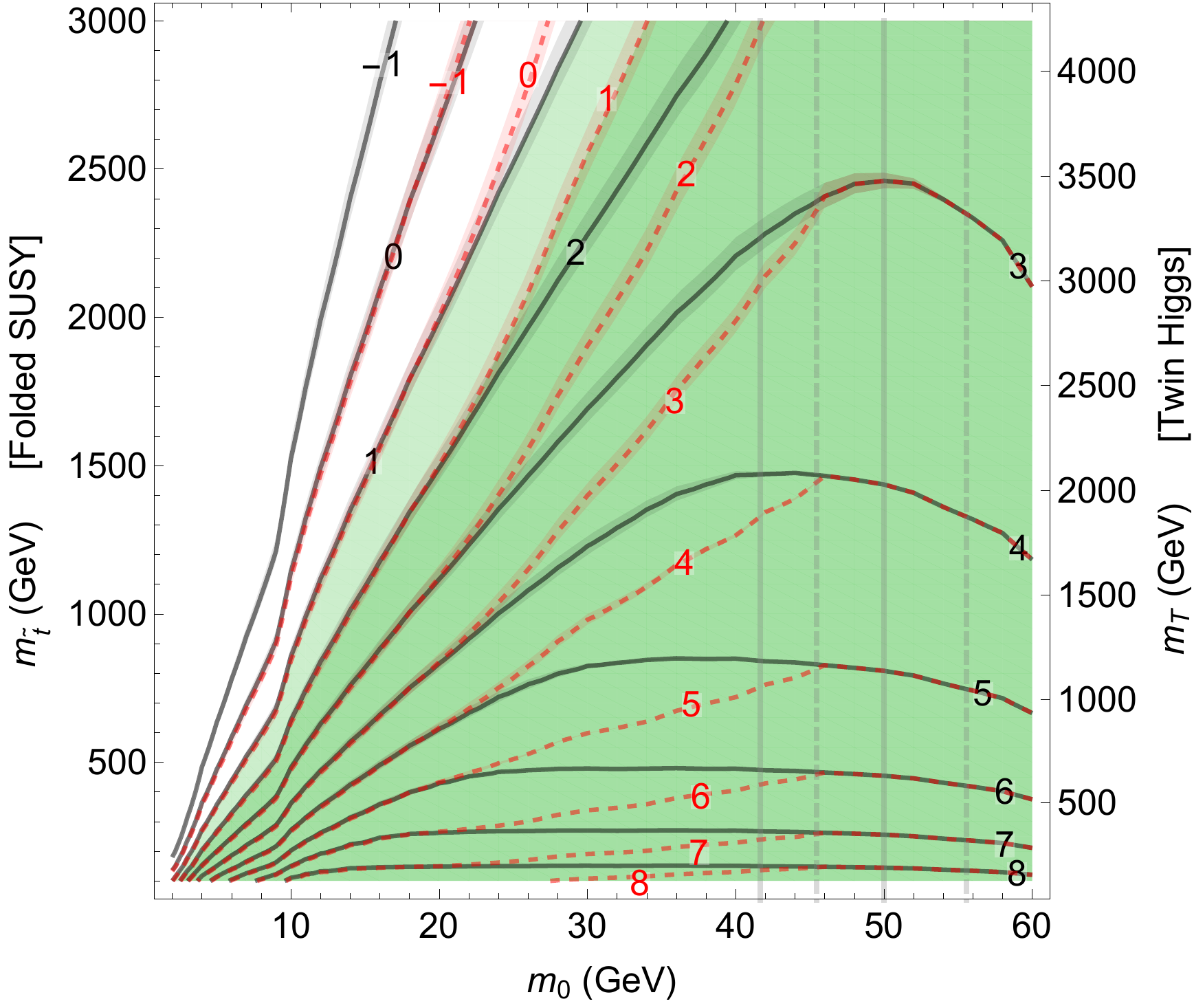}
\\

\begin{sideways} \footnotesize  \phantom{blablablaa} HCAL (barrel) \end{sideways}
&
\includegraphics[width=\tempwidthone]{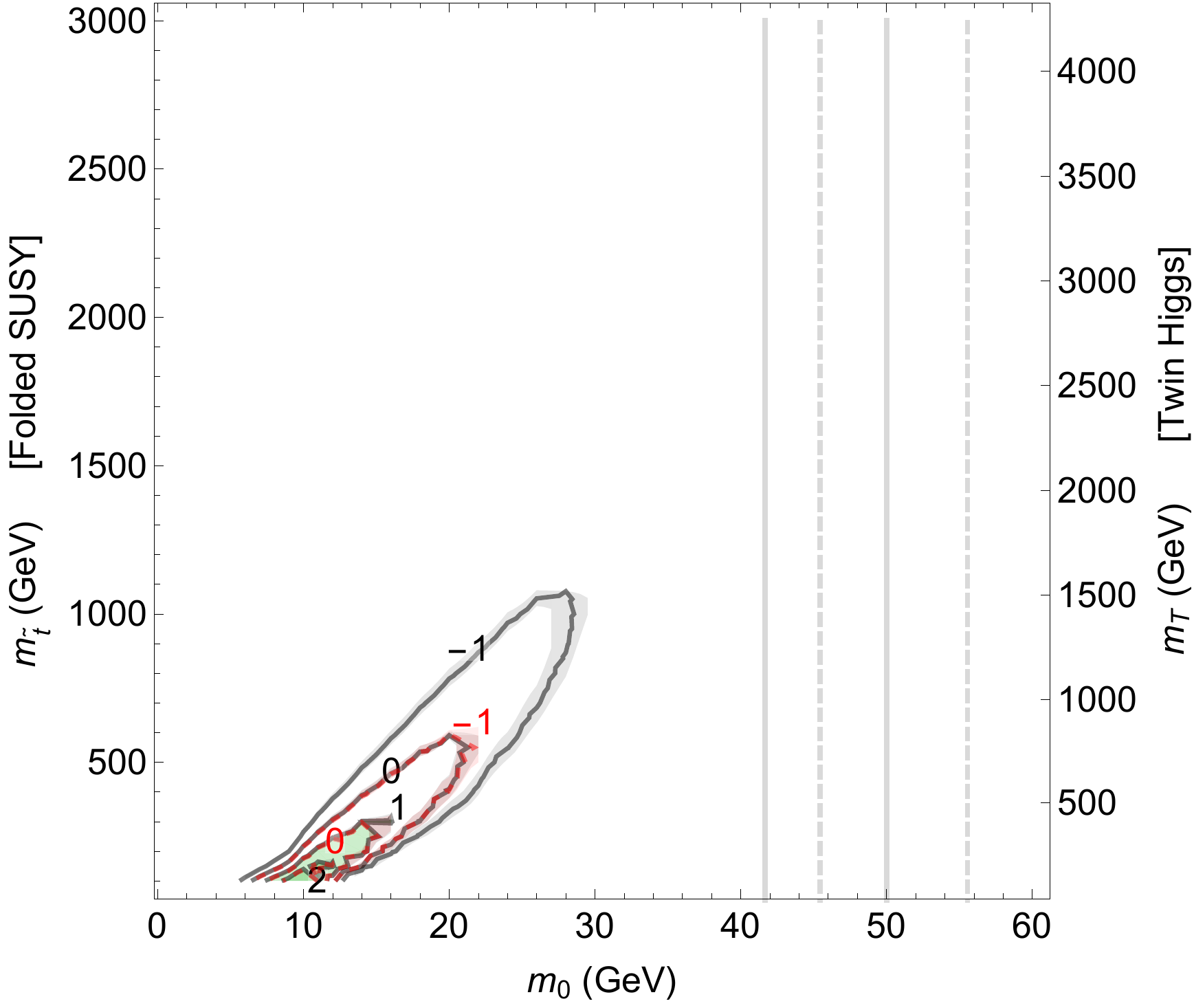}
&&
\includegraphics[width=\tempwidthone]{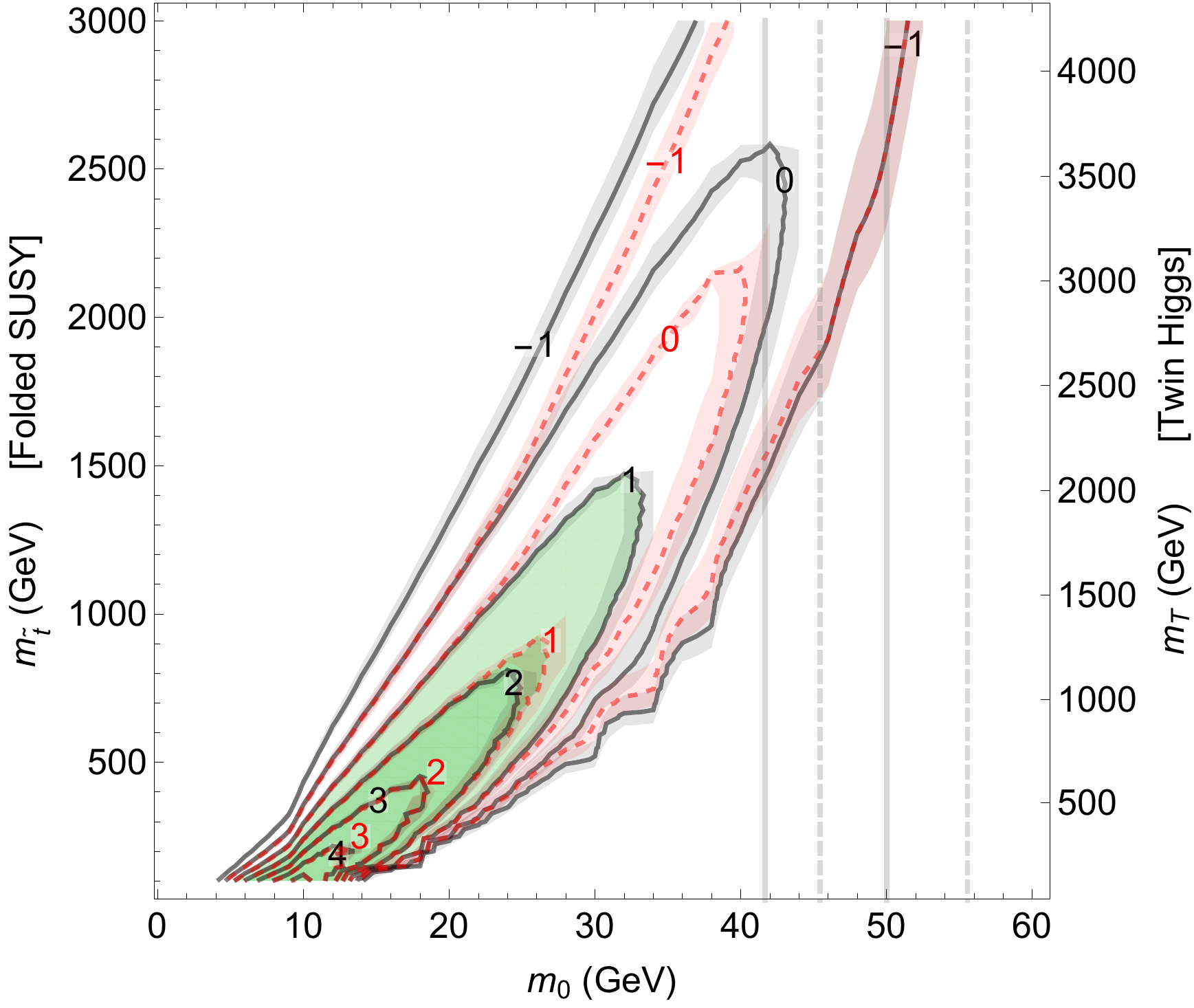}
&&
\includegraphics[width=\tempwidthone]{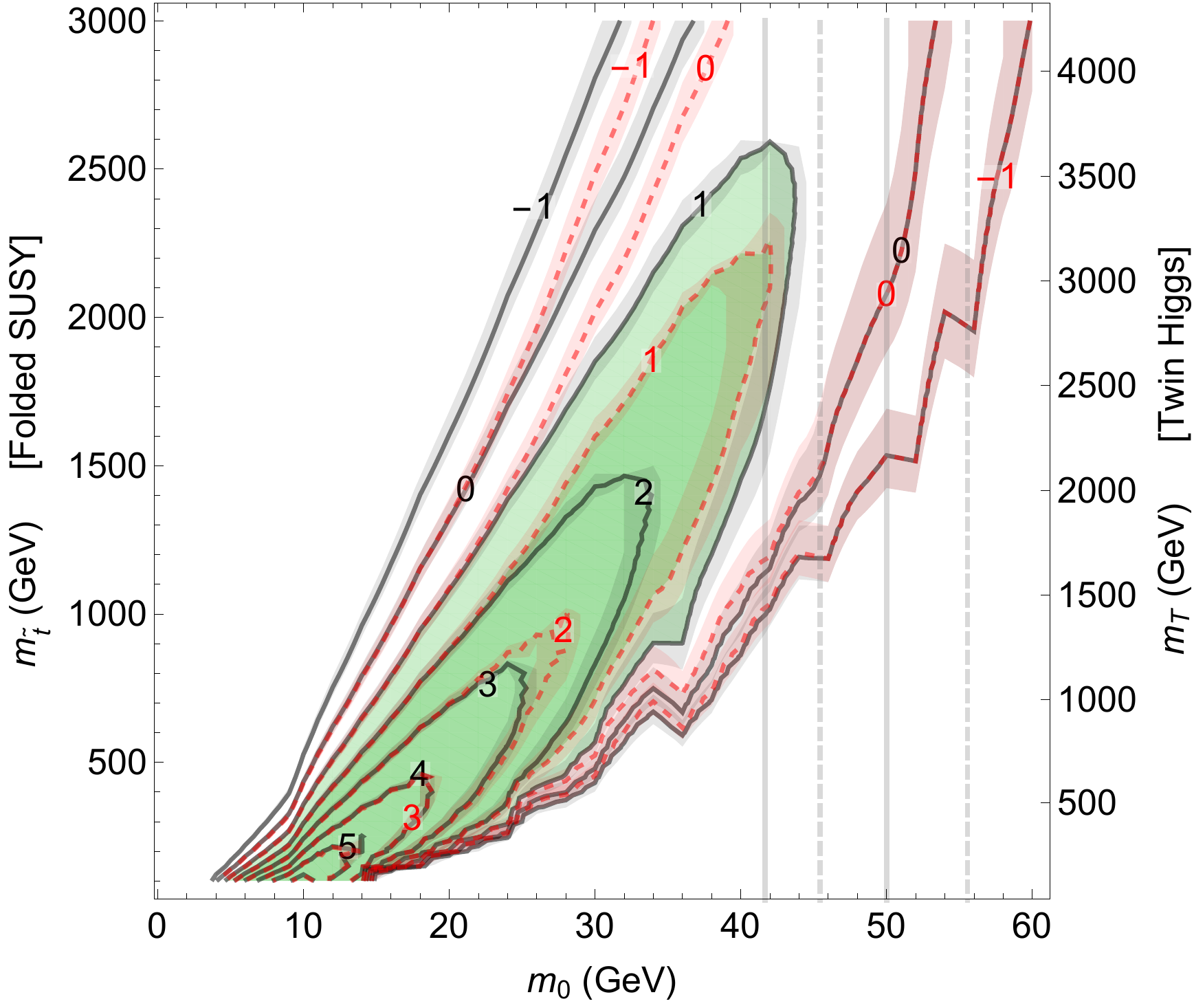}
\\

\begin{sideways} \footnotesize \phantom{blablbl}  Muon System (barrel) \end{sideways}
&
\includegraphics[width=\tempwidthone]{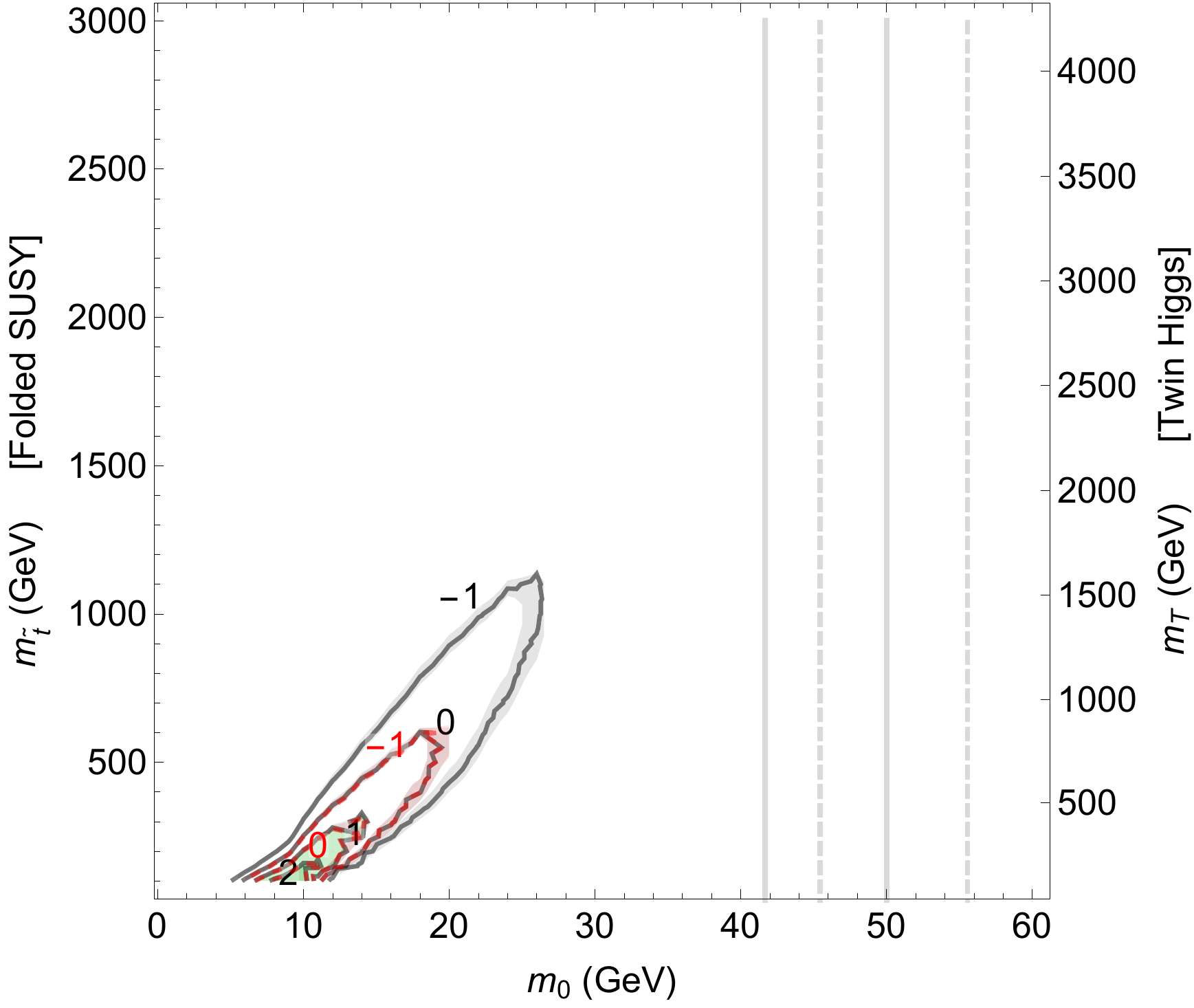}
&&
\includegraphics[width=\tempwidthone]{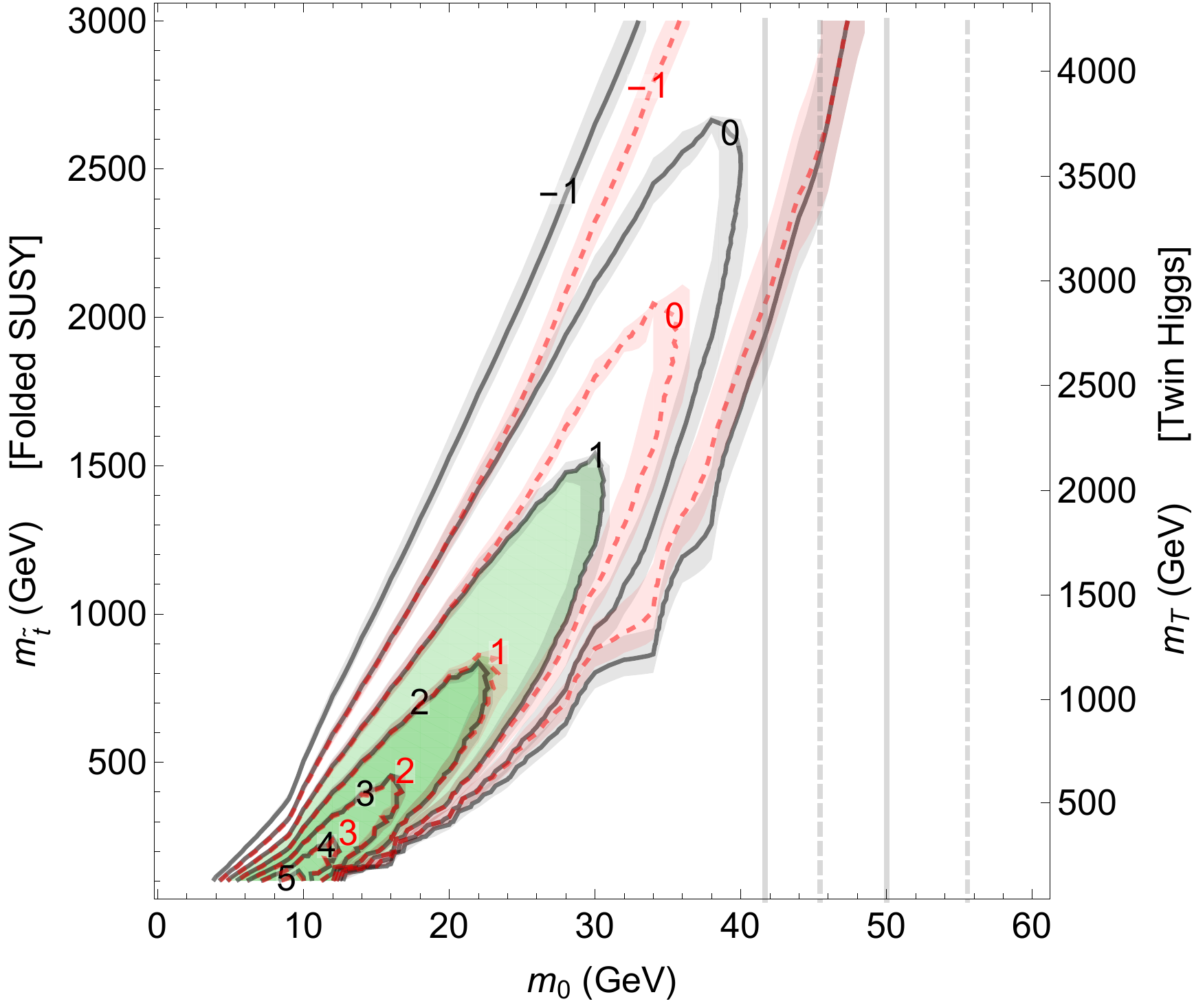}
&&
\includegraphics[width=\tempwidthone]{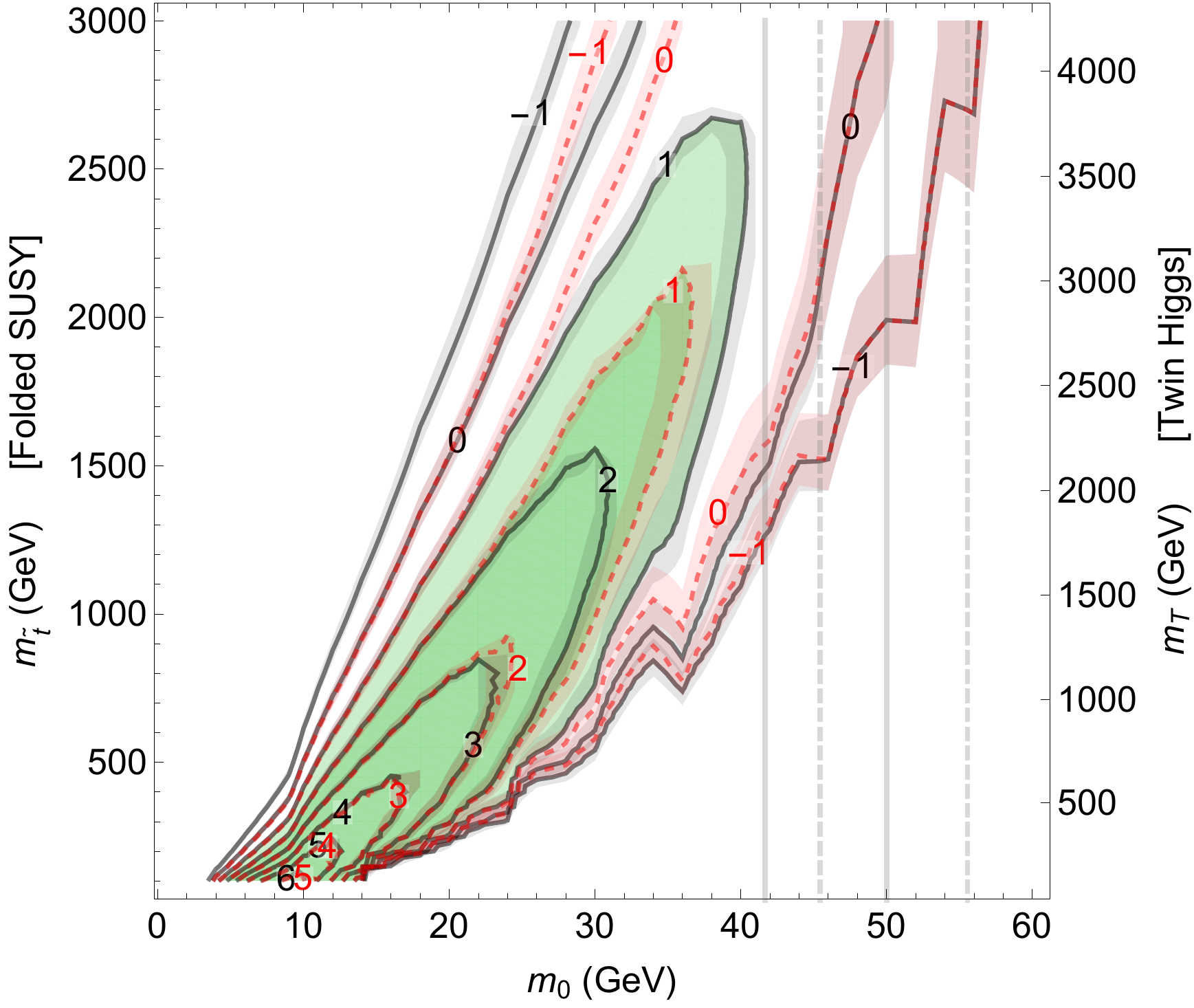}
\end{tabular}
\end{center}
\caption{
Geometrical signal estimates for Number $(N)$ of Higgs production (ggF + VBF) and decay ($h \to 0^{++} 0^{++}$) events where at least one  glueball decays in the tracker (top row) or both glueballs decay in the HCAL or Muon System (middle and bottom row), as defined in \tref{detectorgeometric}. The left, center and right columns correspond to the LHC run 1, HL-LHC and a hypothetical 100 TeV collider, respectively. 
$m_0$ is the mass of the lightest glueball $0^{++}$; the vertical axes correspond to mirror stop mass in Folded SUSY (see \eref{yoverMFoldedSUSY}) and mirror top mass in Twin Higgs (see \eref{yoverMTH}).
Black (dashed red) contours show $\log_{10}N$ for to $\kappa = \kappa_\mathrm{max}$ ($\kappa_\mathrm{min}$) in \eref{brhglueballs}, giving an optimistic (pessimistic) signal estimate under the assumption that $h$ decays dominantly to two glueballs.
Shaded bands around contours indicate effect of the $25\%$ uncertainty in $0^{++}$ lifetime. Vertical solid (dashed) lines show where $\kappa$ might be additionally enhanced (suppressed) due to non-perturbative mixing effects, see \ssref{exclusiveproduction}.
Light (dark) green shaded regions have more than 10 events for $\kappa = \kappa_\mathrm{max}$ ($\kappa_\mathrm{min}$) 
}
\label{f.geometricsignalestimate}
\vspace{4mm}
\end{figure}

We see that reconstructing displaced decays in the tracker is crucial for detecting glueballs above $\sim 30 \gev$. In principle, the first run of the LHC may have access to top partners as heavy as a TeV, with the potential reach exceeding 3 TeV for the HL-LHC. However, as we will see below, triggering and displaced vertex (DV) reconstruction significantly reduce that sensitivity. Even so, the reach will be very relevant for constraining models of uncolored naturalness.

\subsection{Estimated Sensitivity of Searches}

The purely geometrical signal estimate of the previous section suggests that the LHC might have very promising reach for uncolored naturalness. However, arriving at a realistic sensitivity estimate is challenging. A detailed collider study involving DVs, including backgrounds, is  beyond our scope and would be very difficult to validate. 

Fortunately, the ATLAS collaboration recently released two experimental searches \cite{Aad:2015asa, Aad:2015uaa}  for exotic Higgs decays of the form $h\to XX \to 4f$, where $X$ is long-lived and decays with SM-Higgs-like branching ratios to SM-fermions. This is identical to the $h\to 0^{++} 0^{++}$ signature we are focusing on, and the ATLAS analyses contain important lessons that we can use to estimate LHC reach, both beyond run 1 and beyond these two particular searches. 

The first search \cite{Aad:2015asa} used specialized triggers to look for a single displaced decay in the HCAL. A second decay in the HCAL was then required to reduce the background to a very low level, about 20 events. The second search \cite{Aad:2015uaa} followed a similar strategy, using a specialized trigger for displaced decays in the muon system (MS). Two separate offline analyses were performed, which required (a) one DV in the MS and another DV in either the MS or the inner tracker (IT), or (b) one DV in the MS, and at least 4 jets passing stringent $p_T$ cuts.  The requirement of one \emph{fully reconstructed displaced vertex} in addition to either another DV or a hard kinematic cut resulted in, effectively, zero background.

\begin{table}
\begin{center}
\begin{tabular}{|c||c|c|c|}
\hline 
Trigger & 8 TeV & 14 TeV\\
\hline \hline

1 jet & $p_T^{j_1} > 180 \gev$ & $p_T^{j_1} > 290 \gev$ 
\\
\hline

inclusive VBF & 
\begin{tabular}{c}
$|\eta_{j_1,j_2}| > 2$ \\
$\eta_{j_1} \eta_{j_2} < 0$\\
$|\eta_{j_1} - \eta_{j_2}| > 3.6$\\
$m_{j_1j_2} > 600 \gev$
\end{tabular}
&
\begin{tabular}{c}
\\
same\\
\\
$m_{j_1j_2} > 1000 \gev$
\end{tabular}

\\
\hline

VBF
$h\to \bar b b$
&
\begin{tabular}{c}
$p_T^{j_{1,2,3}} >  (70, 50, 35) \gev$\\
$|\eta_{j_1,2,3}| < (5.2, 5.2, 2.6)$\\
$|\eta_{j_1}|$ or $|\eta_{j_2}| < 2.6$
\end{tabular}
&
\begin{tabular}{c}
$p_T^{j_{1,2,3}} >  (112, 80, 56) \gev$\\
\multirow{2}{*}{same}\\
\\
\end{tabular}

\\
\hline \hline

single lepton 
&
\begin{tabular}{c}
one lepton with \\
$p_T > 25 \gev$, $\eta < 2.4$
\end{tabular}

&
same
\\
\hline
\end{tabular}
\end{center}
\caption{
Prompt triggers we explore for displaced vertex (DV) searches of exotic Higgs decays to mirror glueballs. 
\emph{Top three rows:} three representative jet triggers. The VBF $h\to \bar b b$ trigger is modeled on \cite{CMS:2013jda}, the others are representative generic triggers \cite{conversationandy}. The 14 TeV thresholds are derived from the 8 TeV thresholds by a 60\% upscaling, and for illustrative purposes the 100 TeV thresholds are assumed identical to 14 TeV. 
\emph{Bottom row:} single lepton trigger \cite{conversationandy} for DV searches in the $Wh, Zh, t\bar t h$ production channels.
}
\label{t.triggers}
\end{table}

Displaced vertices are a very distinctive signature. What these ATLAS searches suggest is that searches for DVs could be regarded as approximately background-free, provided they look for
\begin{itemize}
\item[(a)] two DVs, or
\item[(b)] one DV, in addition to a stringent non-DV requirement, such as high jet activity or leptons. 
\end{itemize}
We suspect these guidelines to be particularly useful for fully reconstructed DVs in the Muon System or tracker. The absence of track reconstruction in the calorimeters is likely one factor leading to higher (though still very low) background levels in the HCAL search.

The two ATLAS searches looked for particles with decay lengths of about a meter. However, the geometrical signal estimate demonstrated that sensitivity to much shorter decay lengths is required to cover the whole $(m_0, m_{\text{TP}})$ parameter space of uncolored naturalness theories with long-lived glueballs (where $m_\mathrm{TP}$ stands in for the top partner mass in different models). This forces us to utilize strategy (b), since at present there is no way to trigger on only displaced decays in the tracker without other requirements like high $H_T$ \cite{CMS:2014wda}. 
Therefore, we explore the sensitivity of several possible searches that require one DV in the tracker, and additional hadronic or leptonic activity in the event to trigger on. 

A list of prompt trigger candidates, modeled on existing experimental searches, is presented in \tref{triggers}. 
To be conservative we require glueballs to decay to $\bar b b$ in order to pass the VBF $h \to \bar b  b$ trigger. The multijet trigger from \cite{Aad:2015uaa} is not included, since it has very low efficiency for Higgs decays.  
 For jet triggers, we follow \cite{Aad:2015uaa} and assume DV reconstruction down to a minimal impact parameter of $r_\mathrm{min} = 4$cm. We find the VBF $ h\to \bar b b$ trigger to be the most useful above the $\bar b b$ threshold, but other triggers can perform comparably, as we outline in more detail below. Probing parameter regions with relatively heavy glueballs requires sensitivity to even shorter decay lengths, down to $\mathcal{O}(10 \mu \mathrm{m})$. Such DV reconstruction might be possible with a clean enough dataset \cite{conversationandy}. We test this by requiring exotic Higgs decays from $Wh, Zh, \bar t th$ production to pass a single lepton trigger, and assume that DVs can be reconstructed down to $r_\mathrm{min} = 50$ $\mu$m with the same efficiency as for $r_\mathrm{min} = 4$cm. Such an analysis would be no doubt challenging, but our work serves as powerful motivation to pursue this search.

\begin{table}
\begin{center}
\begin{tabular}{| l || c | c |}
\hline
Search & Displaced Vertex requirements & Conventional Trigger \\
\hline \hline
(IT, $r > 50$$\mu$m)$\times$(1L) &
one DV  in IT with $r > 50$$\mu$m
 & 
 single lepton
\\
\hline
(IT, $r > 4$cm)$\times$(VBF) & 
one DV  in IT with $r > 4$cm & best VBF
\\
\hline
(HCAL)$\times$(HCAL) &
two DVs in HCAL barrel or endcap
 & ---
\\
\hline
(MS)$\times$(MS or IT) &
\begin{tabular}{c}
one DV in MS barrel or endcap\\
and an additional DV in\\
either MS or IT ($r > 4$ cm)
\end{tabular}
& ---
\\
\hline
\end{tabular}
\end{center}
\caption{
Summary of explored displaced vertex (DV) searches for exotic Higgs decays to mirror glueballs. The (IT)$\times$(non-DV-trigger) searches in the first two rows are newly suggested searches.
``best VBF'' means that the VBF $h \to \bar b b$ (inclusive VBF) trigger  is used for $m_0$ above (below) the $\bar b b$ threshold.
 (HCAL)$\times$(HCAL) and (MS)$\times$(MS or IT) are recasts of \cite{Aad:2015asa} and \cite{Aad:2015uaa}. Prompt and displaced trigger and reconstruction efficiencies are listed in Tables. \ref{t.triggers} and \ref{t.detectoruseful}. 
}
\label{t.searches}
\end{table}

\begin{table}
\begin{center}
\begin{tabular}{| l ||  c | c | c || c | c |}
\hline 
& $r$ (m) & $|z|$ (m) & $|\eta|$ & $\epsilon_\mathrm{trig}$ & offline\\
\hline \hline
Tracker
& ($r_\mathrm{min}$, 0.3) & (0, 2.7) & $< 2.4$ & --- &   $\epsilon_\mathrm{DV} = 0.10$
\\
\hline\hline
HCAL (barrel)
& (2.1, 3.5)  &(0, 4.3) & --- & 0.22   & 
\multirow{ 2}{*}{$\epsilon_\mathrm{offline} = 0.4$} 
\\
\cline{1-5}
HCAL (endcap)
 & (2.25, 3.5) & (4.3, 5.0)  & $< 3.2$  & 0.07  & 
\\
\hline\hline
Muon System (barrel)
   & (4, 6.5) & --- & $< 1.1$ & 0.40 & $\epsilon_\mathrm{DV} = 0.25$
\\
\hline
Muon System (endcap)
  & --- & (7, 12) & (1.1, 2.4) & 0.25 & $\epsilon_\mathrm{DV} = 0.5$
\\
\hline
\end{tabular}
\end{center}
\caption{
ATLAS detector regions with sensitivity to displaced vertices. $\epsilon_\mathrm{trig}$ is the efficiency to trigger on a single displaced decay in that detector region. 
In the tracker and MS, each displaced decay has offline reconstruction efficiency $\epsilon_\mathrm{DV}$. The overall reconstruction efficiency of an event with two decays in the HCAL that already passed triggers is $\epsilon_\mathrm{offline}$.
Geometrical definition and approximate efficiencies for displaced $h\to XX \to 4f$ decay based on \cite{Aad:2015asa} (HCAL) and \cite{Aad:2015uaa} (Muon System and tracker). For the tracker, \cite{Aad:2015uaa} gives about $r_\mathrm{min} = 4$cm, which we use as well. However, for a clean final state recorded via the lepton trigger we consider $r_\mathrm{min} = 50$ $\mu$m \cite{conversationandy}.
}
\label{t.detectoruseful}
\end{table}

The four searches we investigate are summarized in \tref{searches}: one search of the form (jet activity)$\times$(DV in IT), one search of the form (lepton)$\times$(DV in IT), and the two existing searches (HCAL)$\times$(HCAL) and (MS)$\times$(MS or IT). We assume these searches have close to zero background and estimate sensitivities accordingly, but as we explained above this assumption is likely too optimistic for the HCAL search. 

To arrive at approximately realistic signal estimates, it is  necessary to understand triggering and offline reconstruction efficiencies for displaced decays. 
Fortunately, the two ATLAS searches supply these efficiencies either directly, or  in the form of final event yields, see  \tref{detectoruseful}. 
The displaced decay triggers in the HCAL and MS have triggering efficiencies $\epsilon_\mathrm{trig}$ per decay that can be taken to be approximately constant and (for our purposes) independent of glueball mass in the relevant detector volume. A full displaced vertex in the tracker or MS can be reconstructed with an offline efficiency $\epsilon_\mathrm{DV}$  per vertex \cite{Aad:2015uaa}. For displaced decays in the HCAL, an overall offline efficiency $\epsilon_\mathrm{offline}$ is applied to the event, which reproduces the $\sim 500$ signal events predicted for this search at run 1 of the LHC with $\mathrm{Br}(h \to X X) = 1$  \cite{Aad:2015asa}.

For the prompt lepton trigger, a flat lepton reconstruction efficiency of 85\% is applied. For jet triggers, PGS is used for hadronic object reconstruction. This might not, at first, appear sufficient, since PGS assumes prompt decay. However, when glueball final states are used for triggering the prompt assumption under-estimates trigger efficiency, since it does not take into account collimation of glueball final states decaying on the edge of the tracker, which increases the likelihood of surpassing jet thresholds. Therefore, our simple pipeline is sufficient for a conservative signal estimate.

Our results can be easily rescaled for different DV reconstruction efficiencies. This is especially salient since projecting HL-LHC limits using current ATLAS capabilities may be very conservative. First, even though the CMS displaced dijet search \cite{CMS:2014wda} has no sensitivity to exotic Higgs decays due to a large $H_T$ requirement, it does suggest that CMS may be able to reconstruct DVs in the tracker with significantly higher than 10\% efficiency. Second, both detectors will undergo upgrades as part of the HL-LHC program, which should greatly improve tracking and triggering capabilities\cite{Butler:2020886,Cinca:1711887}.  As a result, the $3000\ifb$ signal may be larger than what we project by an $\mathcal{O}(1)$ factor, but this does not affect our main conclusions.\footnote{It should be noted that our DV + (lepton or jets) searches are robust at the $\mathcal{O}(1)$ level, even under the  pessimistic assumption that most of the unstable glueballs are produced in asymmetric $h\to 0^{++} + X$ decays. The searches requiring two DV's would have to be modified, but in that case it should be possible to combine a single reconstructed vertex in the MS or HCAL with a lepton or jet requirement and recover a similar sensitivity to what we show for the MS and HCAL searches.}

\def\tempwidthone{5cm}
\begin{figure}
\begin{center}
\vspace*{-17mm}
\hspace*{-13mm}
\begin{tabular}{cccccc}

& \footnotesize $\sqrt{s} = 8 \tev$ && \footnotesize  $\sqrt{s} = 14 \tev$ && \footnotesize $\sqrt{s} = 100 \tev$\\
& \footnotesize $\mathcal{L} = 20 \ifb$ && \footnotesize $\mathcal{L} = 3000 \ifb$ && \footnotesize $\mathcal{L} = 3000 \ifb$
\\

\begin{sideways} \footnotesize \phantom{blabll} (IT, $r > 50$ $\mu$m)$\times$(1L)\end{sideways}
&
\includegraphics[width=\tempwidthone]{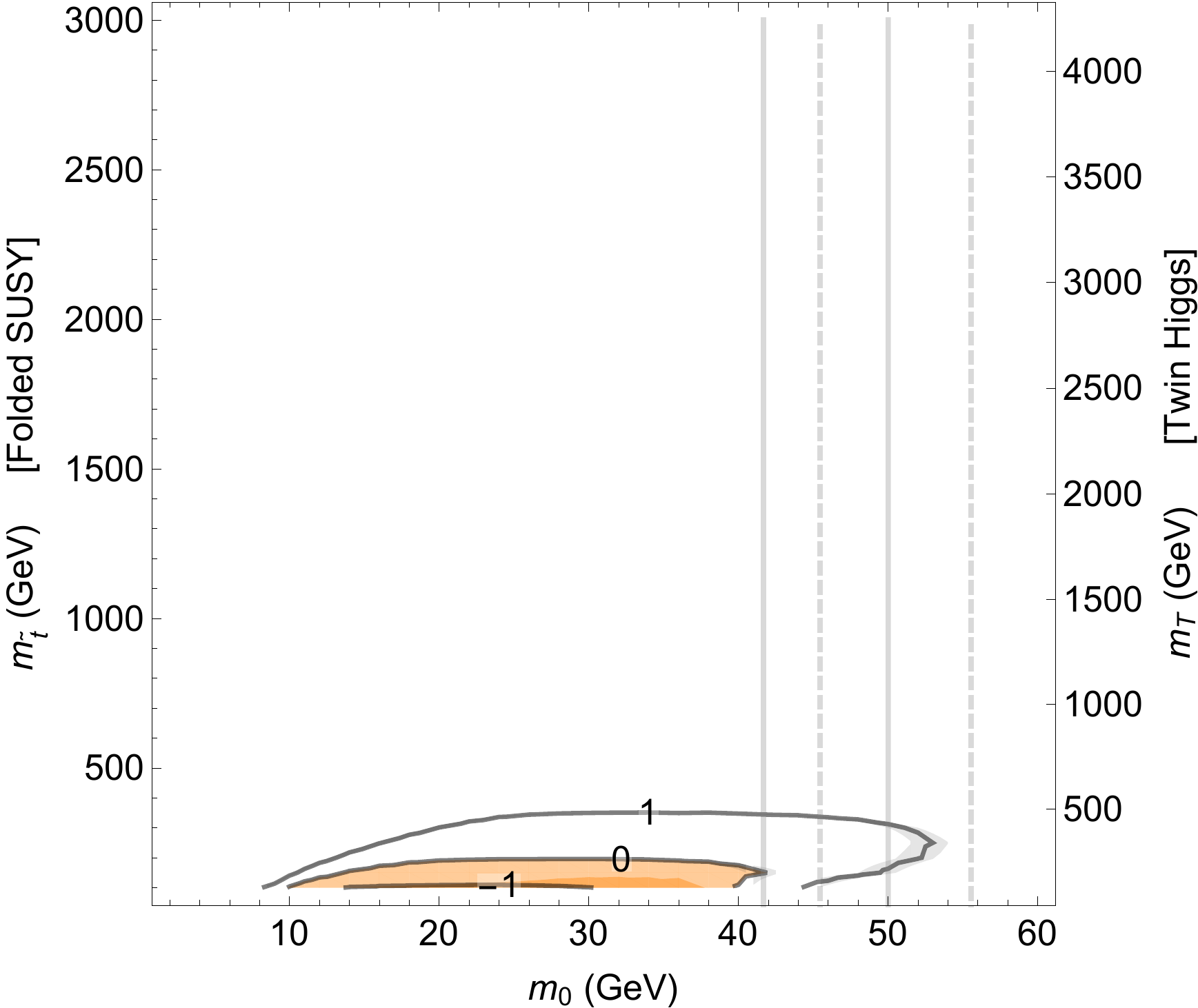}
&&
\includegraphics[width=\tempwidthone]{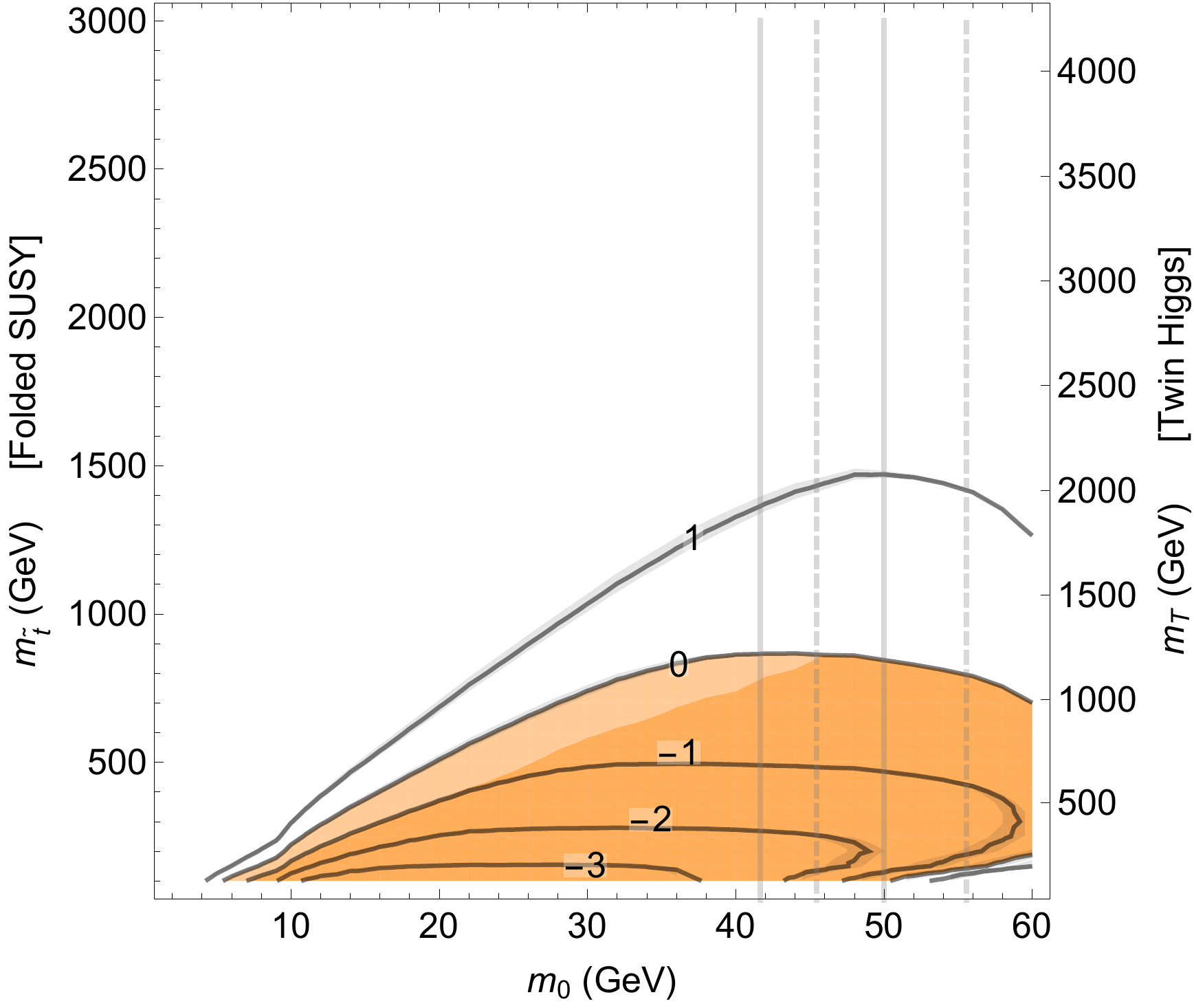}
&&
\includegraphics[width=\tempwidthone]{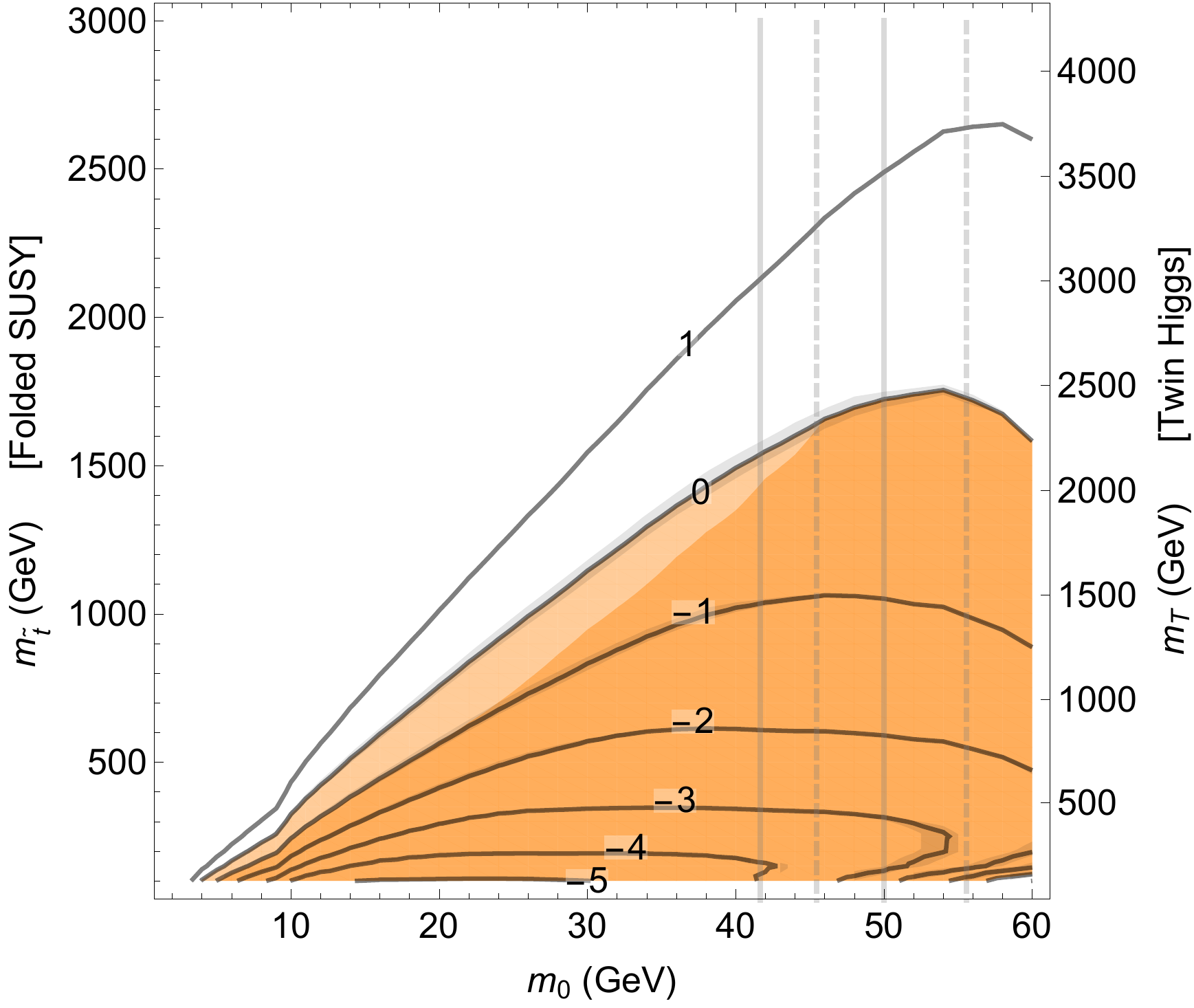}
\\

\begin{sideways} \footnotesize  \phantom{bla} (IT, $r > 4$ cm)$\times$(best VBF)\end{sideways}
&
\includegraphics[width=\tempwidthone]{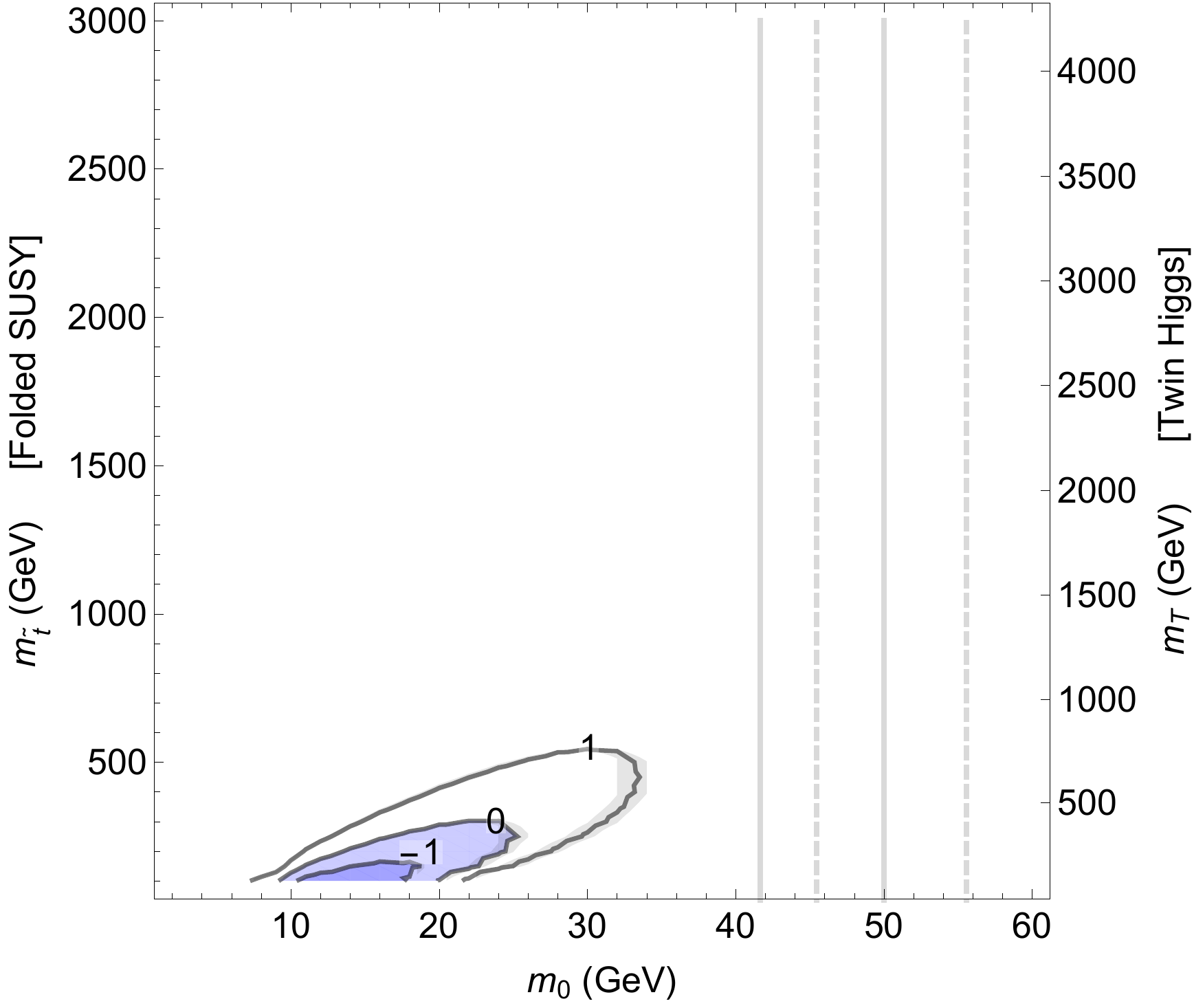}
&&
\includegraphics[width=\tempwidthone]{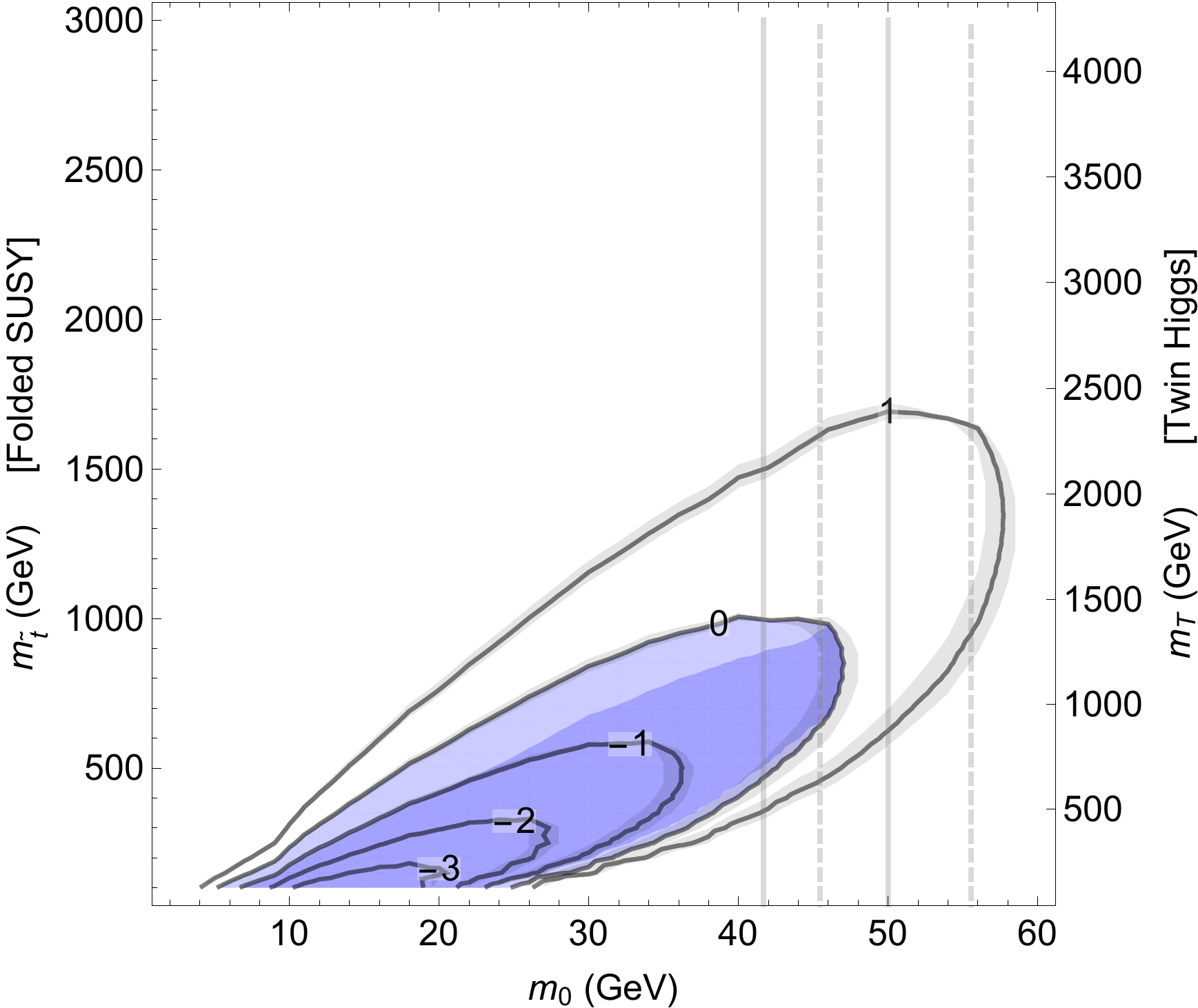}
&&
\includegraphics[width=\tempwidthone]{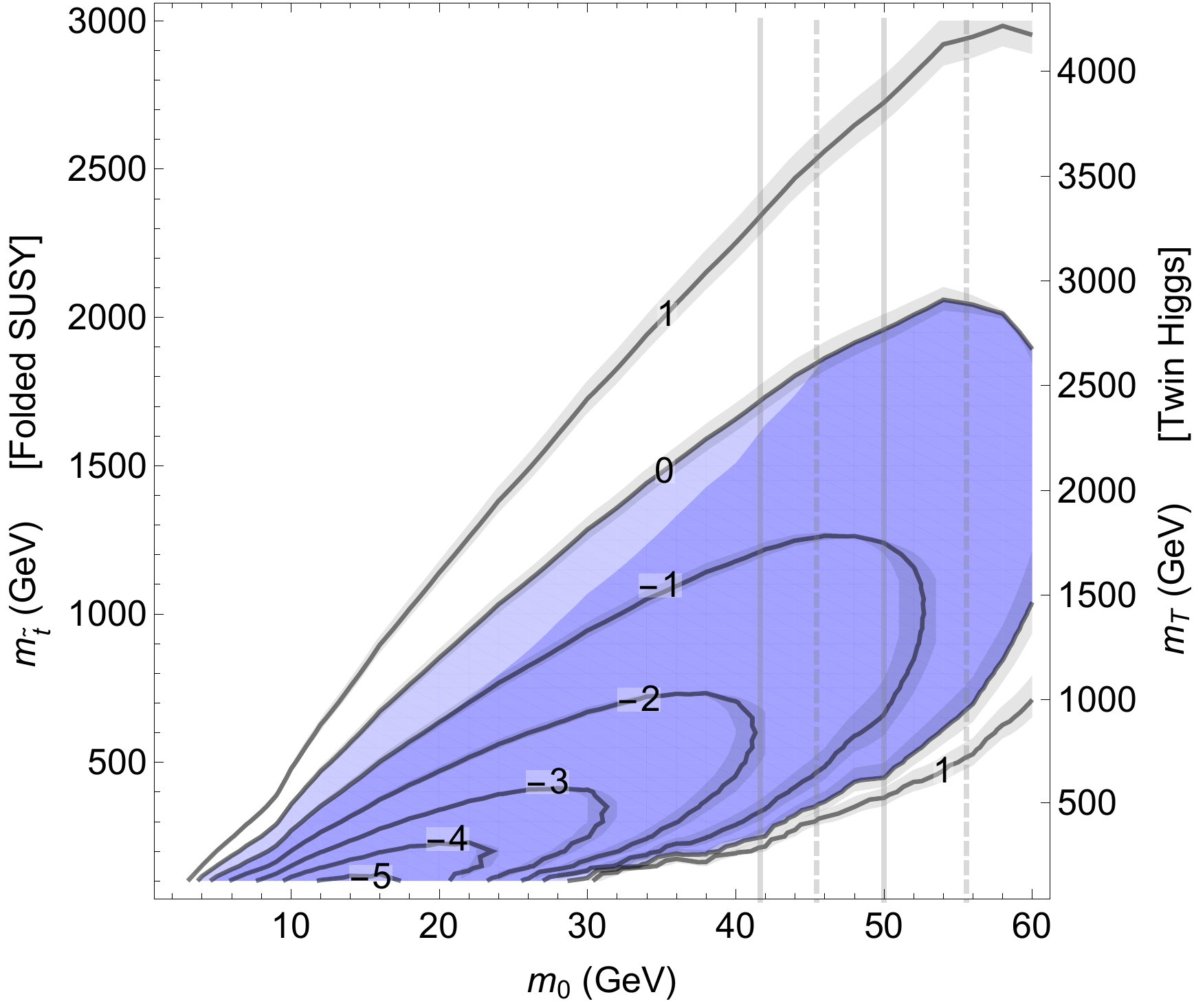}
\\

\begin{sideways} \footnotesize \phantom{blablabl} (HCAL)$\times$(HCAL) \end{sideways}
&
\includegraphics[width=\tempwidthone]{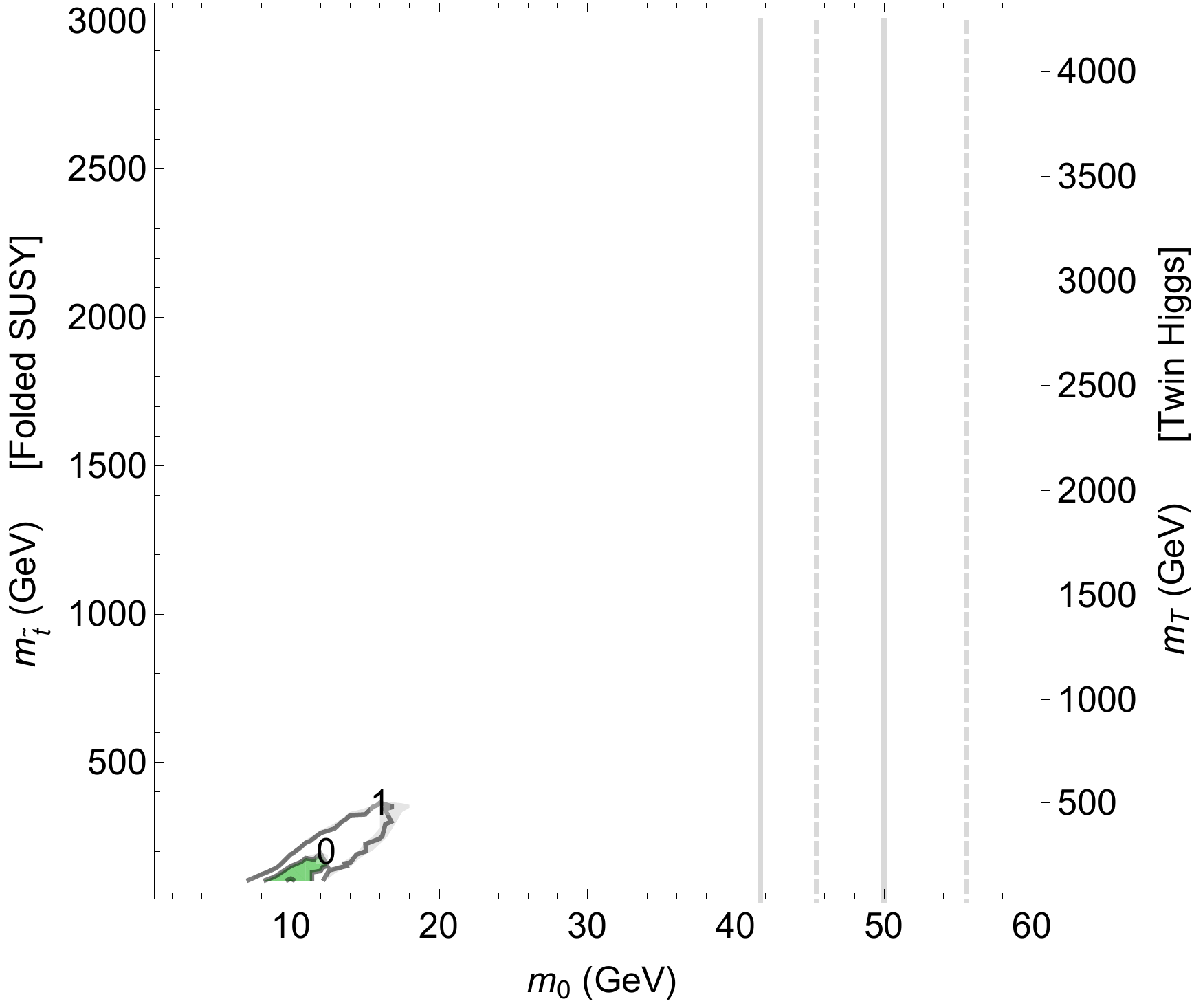}
&&
\includegraphics[width=\tempwidthone]{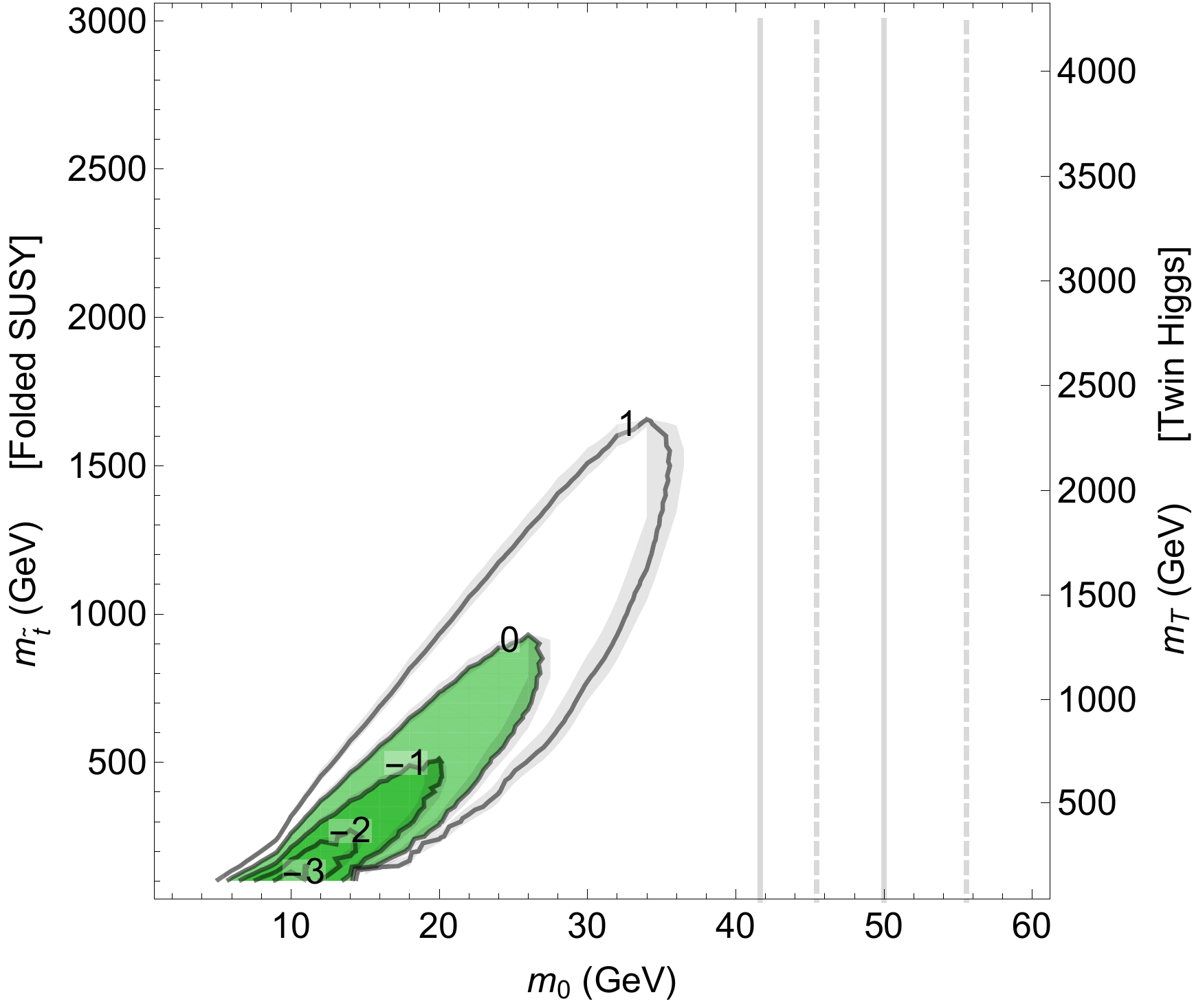}
&&
\includegraphics[width=\tempwidthone]{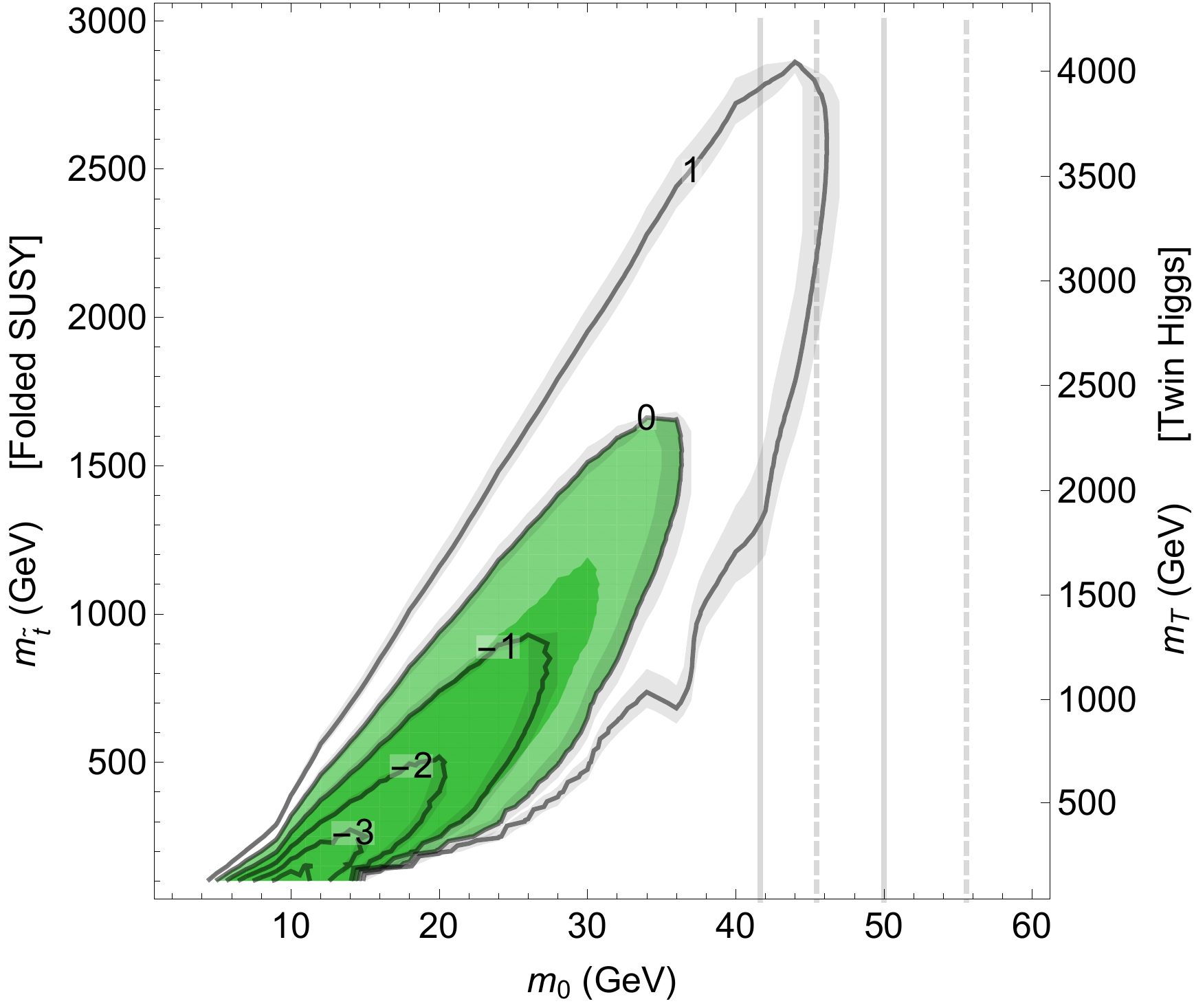}
\\

\begin{sideways} \footnotesize \phantom{blablabl} (MS)$\times$(MS or IT) \end{sideways}
&
\includegraphics[width=\tempwidthone]{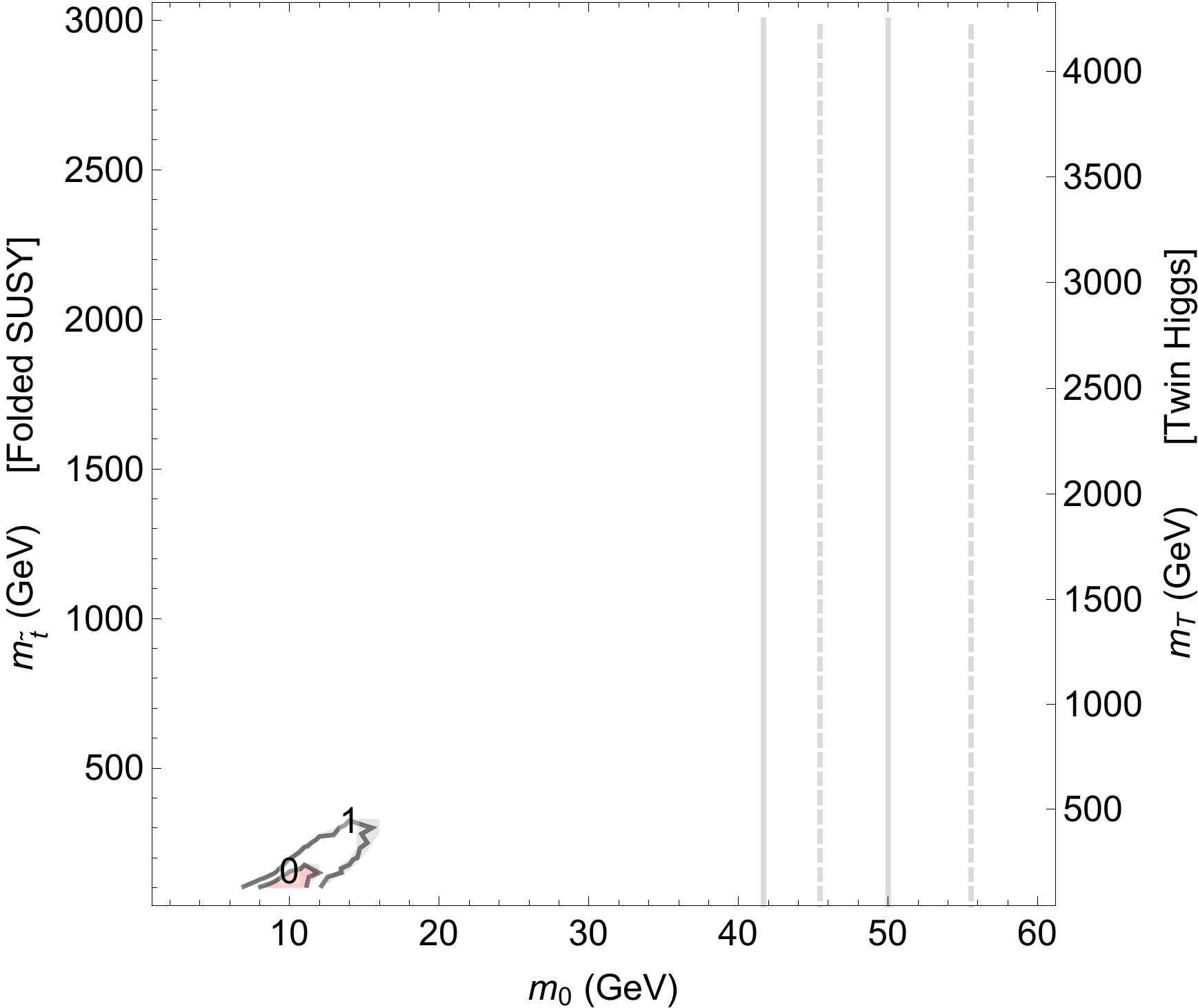}
&&
\includegraphics[width=\tempwidthone]{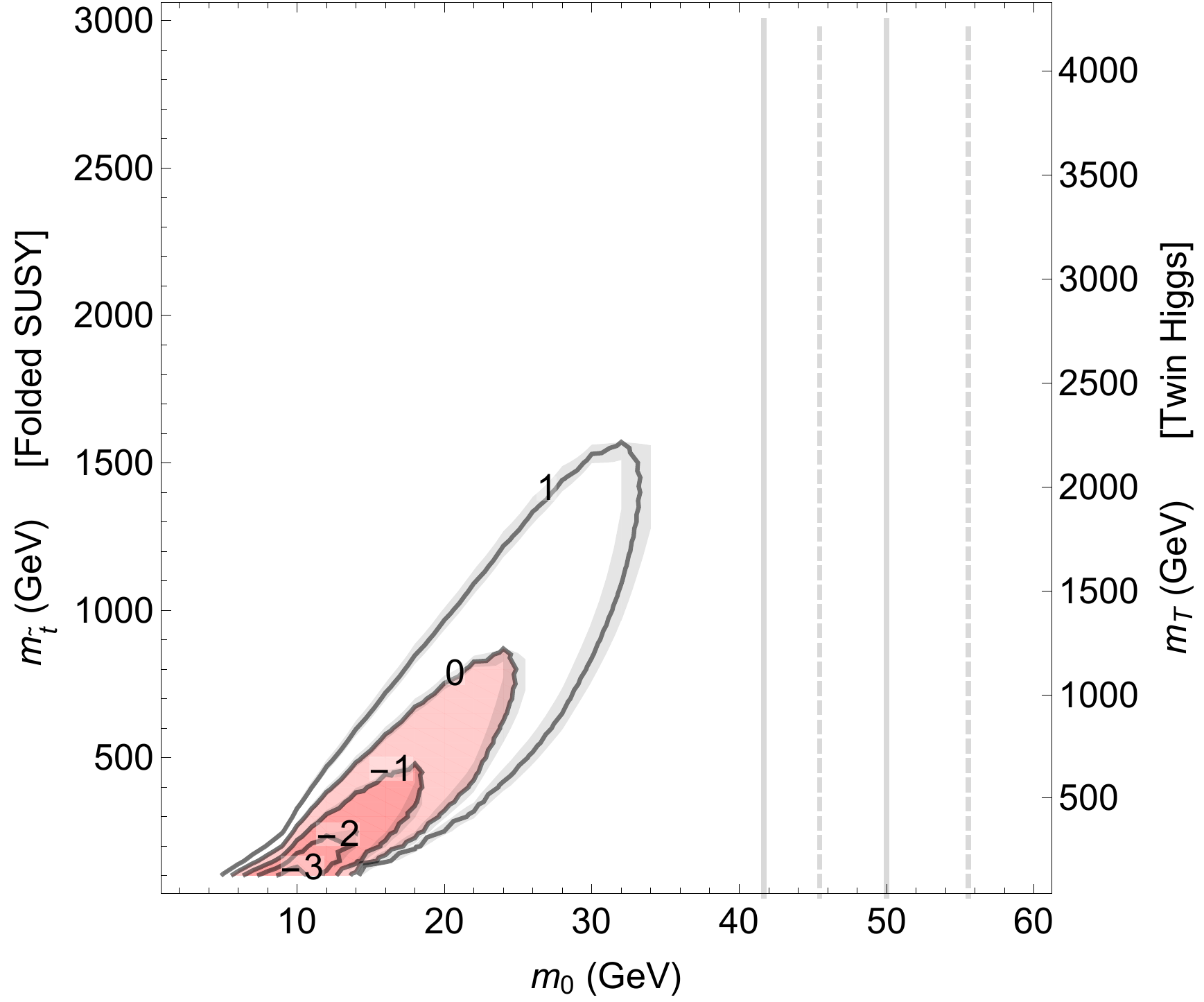}
&&
\includegraphics[width=\tempwidthone]{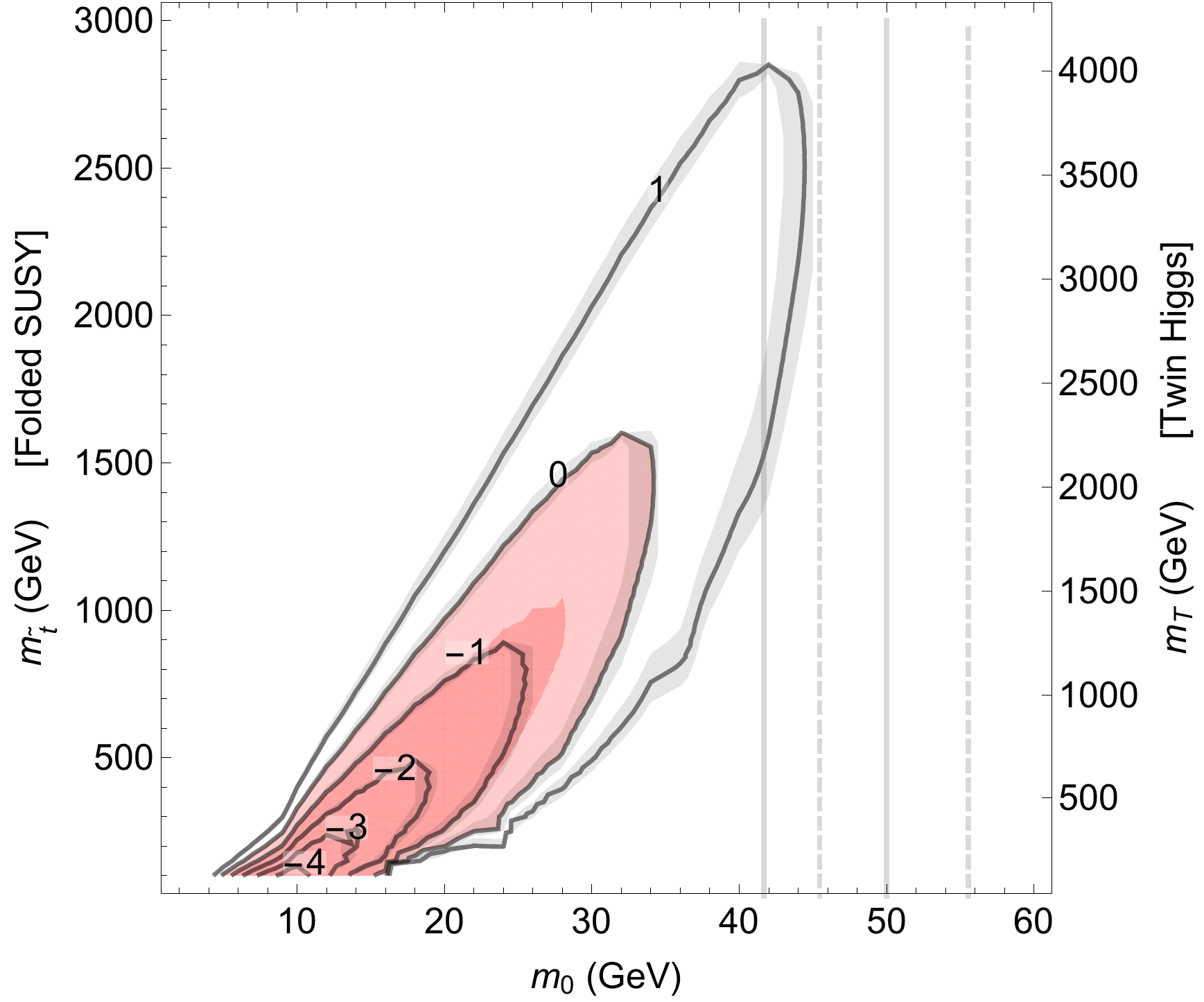}
\vspace*{-3mm}
\end{tabular}
\end{center}
\caption{
Contours show excludable (or discoverable) values of for $\log_{10} \kappa$, the overall factor in \eref{brhglueballs} for $\mathrm{Br}(h\to 0^{++} 0^{++})$,  if $N = 10$ events can be excluded (or discovered) in the four searches (top to bottom row) of \tref{searches}.  Light (dark) shaded colored regions correspond to exclusions on uncolored naturalness models if $\kappa = \kappa_\mathrm{max}$ ($\kappa_\mathrm{min}$), corresponding to optimistic (pessimistic) signal estimate under the assumption that $h$ decays dominantly to two glueballs. 
Shaded bands around contours indicate effect of the $25\%$ uncertainty in $0^{++}$ lifetime. Vertical solid (dashed) lines show where $\kappa$ might be enhanced (suppressed) due to non-perturbative mixing effects, see \ssref{exclusiveproduction}. All other formatting same as \fref{geometricsignalestimate}. 
}
\label{f.realisticsignalestimate}
\end{figure}

\begin{figure}
\begin{center}
\hspace*{-1.7cm}
\begin{tabular}{c}
\includegraphics[width=17cm]{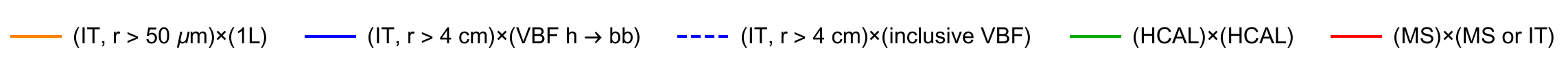}
\\
\begin{tabular}{cc}
\includegraphics[width=9cm]{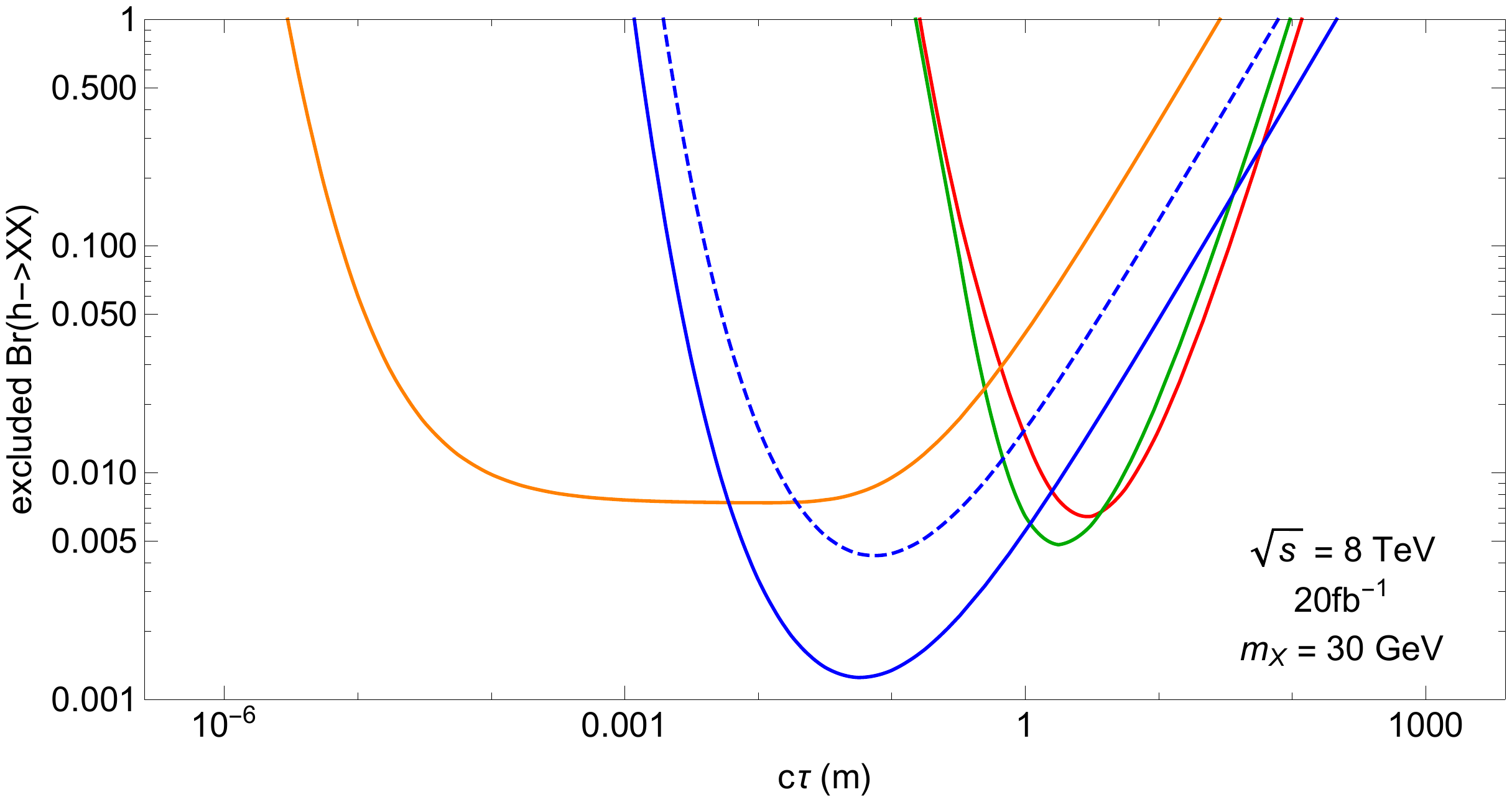}
& 
\includegraphics[width=9cm]{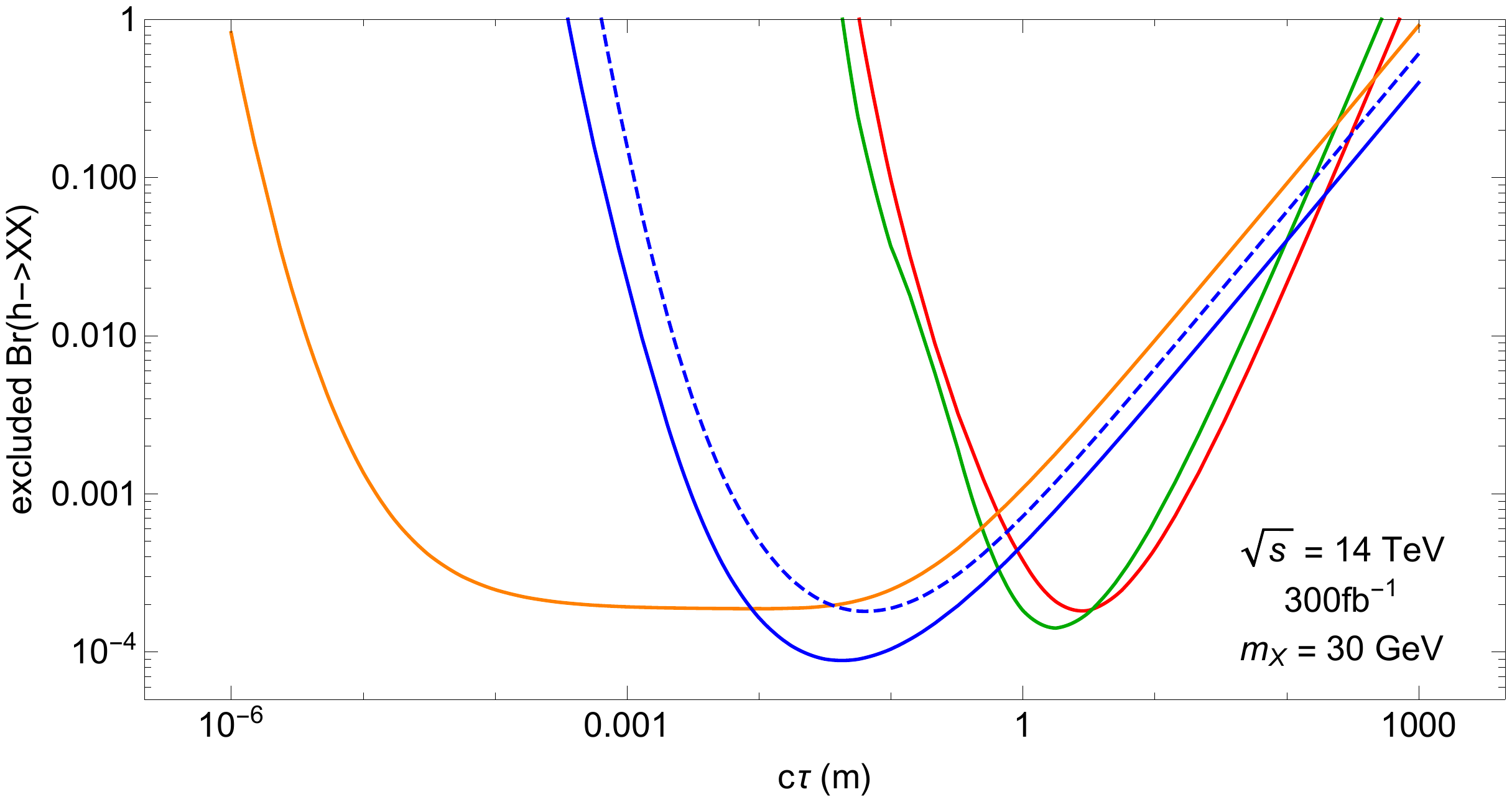}
\\
\includegraphics[width=9cm]{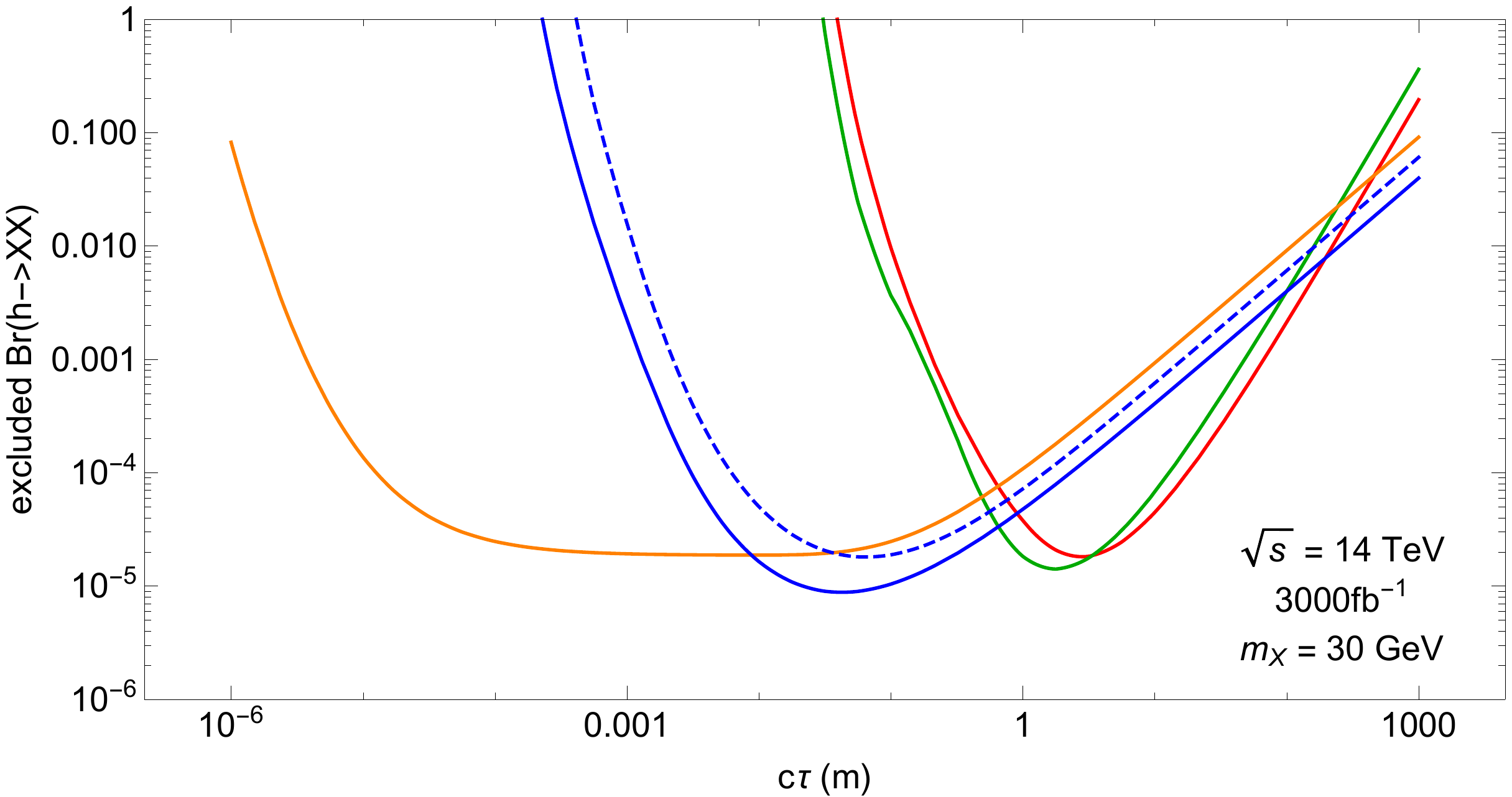}
 &
\includegraphics[width=9cm]{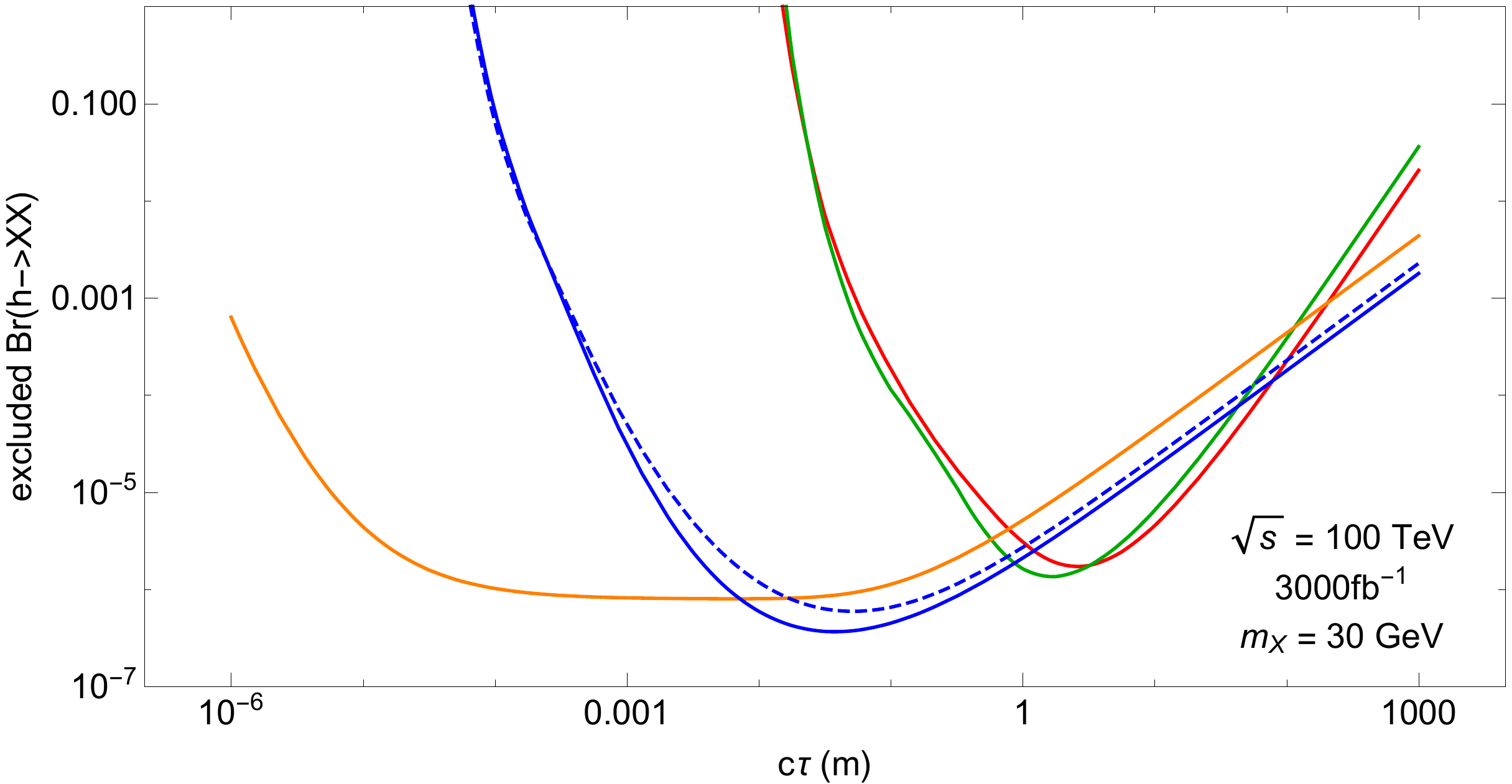}
\end{tabular}
\end{tabular}
\end{center}
\caption{
Projected sensitivities of the displaced decay searches listed in \tref{searches}, expressed model-independently as limits on the exotic Higgs decay branching ratio $\mathrm{Br}(h \to X X)$ as a function of the proper lifetime $c \tau$ of X, for $m_X = 30 \gev$. Zero background is assumed, which is not likely to be realistic for the HCAL search.  Decreasing (increasing) $m_X$ shifts the curves slightly to the left (right).
}
\label{f.brlimits}
\end{figure}

The results are presented in \fref{realisticsignalestimate}.
The dark (light) shaded colored regions indicate which regions of the  $(m_0, m_\text{TP})$ parameter space give more than $N = 10$ signal detected signal events for $\kappa = \kappa_\mathrm{max}$ ($\kappa_\mathrm{min}$), which for a relatively background-free search should approximate discovery or exclusion potential.  The black contours indicate which value of $\kappa$ is required to give $N = 10$ across the whole parameter space. This allows for an easy rescaling of the actual parameter space exclusions for different hypotheses of what $\kappa(m_0)$ might be. It also makes clear that our conclusions are robust even assuming $\mathcal{O}(1)$ uncertainties for our signal estimate (as might be the case if additional cuts are required to reduce background to zero in a realistic analysis). 

We also show the sensitivity of these searches in a model-independent way, as projected limits on the exotic Higgs decay branching ratio $\mathrm{Br}(h \to X X)$ as a function of lifetime for $m_X = 30 \gev$ in \fref{brlimits}. 

The 8 TeV (HCAL)$\times$(HCAL) and  (MS)$\times$(MS or IT) searches have very little sensitivity to uncolored naturalness, and only probe a small part of parameter space with very light top partners.\footnote{That region is already probed by $h\to \gamma \gamma$ signal strength measurements and other Higgs coupling measurements for the case of Folded SUSY and Twin Higgs respectively~\cite{Burdman:2014zta}.}
The MS limits in \fref{brlimits} agree well with the experimental exclusions \cite{Aad:2015uaa}, while our background-free assumption overestimates the sensitivity of the HCAL compared to the published limits \cite{Aad:2015asa}, as expected.
If the two new searches with a single DV in the IT were performed on the run 1 dataset, mirror glueballs lighter than about 40 GeV could be probed for top partner masses of about 100 - 300 GeV.

At the HL-LHC, the most coverage is achieved by looking for one DV in association with a lepton. For lighter glueballs, jets + one DV or two DVs in the muon system provide additional coverage. The jets + one DV search could additionally cover much of the same parameter space as the lepton + one DV search if DV reconstruction down to $r_\mathrm{min} = 50$ $\mu$m was possible for that search as well. Overall, the HL-LHC should be able to probe uncolored naturalness via Higgs to glueball decay with top partner masses up to about a TeV for a wide range of theoretically motivated glueball masses.

The reach at the 14 TeV LHC with only 300 $\ifb$ can be easily read off from \fref{realisticsignalestimate} by shifting the exclusions one $\log_{10}\kappa$ contour inwards, corresponding to a factor of 10 reduction in signal compared to the HL-LHC. Most of the glueball masses are covered, and top partners up to 500 - 700 GeV can be probed, though the lepton + one DV search looses sensitivity for $m_0 \sim 60 \gev, m_{\text{TP}} \sim 200 \gev$. 

The 100 TeV results are meant to be illustrative only, since the assumptions we used are driven by the limitations of present-day experiments. By the time the next collider is built, it is likely that full track reconstruction can be used for low-level triggering, if triggering is needed at all. Displaced vertex reconstruction capabilities might be superior as well. That being said, new backgrounds e.g. from $B$ and $D$ decays, which are not essential at current energies and luminosities, may play a role at 100 TeV. Even so, our estimates serve to demonstrate an enticing potential sensitivity to exotic Higgs decays with glueballs in the final state. This provides strong motivation to make sure such relatively soft signatures, which in this case give access to multi-TeV scale top partner masses, are not missed in the detector design of future machines.

We have used the VBF $h \to \bar b b$ and the inclusive VBF trigger for our example of a (jets)$\times$(IT) type search.  In \fref{jetcomparison} we show how sensitivity depends on the type of jet trigger utilized for a search at the HL-LHC. All the jet triggers have roughly comparable performance. 

Finally, it is possible that other glueball decays provide additional sensitivity in certain parts of parameter space. For example, the $2^{++}$ glueball has a mass of $\approx 1.4 m_0$ and a lifetime several orders of magnitude longer than the $0^{++}$ state \cite{Juknevich:2009gg}. Depending on the details of mirror hadronization, $h \to 2^{++} + X$ decays may therefore produce DV's in the tracker or MS for $30 - 40 \gev \lesssim m_0 < 52 \gev$, when decays to $2^{++}$ are kinematically allowed, and $0^{++}$ is relatively short-lived.

The overall lesson of our investigation is clear. The LHC has great potential to probe uncolored naturalness. Exotic Higgs decays to mirror glueballs, with reconstruction of the subsequent displaced decay down to $50$ $\mu$m from the interaction point, gives sensitivity to glueballs across the theoretically preferred 12 - 60 GeV range, with uncolored top partner masses up to around a TeV.

\begin{figure}
\begin{center}
\includegraphics[width=7cm]{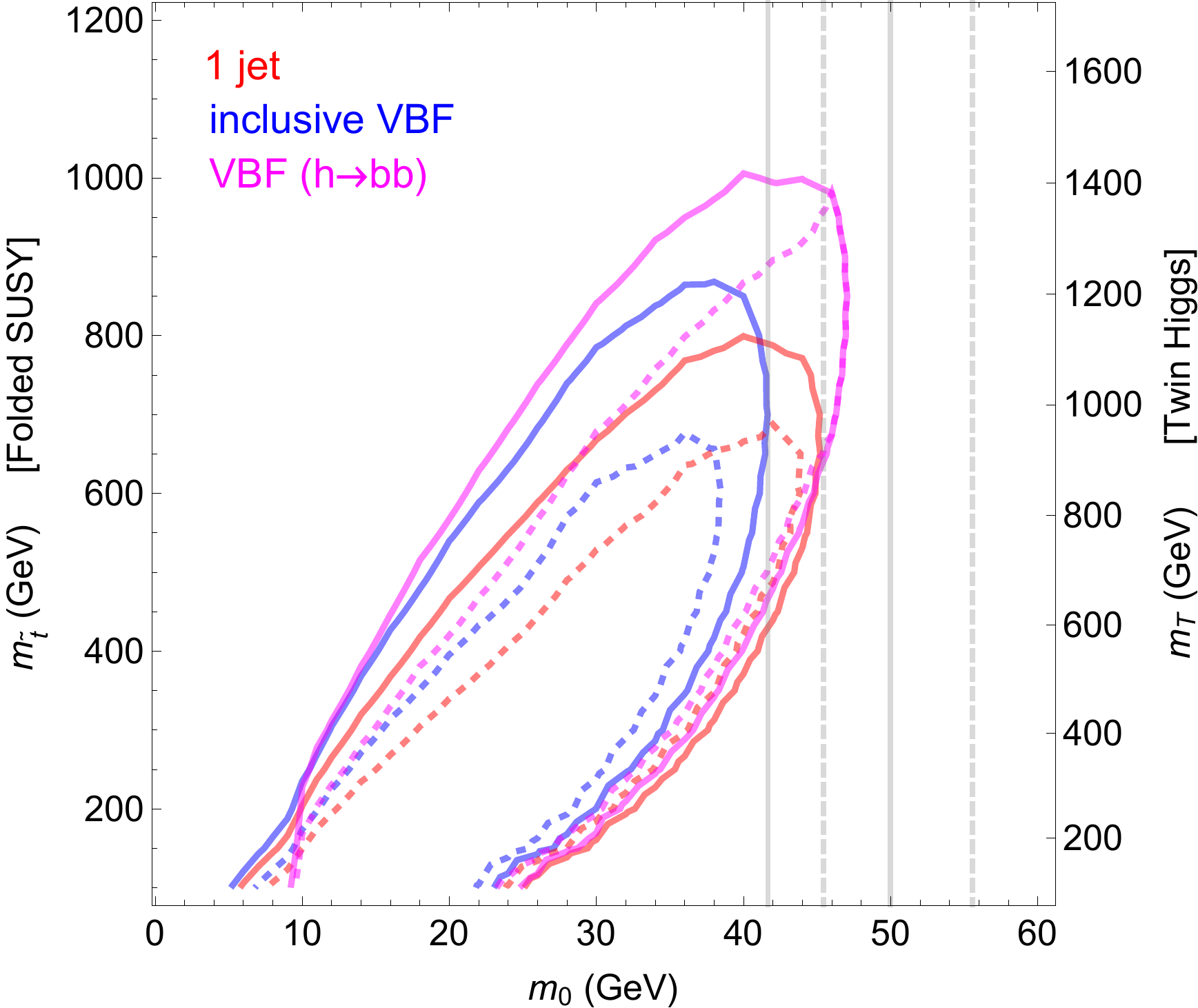}
\end{center}
\caption{
Comparison of different (IT, $r > 4$cm)$\times$(jet trigger) searches for $\sqrt{s} = 14 \tev$ with $3000 \ifb$. The solid (dashed) lines bound regions with more than 10 signal events for $\kappa = \kappa_\mathrm{max}$ ($\kappa_\mathrm{min}$). For the purpose of this comparison, glueball lifetime uncertainties are not shown.
}
\label{f.jetcomparison}
\end{figure}

\section{Glueball Production through Mirror Partner Annihilation}\label{s.toppartners}

Exotic Higgs decays are not the only production mode for mirror glueballs. Top partners, or any mirror particles, can be pair-produced directly via their coupling to the SM-like Higgs or (in the case of EW-charged hidden sectors) DY-like processes. If the partner pair is charged under mirror QCD it will form a stable excited `quirky' bound state that de-excites via soft glueball and/or photon emission \cite{Kang:2008ea,Burdman:2008ek,Harnik:2008ax,Harnik:2011mv}  until it annihilates into mirror gluons, Higgs bosons, and possibly other particles in the visible or mirror sectors. 

Mirror gluons resulting from top partner annihilation undergo a perturbative pure-gauge parton-shower, then hadronize into jets of mirror glueballs. One would reasonably expect at least $\sim 1/10$ of these glueballs to be the lightest, $0^{++}$, state, with total glueball multiplicities per jet $\mathcal{O}(m_\text{TP}/m_0)$ for $m_\text{TP} \lesssim \mathcal{O}(10 m_0)$.\footnote{By analogy to hadron multiplicity for jets from $e^+ e^-$ annihilation with $\sqrt{s} \lesssim 10 \gev$ \cite{pdg}.} In much of the parameter space of uncolored naturalness, these $0^{++}$ decay within the tracker volume, which allows mirror partner pair production events to be tagged by either reconstructing DVs, or observing a very $b$- and $\tau$-rich final state from $0^{++}$ decay.\footnote{For some related signatures, see \cite{Schwaller:2015gea, Cohen:2015toa}.} For DY production of EW-charged top partners, rudimentary signal estimates indicate that such channels may be competitive with exotic Higgs decays for mirror glueball discovery.

For large regions of their parameter space, mirror partners annihilate into $hh$ about $1/10$ as often as into mirror gluons~\cite{Martin:2008sv,Batell:2015zla}. Including reconstruction and $b$-tagging efficiencies, we expect the number of detectable $hh$ events in the $4b$ final state \cite{Aad:2015uka} to be an $\mathcal{O}(10)$ factor smaller than the number of events with mirror glueballs, but the $hh$ final state may be the best avenue for measuring quirkonium mass (and hence the top partner masses). This could also be possible with the glueball jet final state, but it is unclear to what extent the larger signal rate compensates for greater difficulty in reconstructing the initial particle mass in a single event.

Additional complications may also arise if the produced partner-pair beta-decays before annihilating. In this case the emitted lepton, if it is in the visible sector as in Folded SUSY or Quirkly Little Higgs, may contain information on the mirror sector spectrum, but for the Twin Higgs models a mass measurement would become much more difficult. 

The prospect of using top partner pair production to discover uncolored naturalness is an important alternative avenue to be explored. It is made even more attractive by the possibility of measuring masses and couplings in the mirror sector, both directly (as described above) and by correlation with measurements of exotic Higgs decays and, perhaps, glueball lifetime measurements. Following discovery of mirror glueballs, this would allow uncolored naturalness models to be distinguished from generic Hidden Valleys, and could confirm that the little hierarchy problem is solved by uncolored top partners. We will investigate partner annihilation signatures in future work.

\begin{figure}
\begin{center}
\begin{tabular}{ccc}
\includegraphics[height=6.6cm]{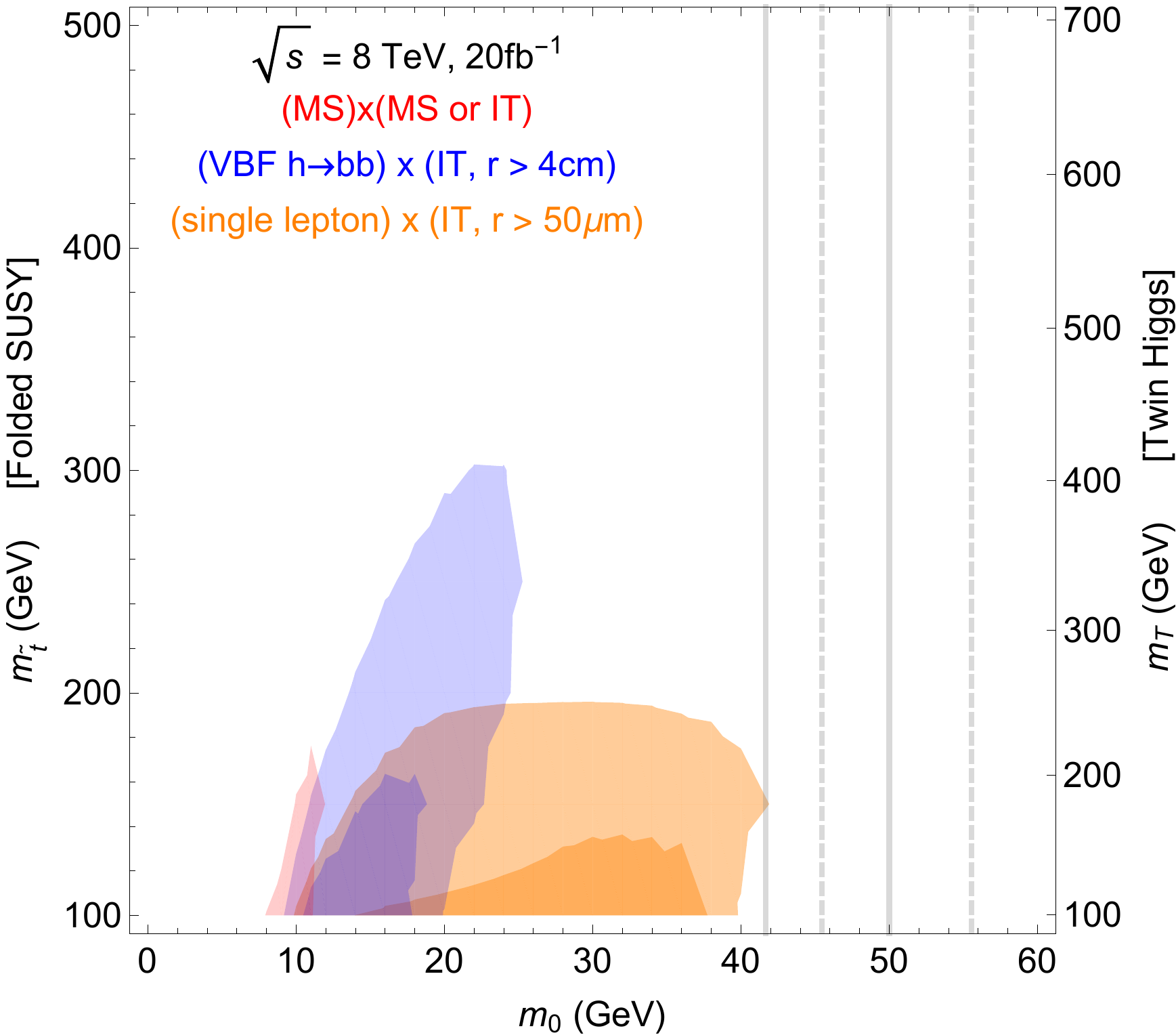} 
& &
\includegraphics[height=6.6cm]{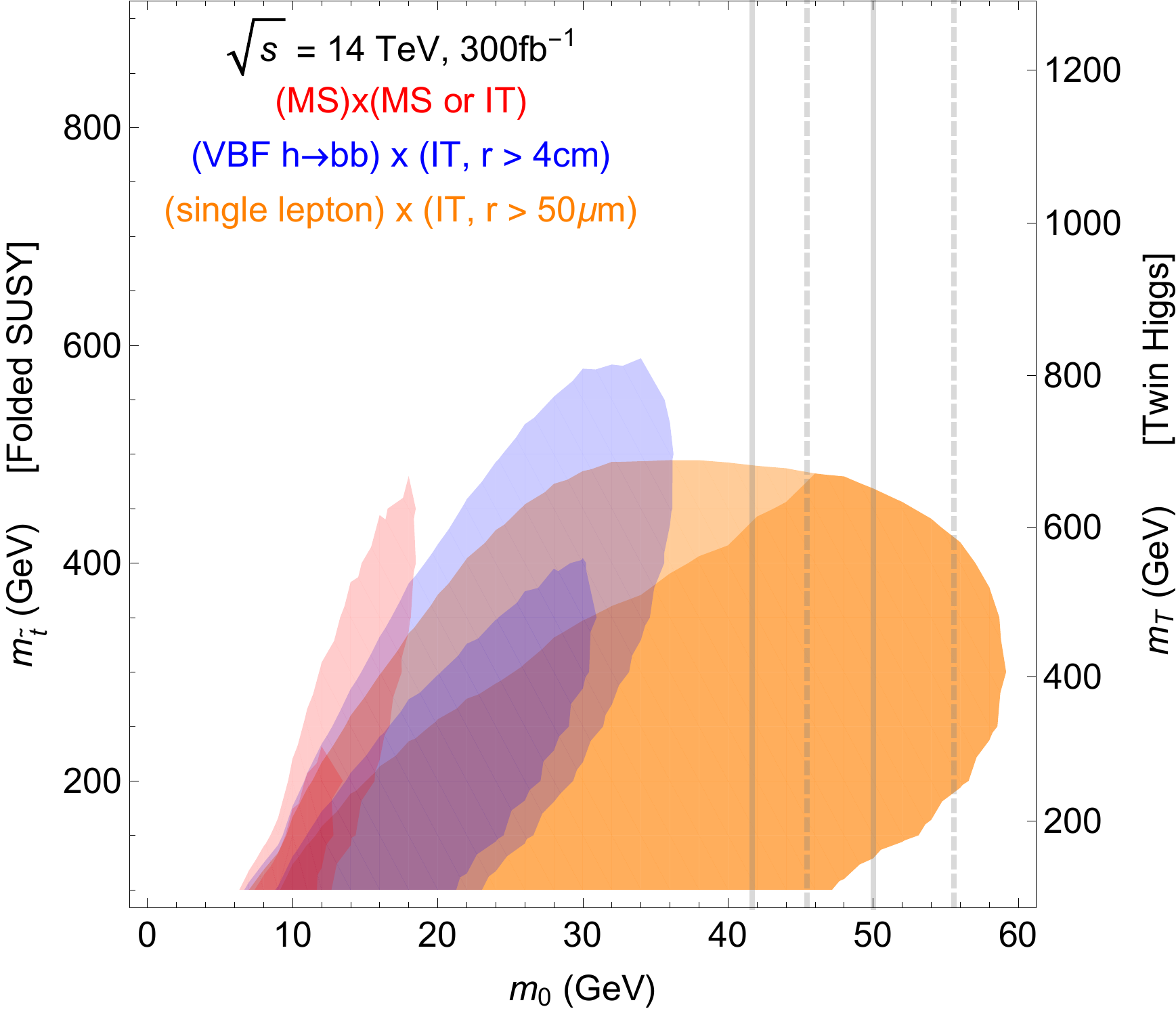} 
\\ \\
\includegraphics[height=6.6cm]{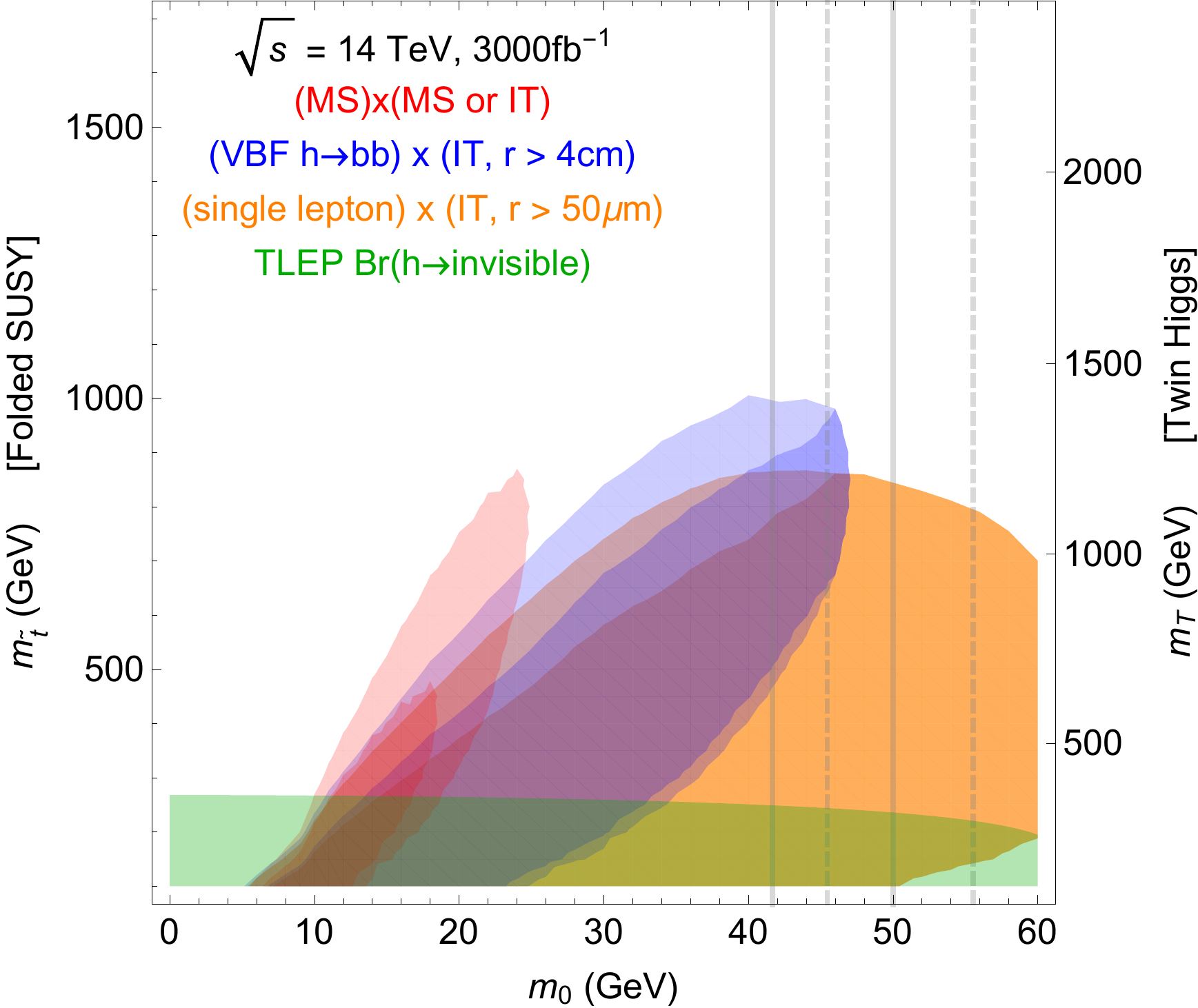}
&&
\includegraphics[height=6.6cm]{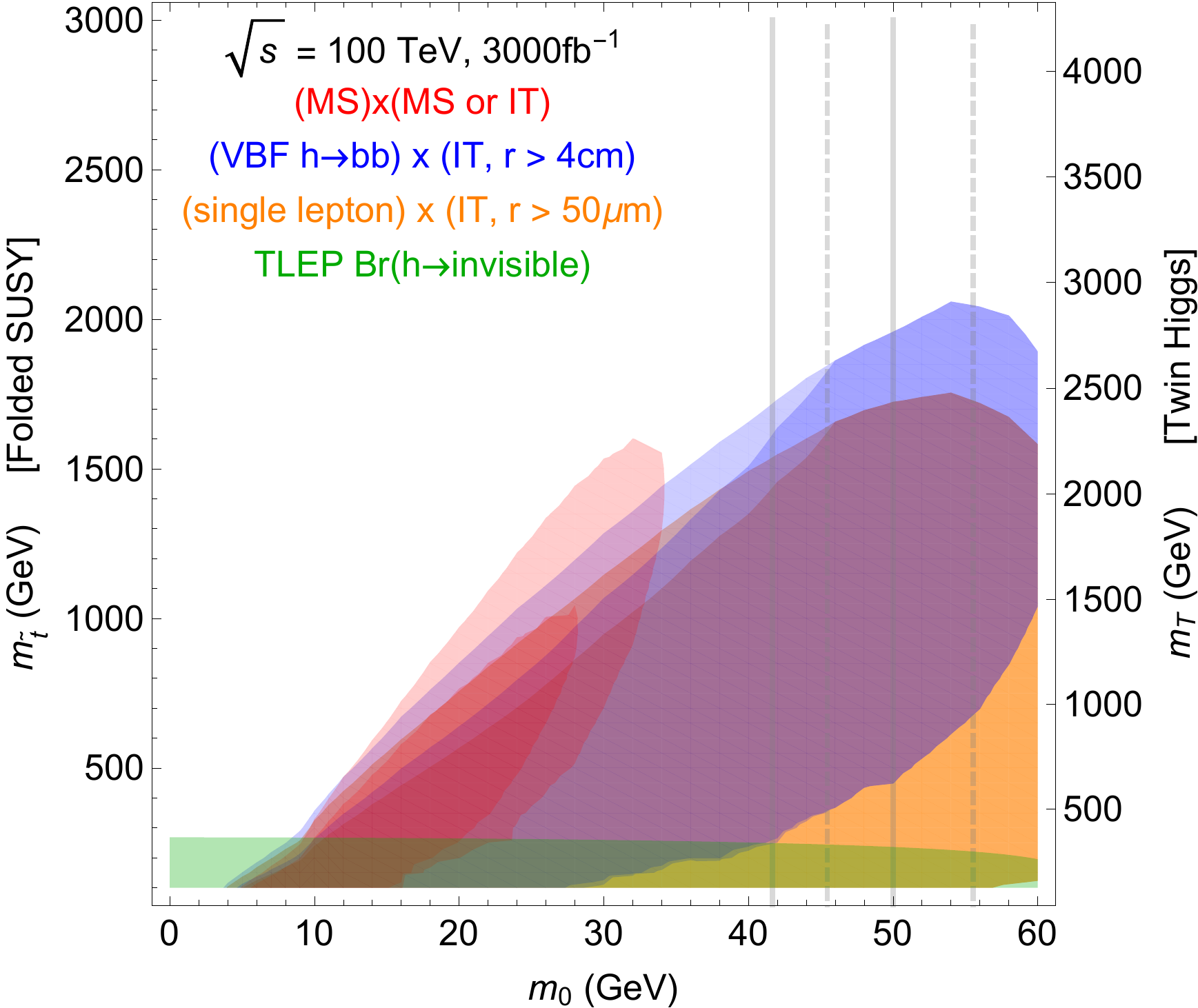}
\end{tabular}
\end{center}
\caption{
Summary of discovery potential at LHC run 1, LHC14 with 300$\ifb$, HL-LHC and 100 TeV if the searches in \tref{searches}, or similar, are approximately background-free, and $\sim$ 10 events allow for discovery.  We omit the HCAL search since it likely is not background-free.
Note different scaling of vertical axes. 
For comparison, the inclusive TLEP $h\to \mathrm{invisible}$ limit, as applied to the perturbative prediction for $\mathrm{Br}(h\to \mathrm{all\ glueballs})$, is shown for future searches as well. 
Lighter and darker shading correspond to the optimistic (pessimistic) signal estimates $\kappa = \kappa_\mathrm{max}$ ($\kappa_\mathrm{min}$), under the assumption that $h$ decays dominantly to two glueballs.
Effect of glueball lifetime uncertainty is small and not shown. 
$m_0$ is the mass of the lightest glueball $0^{++}$; the vertical axes correspond to mirror stop mass in Folded SUSY (see \eref{yoverMFoldedSUSY}) and mirror top mass in Twin Higgs and Quirky Little Higgs (see \eref{yoverMTH}).
Vertical solid (dashed) lines show where $\kappa$ might be enhanced (suppressed) due to non-perturbative mixing effects, see \ssref{exclusiveproduction}.
}
\label{f.summaryplot}
\end{figure}

\section{Conclusion}\label{s.conclusion}

Theories of uncolored naturalness offer a compelling alternative to standard SUSY and composite Higgs theories.
They allow for a natural solution to the little hierarchy problem with phenomenology that is completely distinct from colored top partner scenarios. Far from being undetectable at the LHC, these signatures, while exotic, may still be effectively probed in the near future. We place the experimental signals of uncolored top partner theories in broader context, with Folded SUSY, Quirky Little Higgs, and Twin Higgs theories considered as representatives in a bottom-up theory space, see \tref{modeltable}. This makes clear that the signature of exotic Higgs decays to unstable mirror glueballs, first pointed out explicitly in \cite{Craig:2015pha} with a primary focus on the Fraternal Twin Higgs model, is in fact the smoking gun for models with electroweak-charged mirror sectors. 

In this paper we present the first detailed phenomenological analysis of these exotic Higgs decays to mirror glueballs. RG arguments suggest that the lightest and most unstable glueball, which is a prediction of Folded SUSY and Quirky Little Higgs models and a possible outcome of the Twin Higgs scenario, has a mass in the $\sim 10 - 60 \gev$ range, which makes the discovered 125 GeV SM-like Higgs a powerful probe of uncolored naturalness. A careful treatment of hadronization and non-perturbative uncertainties of the mirror sector allows us to perform an explicit signal estimate for the 8 and 14 TeV LHC, as well as a hypothetical 100 TeV machine. This is greatly aided by efficiency tables for displaced vertex reconstruction released by the ATLAS collaboration \cite{Aad:2015asa, Aad:2015uaa}. We suggest several new searches requiring one displaced vertex in the tracker along with jet or lepton triggers. This approach is vital to probe glueballs with $\gtrsim 30 \gev$ masses, and strongly motivates the reconstruction of displaced vertices with decay lengths $\sim 50 \mu$m.

The discovery potential of various searches is summarized in  \fref{summaryplot}.
Our approach for estimating signal has been conservative, both by assuming only two-body production of $0^{++}$ from Higgs decays and by assuming present-day ATLAS detector capabilities for all future projections.
Even so, the achievable reach at the LHC across the whole range of considered glueball masses  is impressive. Folded SUSY stops could be discovered with masses up to 600 (1100) GeV at the LHC with 300 (3000) $\ifb$ of data, while Twin Higgs top partners could be accessible with masses up to 900 (1500) GeV.
At a 100 TeV collider, top partner masses in excess of 2 TeV are easily probed. This allows for an exciting complementarity between measuring the low-energy consequences of uncolored naturalness and directly probing details of the UV completion required by such theories.

\fref{summaryplot} also shows the TLEP limit \cite{Dawson:2013bba} on $\mathrm{Br}(h \to \mathrm{invisible})$, as applied to the perturbative prediction for $\mathrm{Br}(h\to \mathrm{all\ glueballs})$.  However, lepton colliders could also set powerful limits by directly looking for prompt or displaced $h\to 4b$ decays. This channel  deserves future study. 
Indirect constraints on Folded SUSY from Higgs coupling measurements will only constrain for top partner masses $\lesssim 350 \gev$ \cite{Burdman:2014zta}.
Direct production of first and second generation Folded SUSY mirror squarks would be a distinctive signal, but their mass is not tied directly to the little hierarchy problem and they could easily escape detection in a natural theory \cite{Burdman:2014zta}. 
In the Twin and Quirky Little Higgs models, precision Higgs measurements with $3000\ifb$ of data may probe top partners up to $\lesssim 800 \gev$ (depending on the cutoff). Exotic Higgs decay searches could easily surpass this sensitivity.

All the searches we examine lose sensitivity if the glueball mass is below the $\bar b b$ threshold. This leads to very long glueball lifetimes and very few decays in the detector. While theories with electroweak top partners (FSUSY, Quirky Little Higgs) motivate glueballs heavier than 12 GeV, such light and long-lived glueballs can easily be realized in Twin Higgs models, and developing searches with sensitivity to these scenarios is highly motivated.

It has long been understood that future lepton colliders allow for the detection of all known natural Twin Higgs scenarios through precision Higgs coupling measurements. We demonstrated that the LHC has sensitivity to TeV-scale top partners through exotic Higgs decays to mirror glueballs, which covers Folded SUSY and Quirky Little Higgs theories. This complementarity puts  all known uncolored top partner theories within our reach, and might allow us to eventually probe naturalness in all its forms.

\acknowledgments
We are grateful to Nima Arkani-Hamed, Zackaria Chacko, Andrea Coccaro, Nathaniel Craig, Tim Cohen, Andy Haas, Roni Harnik, Jos\'e Juknevich, Andrey Katz, Henry Lubatti, Heather Russell, Matthew Strassler, and Raman Sundrum for useful discussions. 
D.C. and C.V. are supported by the National Science Foundation Grant No. PHY-1315155 and by the Maryland Center for
Fundamental Physics.


\bibliography{detectcolorlesstops}

\providecommand{\href}[2]{#2}\begingroup\raggedright\begin{thebibliography}{10}

\bibitem{Aad:2012tfa}
{\bf ATLAS} Collaboration, G.~Aad et~al., {\it {Observation of a new particle
  in the search for the Standard Model Higgs boson with the ATLAS detector at
  the LHC}},  {\em Phys.Lett.} {\bf B716} (2012) 1--29,
  [\href{http://arxiv.org/abs/1207.7214}{{\tt arXiv:1207.7214}}].

\bibitem{Chatrchyan:2012ufa}
{\bf CMS} Collaboration, S.~Chatrchyan et~al., {\it {Observation of a new boson
  at a mass of 125 GeV with the CMS experiment at the LHC}},  {\em Phys.Lett.}
  {\bf B716} (2012) 30--61, [\href{http://arxiv.org/abs/1207.7235}{{\tt
  arXiv:1207.7235}}].

\bibitem{Graham:2015cka}
P.~W. Graham, D.~E. Kaplan, and S.~Rajendran, {\it {Cosmological Relaxation of
  the Electroweak Scale}},  \href{http://arxiv.org/abs/1504.07551}{{\tt
  arXiv:1504.07551}}.

\bibitem{Martin:1997ns}
S.~P. Martin, {\it {A Supersymmetry primer}},  {\em Adv.Ser.Direct.High Energy
  Phys.} {\bf 21} (2010) 1--153,
  [\href{http://arxiv.org/abs/hep-ph/9709356}{{\tt hep-ph/9709356}}].

\bibitem{Bellazzini:2014yua}
B.~Bellazzini, C.~Csáki, and J.~Serra, {\it {Composite Higgses}},  {\em
  Eur.Phys.J.} {\bf C74} (2014), no.~5 2766,
  [\href{http://arxiv.org/abs/1401.2457}{{\tt arXiv:1401.2457}}].

\bibitem{ArkaniHamed:2001nc}
N.~Arkani-Hamed, A.~G. Cohen, and H.~Georgi, {\it {Electroweak symmetry
  breaking from dimensional deconstruction}},  {\em Phys.Lett.} {\bf B513}
  (2001) 232--240, [\href{http://arxiv.org/abs/hep-ph/0105239}{{\tt
  hep-ph/0105239}}].

\bibitem{ArkaniHamed:2002qx}
N.~Arkani-Hamed, A.~Cohen, E.~Katz, A.~Nelson, T.~Gregoire, et~al., {\it {The
  Minimal moose for a little Higgs}},  {\em JHEP} {\bf 0208} (2002) 021,
  [\href{http://arxiv.org/abs/hep-ph/0206020}{{\tt hep-ph/0206020}}].

\bibitem{ArkaniHamed:2002qy}
N.~Arkani-Hamed, A.~Cohen, E.~Katz, and A.~Nelson, {\it {The Littlest Higgs}},
  {\em JHEP} {\bf 0207} (2002) 034,
  [\href{http://arxiv.org/abs/hep-ph/0206021}{{\tt hep-ph/0206021}}].

\bibitem{Schmaltz:2004de}
M.~Schmaltz, {\it {The Simplest little Higgs}},  {\em JHEP} {\bf 0408} (2004)
  056, [\href{http://arxiv.org/abs/hep-ph/0407143}{{\tt hep-ph/0407143}}].

\bibitem{Contino:2003ve}
R.~Contino, Y.~Nomura, and A.~Pomarol, {\it {Higgs as a holographic
  pseudoGoldstone boson}},  {\em Nucl.Phys.} {\bf B671} (2003) 148--174,
  [\href{http://arxiv.org/abs/hep-ph/0306259}{{\tt hep-ph/0306259}}].

\bibitem{Agashe:2004rs}
K.~Agashe, R.~Contino, and A.~Pomarol, {\it {The Minimal composite Higgs
  model}},  {\em Nucl.Phys.} {\bf B719} (2005) 165--187,
  [\href{http://arxiv.org/abs/hep-ph/0412089}{{\tt hep-ph/0412089}}].

\bibitem{Contino:2006qr}
R.~Contino, L.~Da~Rold, and A.~Pomarol, {\it {Light custodians in natural
  composite Higgs models}},  {\em Phys.Rev.} {\bf D75} (2007) 055014,
  [\href{http://arxiv.org/abs/hep-ph/0612048}{{\tt hep-ph/0612048}}].

\bibitem{Chatrchyan:2013xna}
{\bf CMS} Collaboration, S.~Chatrchyan et~al., {\it {Search for top-squark pair
  production in the single-lepton final state in pp collisions at $\sqrt{s}$ =
  8 TeV}},  {\em Eur.Phys.J.} {\bf C73} (2013), no.~12 2677,
  [\href{http://arxiv.org/abs/1308.1586}{{\tt arXiv:1308.1586}}].

\bibitem{CMS:2014wsa}
{\bf CMS} Collaboration, C{MS Collaboration}, {\it {Exclusion limits on gluino
  and top-squark pair production in natural SUSY scenarios with inclusive razor
  and exclusive single-lepton searches at 8 TeV.}}, .

\bibitem{Aad:2014bva}
{\bf ATLAS} Collaboration, G.~Aad et~al., {\it {Search for direct pair
  production of the top squark in all-hadronic final states in proton-proton
  collisions at $\sqrt{s}=8$ TeV with the ATLAS detector}},  {\em JHEP} {\bf
  1409} (2014) 015, [\href{http://arxiv.org/abs/1406.1122}{{\tt
  arXiv:1406.1122}}].

\bibitem{Aad:2014kra}
{\bf ATLAS} Collaboration, G.~Aad et~al., {\it {Search for top squark pair
  production in final states with one isolated lepton, jets, and missing
  transverse momentum in $\sqrt s =$8 TeV $pp$ collisions with the ATLAS
  detector}},  {\em JHEP} {\bf 1411} (2014) 118,
  [\href{http://arxiv.org/abs/1407.0583}{{\tt arXiv:1407.0583}}].

\bibitem{Fan:2011yu}
J.~Fan, M.~Reece, and J.~T. Ruderman, {\it {Stealth Supersymmetry}},  {\em
  JHEP} {\bf 1111} (2011) 012, [\href{http://arxiv.org/abs/1105.5135}{{\tt
  arXiv:1105.5135}}].

\bibitem{Martin:2007gf}
S.~P. Martin, {\it {Compressed supersymmetry and natural neutralino dark matter
  from top squark-mediated annihilation to top quarks}},  {\em Phys.Rev.} {\bf
  D75} (2007) 115005, [\href{http://arxiv.org/abs/hep-ph/0703097}{{\tt
  hep-ph/0703097}}].

\bibitem{Martin:2008sv}
S.~P. Martin, {\it {Diphoton decays of stoponium at the Large Hadron
  Collider}},  {\em Phys.Rev.} {\bf D77} (2008) 075002,
  [\href{http://arxiv.org/abs/0801.0237}{{\tt arXiv:0801.0237}}].

\bibitem{LeCompte:2011fh}
T.~J. LeCompte and S.~P. Martin, {\it {Compressed supersymmetry after 1/fb at
  the Large Hadron Collider}},  {\em Phys.Rev.} {\bf D85} (2012) 035023,
  [\href{http://arxiv.org/abs/1111.6897}{{\tt arXiv:1111.6897}}].

\bibitem{Belanger:2012mk}
G.~Belanger, M.~Heikinheimo, and V.~Sanz, {\it {Model-Independent Bounds on
  Squarks from Monophoton Searches}},  {\em JHEP} {\bf 1208} (2012) 151,
  [\href{http://arxiv.org/abs/1205.1463}{{\tt arXiv:1205.1463}}].

\bibitem{Rolbiecki:2012gn}
K.~Rolbiecki and K.~Sakurai, {\it {Constraining compressed supersymmetry using
  leptonic signatures}},  {\em JHEP} {\bf 1210} (2012) 071,
  [\href{http://arxiv.org/abs/1206.6767}{{\tt arXiv:1206.6767}}].

\bibitem{Curtin:2014zua}
D.~Curtin, P.~Meade, and P.-J. Tien, {\it {Natural SUSY in Plain Sight}},  {\em
  Phys.Rev.} {\bf D90} (2014), no.~11 115012,
  [\href{http://arxiv.org/abs/1406.0848}{{\tt arXiv:1406.0848}}].

\bibitem{Kim:2014eva}
J.~S. Kim, K.~Rolbiecki, K.~Sakurai, and J.~Tattersall, {\it {'Stop' that
  ambulance! New physics at the LHC?}},  {\em JHEP} {\bf 1412} (2014) 010,
  [\href{http://arxiv.org/abs/1406.0858}{{\tt arXiv:1406.0858}}].

\bibitem{Czakon:2014fka}
M.~Czakon, A.~Mitov, M.~Papucci, J.~T. Ruderman, and A.~Weiler, {\it {Closing
  the stop gap}},  {\em Phys.Rev.Lett.} {\bf 113} (2014), no.~20 201803,
  [\href{http://arxiv.org/abs/1407.1043}{{\tt arXiv:1407.1043}}].

\bibitem{CMS:2014exa}
{\bf CMS} Collaboration, V.~Khachatryan et~al., {\it {Search for stealth
  supersymmetry in events with jets, either photons or leptons, and low missing
  transverse momentum in pp collisions at 8 TeV}},  {\em Phys.Lett.} {\bf B743}
  (2015) 503--525, [\href{http://arxiv.org/abs/1411.7255}{{\tt
  arXiv:1411.7255}}].

\bibitem{Rolbiecki:2015lsa}
K.~Rolbiecki and J.~Tattersall, {\it {Refining light stop exclusion limits with
  $W^+W^-$ cross sections}},  {\em Phys. Lett.} {\bf B750} (2015) 247--251,
  [\href{http://arxiv.org/abs/1505.05523}{{\tt arXiv:1505.05523}}].

\bibitem{An:2015uwa}
H.~An and L.-T. Wang, {\it {Opening up the compressed region of stop searches
  at 13 TeV LHC}},  \href{http://arxiv.org/abs/1506.00653}{{\tt
  arXiv:1506.00653}}.

\bibitem{Chacko:2005pe}
Z.~Chacko, H.-S. Goh, and R.~Harnik, {\it {The Twin Higgs: Natural electroweak
  breaking from mirror symmetry}},  {\em Phys.Rev.Lett.} {\bf 96} (2006)
  231802, [\href{http://arxiv.org/abs/hep-ph/0506256}{{\tt hep-ph/0506256}}].

\bibitem{Burdman:2006tz}
G.~Burdman, Z.~Chacko, H.-S. Goh, and R.~Harnik, {\it {Folded supersymmetry and
  the LEP paradox}},  {\em JHEP} {\bf 0702} (2007) 009,
  [\href{http://arxiv.org/abs/hep-ph/0609152}{{\tt hep-ph/0609152}}].

\bibitem{Cai:2008au}
H.~Cai, H.-C. Cheng, and J.~Terning, {\it {A Quirky Little Higgs Model}},  {\em
  JHEP} {\bf 0905} (2009) 045, [\href{http://arxiv.org/abs/0812.0843}{{\tt
  arXiv:0812.0843}}].

\bibitem{Morningstar:1999rf}
C.~J. Morningstar and M.~J. Peardon, {\it {The Glueball spectrum from an
  anisotropic lattice study}},  {\em Phys.Rev.} {\bf D60} (1999) 034509,
  [\href{http://arxiv.org/abs/hep-lat/9901004}{{\tt hep-lat/9901004}}].

\bibitem{Juknevich:2009ji}
J.~E. Juknevich, D.~Melnikov, and M.~J. Strassler, {\it {A Pure-Glue Hidden
  Valley I. States and Decays}},  {\em JHEP} {\bf 0907} (2009) 055,
  [\href{http://arxiv.org/abs/0903.0883}{{\tt arXiv:0903.0883}}].

\bibitem{Juknevich:2009gg}
J.~E. Juknevich, {\it {Pure-glue hidden valleys through the Higgs portal}},
  {\em JHEP} {\bf 1008} (2010) 121, [\href{http://arxiv.org/abs/0911.5616}{{\tt
  arXiv:0911.5616}}].

\bibitem{Strassler:2006im}
M.~J. Strassler and K.~M. Zurek, {\it {Echoes of a hidden valley at hadron
  colliders}},  {\em Phys.Lett.} {\bf B651} (2007) 374--379,
  [\href{http://arxiv.org/abs/hep-ph/0604261}{{\tt hep-ph/0604261}}].

\bibitem{Strassler:2006ri}
M.~J. Strassler and K.~M. Zurek, {\it {Discovering the Higgs through
  highly-displaced vertices}},  {\em Phys.Lett.} {\bf B661} (2008) 263--267,
  [\href{http://arxiv.org/abs/hep-ph/0605193}{{\tt hep-ph/0605193}}].

\bibitem{Strassler:2006qa}
M.~J. Strassler, {\it {Possible effects of a hidden valley on supersymmetric
  phenomenology}},  \href{http://arxiv.org/abs/hep-ph/0607160}{{\tt
  hep-ph/0607160}}.

\bibitem{Han:2007ae}
T.~Han, Z.~Si, K.~M. Zurek, and M.~J. Strassler, {\it {Phenomenology of hidden
  valleys at hadron colliders}},  {\em JHEP} {\bf 0807} (2008) 008,
  [\href{http://arxiv.org/abs/0712.2041}{{\tt arXiv:0712.2041}}].

\bibitem{Strassler:2008bv}
M.~J. Strassler, {\it {Why Unparticle Models with Mass Gaps are Examples of
  Hidden Valleys}},  \href{http://arxiv.org/abs/0801.0629}{{\tt
  arXiv:0801.0629}}.

\bibitem{Strassler:2008fv}
M.~J. Strassler, {\it {On the Phenomenology of Hidden Valleys with Heavy
  Flavor}},  \href{http://arxiv.org/abs/0806.2385}{{\tt arXiv:0806.2385}}.

\bibitem{Craig:2014aea}
N.~Craig, S.~Knapen, and P.~Longhi, {\it {Neutral Naturalness from Orbifold
  Higgs Models}},  {\em Phys.Rev.Lett.} {\bf 114} (2015), no.~6 061803,
  [\href{http://arxiv.org/abs/1410.6808}{{\tt arXiv:1410.6808}}].

\bibitem{Craig:2014roa}
N.~Craig, S.~Knapen, and P.~Longhi, {\it {The Orbifold Higgs}},  {\em JHEP}
  {\bf 1503} (2015) 106, [\href{http://arxiv.org/abs/1411.7393}{{\tt
  arXiv:1411.7393}}].

\bibitem{Craig:2013fga}
N.~Craig and K.~Howe, {\it {Doubling down on naturalness with a supersymmetric
  twin Higgs}},  {\em JHEP} {\bf 1403} (2014) 140,
  [\href{http://arxiv.org/abs/1312.1341}{{\tt arXiv:1312.1341}}].

\bibitem{Geller:2014kta}
M.~Geller and O.~Telem, {\it {A Holographic Twin Higgs Model}},  {\em
  Phys.Rev.Lett.} {\bf 114} (2015), no.~19 191801,
  [\href{http://arxiv.org/abs/1411.2974}{{\tt arXiv:1411.2974}}].

\bibitem{Batra:2008jy}
P.~Batra and Z.~Chacko, {\it {A Composite Twin Higgs Model}},  {\em Phys.Rev.}
  {\bf D79} (2009) 095012, [\href{http://arxiv.org/abs/0811.0394}{{\tt
  arXiv:0811.0394}}].

\bibitem{Barbieri:2015lqa}
R.~Barbieri, D.~Greco, R.~Rattazzi, and A.~Wulzer, {\it {The Composite Twin
  Higgs scenario}},  {\em JHEP} {\bf 08} (2015) 161,
  [\href{http://arxiv.org/abs/1501.07803}{{\tt arXiv:1501.07803}}].

\bibitem{Low:2015nqa}
M.~Low, A.~Tesi, and L.-T. Wang, {\it {Twin Higgs mechanism and a composite
  Higgs boson}},  {\em Phys.Rev.} {\bf D91} (2015), no.~9 095012,
  [\href{http://arxiv.org/abs/1501.07890}{{\tt arXiv:1501.07890}}].

\bibitem{Craig:2014fka}
N.~Craig and H.~K. Lou, {\it {Scherk-Schwarz Supersymmetry Breaking in 4D}},
  {\em JHEP} {\bf 1412} (2014) 184, [\href{http://arxiv.org/abs/1406.4880}{{\tt
  arXiv:1406.4880}}].

\bibitem{Garcia:2015loa}
I.~G. García, R.~Lasenby, and J.~March-Russell, {\it {Twin Higgs WIMP Dark
  Matter}},  \href{http://arxiv.org/abs/1505.07109}{{\tt arXiv:1505.07109}}.

\bibitem{Craig:2015xla}
N.~Craig and A.~Katz, {\it {The Fraternal WIMP Miracle}},
  \href{http://arxiv.org/abs/1505.07113}{{\tt arXiv:1505.07113}}.

\bibitem{Garcia:2015toa}
I.~García~García, R.~Lasenby, and J.~March-Russell, {\it {Twin Higgs Asymmetric
  Dark Matter}},  {\em Phys. Rev. Lett.} {\bf 115} (2015), no.~12 121801,
  [\href{http://arxiv.org/abs/1505.07410}{{\tt arXiv:1505.07410}}].

\bibitem{Farina:2015uea}
M.~Farina, {\it {Asymmetric Twin Dark Matter}},
  \href{http://arxiv.org/abs/1506.03520}{{\tt arXiv:1506.03520}}.

\bibitem{Schwaller:2015tja}
P.~Schwaller, {\it {Gravitational Waves From a Dark (Twin) Phase Transition}},
  \href{http://arxiv.org/abs/1504.07263}{{\tt arXiv:1504.07263}}.

\bibitem{Batell:2015aha}
B.~Batell and M.~McCullough, {\it {Neutrino Masses from Neutral Top Partners}},
   \href{http://arxiv.org/abs/1504.04016}{{\tt arXiv:1504.04016}}.

\bibitem{Burdman:2008ek}
G.~Burdman, Z.~Chacko, H.-S. Goh, R.~Harnik, and C.~A. Krenke, {\it {The Quirky
  Collider Signals of Folded Supersymmetry}},  {\em Phys.Rev.} {\bf D78} (2008)
  075028, [\href{http://arxiv.org/abs/0805.4667}{{\tt arXiv:0805.4667}}].

\bibitem{Burdman:2014zta}
G.~Burdman, Z.~Chacko, R.~Harnik, L.~de~Lima, and C.~B. Verhaaren, {\it
  {Colorless Top Partners, a 125 GeV Higgs, and the Limits on Naturalness}},
  {\em Phys.Rev.} {\bf D91} (2015) 055007,
  [\href{http://arxiv.org/abs/1411.3310}{{\tt arXiv:1411.3310}}].

\bibitem{Craig:2015pha}
N.~Craig, A.~Katz, M.~Strassler, and R.~Sundrum, {\it {Naturalness in the Dark
  at the LHC}},  {\em JHEP} {\bf 07} (2015) 105,
  [\href{http://arxiv.org/abs/1501.05310}{{\tt arXiv:1501.05310}}].

\bibitem{Dawson:2013bba}
S.~Dawson, A.~Gritsan, H.~Logan, J.~Qian, C.~Tully, et~al., {\it {Working Group
  Report: Higgs Boson}},  \href{http://arxiv.org/abs/1310.8361}{{\tt
  arXiv:1310.8361}}.

\bibitem{Beringer:1900zz}
{\bf Particle Data Group} Collaboration, J.~Beringer et~al., {\it {Review of
  Particle Physics (RPP)}},  {\em Phys.Rev.} {\bf D86} (2012) 010001.

\bibitem{Curtin:2014jma}
D.~Curtin, P.~Meade, and C.-T. Yu, {\it {Testing Electroweak Baryogenesis with
  Future Colliders}},  {\em JHEP} {\bf 1411} (2014) 127,
  [\href{http://arxiv.org/abs/1409.0005}{{\tt arXiv:1409.0005}}].

\bibitem{Craig:2014lda}
N.~Craig, H.~K. Lou, M.~McCullough, and A.~Thalapillil, {\it {The Higgs Portal
  Above Threshold}},  \href{http://arxiv.org/abs/1412.0258}{{\tt
  arXiv:1412.0258}}.

\bibitem{Curtin:2015bka}
D.~Curtin and P.~Saraswat, {\it {Towards a No-Lose Theorem for Naturalness}},
  \href{http://arxiv.org/abs/1509.04284}{{\tt arXiv:1509.04284}}.

\bibitem{Craig:2013xia}
N.~Craig, C.~Englert, and M.~McCullough, {\it {New Probe of Naturalness}},
  {\em Phys.Rev.Lett.} {\bf 111} (2013), no.~12 121803,
  [\href{http://arxiv.org/abs/1305.5251}{{\tt arXiv:1305.5251}}].

\bibitem{Poland:2008ev}
D.~Poland and J.~Thaler, {\it {The Dark Top}},  {\em JHEP} {\bf 0811} (2008)
  083, [\href{http://arxiv.org/abs/0808.1290}{{\tt arXiv:0808.1290}}].

\bibitem{Aad:2015asa}
{\bf ATLAS} Collaboration, G.~Aad et~al., {\it {Search for pair-produced
  long-lived neutral particles decaying in the ATLAS hadronic calorimeter in
  $pp$ collisions at $\sqrt{s}$ = 8 TeV}},  {\em Phys.Lett.} {\bf B743} (2015)
  15--34, [\href{http://arxiv.org/abs/1501.04020}{{\tt arXiv:1501.04020}}].

\bibitem{Aad:2015uaa}
{\bf ATLAS} Collaboration, G.~Aad et~al., {\it {Search for long-lived, weakly
  interacting particles that decay to displaced hadronic jets in proton-proton
  collisions at $\sqrt{s}=8$ TeV with the ATLAS detector}},  {\em Phys. Rev.}
  {\bf D92} (2015), no.~1 012010, [\href{http://arxiv.org/abs/1504.03634}{{\tt
  arXiv:1504.03634}}].

\bibitem{Halyo:2013yfa}
V.~Halyo, H.~K. Lou, P.~Lujan, and W.~Zhu, {\it {Data driven search in the
  displaced $b\bar{b}$ pair channel for a Higgs boson decaying to long-lived
  neutral particles}},  {\em JHEP} {\bf 01} (2014) 140,
  [\href{http://arxiv.org/abs/1308.6213}{{\tt arXiv:1308.6213}}].

\bibitem{Curtin:2013fra}
D.~Curtin, R.~Essig, S.~Gori, P.~Jaiswal, A.~Katz, et~al., {\it {Exotic decays
  of the 125 GeV Higgs boson}},  {\em Phys.Rev.} {\bf D90} (2014), no.~7
  075004, [\href{http://arxiv.org/abs/1312.4992}{{\tt arXiv:1312.4992}}].

\bibitem{Scherk:1978ta}
J.~Scherk and J.~H. Schwarz, {\it {Spontaneous Breaking of Supersymmetry
  Through Dimensional Reduction}},  {\em Phys.Lett.} {\bf B82} (1979) 60.

\bibitem{Chen:2005mg}
Y.~Chen, A.~Alexandru, S.~Dong, T.~Draper, I.~Horvath, et~al., {\it {Glueball
  spectrum and matrix elements on anisotropic lattices}},  {\em Phys.Rev.} {\bf
  D73} (2006) 014516, [\href{http://arxiv.org/abs/hep-lat/0510074}{{\tt
  hep-lat/0510074}}].

\bibitem{Machacek:1983tz}
M.~E. Machacek and M.~T. Vaughn, {\it {Two Loop Renormalization Group Equations
  in a General Quantum Field Theory. 1. Wave Function Renormalization}},  {\em
  Nucl.Phys.} {\bf B222} (1983) 83.

\bibitem{Carmi:2012yp}
D.~Carmi, A.~Falkowski, E.~Kuflik, and T.~Volansky, {\it {Interpreting LHC
  Higgs Results from Natural New Physics Perspective}},  {\em JHEP} {\bf 1207}
  (2012) 136, [\href{http://arxiv.org/abs/1202.3144}{{\tt arXiv:1202.3144}}].

\bibitem{Djouadi:1997yw}
A.~Djouadi, J.~Kalinowski, and M.~Spira, {\it {HDECAY: A Program for Higgs
  boson decays in the standard model and its supersymmetric extension}},  {\em
  Comput.Phys.Commun.} {\bf 108} (1998) 56--74,
  [\href{http://arxiv.org/abs/hep-ph/9704448}{{\tt hep-ph/9704448}}].

\bibitem{Meyer:2008tr}
H.~B. Meyer, {\it {Glueball matrix elements: A Lattice calculation and
  applications}},  {\em JHEP} {\bf 0901} (2009) 071,
  [\href{http://arxiv.org/abs/0808.3151}{{\tt arXiv:0808.3151}}].

\bibitem{JuknevichPhD}
J.~Juknevich, {\em {Phenomenology of pure-gauge hidden valleys at Hadron
  colliders}}.
\newblock PhD thesis, Rutgers U., Piscataway.

\bibitem{Curtin:2014cca}
D.~Curtin, R.~Essig, S.~Gori, and J.~Shelton, {\it {Illuminating Dark Photons
  with High-Energy Colliders}},  {\em JHEP} {\bf 1502} (2015) 157,
  [\href{http://arxiv.org/abs/1412.0018}{{\tt arXiv:1412.0018}}].

\bibitem{Alwall:2011uj}
J.~Alwall, M.~Herquet, F.~Maltoni, O.~Mattelaer, and T.~Stelzer, {\it {MadGraph
  5 : Going Beyond}},  {\em JHEP} {\bf 1106} (2011) 128,
  [\href{http://arxiv.org/abs/1106.0522}{{\tt arXiv:1106.0522}}].

\bibitem{Sjostrand:2007gs}
T.~Sjostrand, S.~Mrenna, and P.~Z. Skands, {\it {A Brief Introduction to PYTHIA
  8.1}},  {\em Comput.Phys.Commun.} {\bf 178} (2008) 852--867,
  [\href{http://arxiv.org/abs/0710.3820}{{\tt arXiv:0710.3820}}].

\bibitem{HWG}
L{HC Higgs Cross Section Working Group}.
  \href{https://twiki.cern.ch/twiki/bin/view/LHCPhysics/HiggsEuropeanStrategy}
  {https://twiki.cern.ch/twiki/bin/view/LHCPhysics/HiggsEuropeanStrategy},
  2014.

\bibitem{CMS:2013jda}
{\bf CMS} Collaboration, C{MS Collaboration}, {\it {Higgs to bb in the VBF
  channel}}, .

\bibitem{conversationandy}
{\it {Private conversation with Andy Haas}}, .

\bibitem{CMS:2014wda}
{\bf CMS} Collaboration, V.~Khachatryan et~al., {\it {Search for long-lived
  neutral particles decaying to quark-antiquark pairs in proton-proton
  collisions at $\sqrt{s} =$ 8 TeV}},  {\em Phys.Rev.} {\bf D91} (2015), no.~1
  012007, [\href{http://arxiv.org/abs/1411.6530}{{\tt arXiv:1411.6530}}].

\bibitem{Butler:2020886}
J.~Butler, D.~Contardo, M.~Klute, J.~Mans, and L.~Silvestris, {\it {Technical
  Proposal for the Phase-II Upgrade of the CMS Detector}},  Tech. Rep.
  CERN-LHCC-2015-010. LHCC-P-008, CERN, Geneva, Jun, 2015.

\bibitem{Cinca:1711887}
D.~Cinca, {\it {ATLAS Upgrades Towards the High Luminosity LHC}},  Tech. Rep.
  ATL-UPGRADE-PROC-2014-001, CERN, Geneva, Jun, 2014.

\bibitem{Kang:2008ea}
J.~Kang and M.~A. Luty, {\it {Macroscopic Strings and 'Quirks' at Colliders}},
  {\em JHEP} {\bf 0911} (2009) 065, [\href{http://arxiv.org/abs/0805.4642}{{\tt
  arXiv:0805.4642}}].

\bibitem{Harnik:2008ax}
R.~Harnik and T.~Wizansky, {\it {Signals of New Physics in the Underlying
  Event}},  {\em Phys.Rev.} {\bf D80} (2009) 075015,
  [\href{http://arxiv.org/abs/0810.3948}{{\tt arXiv:0810.3948}}].

\bibitem{Harnik:2011mv}
R.~Harnik, G.~D. Kribs, and A.~Martin, {\it {Quirks at the Tevatron and
  Beyond}},  {\em Phys.Rev.} {\bf D84} (2011) 035029,
  [\href{http://arxiv.org/abs/1106.2569}{{\tt arXiv:1106.2569}}].

\bibitem{pdg}
{\bf Particle Data Group} Collaboration, K.~Olive et~al., {\it {Review of
  Particle Physics (RPP)}},  {\em Chin.Phys.} {\bf C38} (2014) 090001.

\bibitem{Schwaller:2015gea}
P.~Schwaller, D.~Stolarski, and A.~Weiler, {\it {Emerging Jets}},  {\em JHEP}
  {\bf 1505} (2015) 059, [\href{http://arxiv.org/abs/1502.05409}{{\tt
  arXiv:1502.05409}}].

\bibitem{Cohen:2015toa}
T.~Cohen, M.~Lisanti, and H.~K. Lou, {\it {Semi-visible Jets: Dark Matter
  Undercover at the LHC}},  \href{http://arxiv.org/abs/1503.00009}{{\tt
  arXiv:1503.00009}}.

\bibitem{Batell:2015zla}
B.~Batell and S.~Jung, {\it {Probing Light Stops with Stoponium}},  {\em JHEP}
  {\bf 07} (2015) 061, [\href{http://arxiv.org/abs/1504.01740}{{\tt
  arXiv:1504.01740}}].

\bibitem{Aad:2015uka}
{\bf ATLAS} Collaboration, G.~Aad et~al., {\it {Search for Higgs boson pair
  production in the $b\bar{b} b\bar{b}$ final state from $pp$ collisions at
  $\sqrt{s} = 8$ TeV with the ATLAS detector}},  {\em Eur. Phys. J.} {\bf C75}
  (2015), no.~9 412, [\href{http://arxiv.org/abs/1506.00285}{{\tt
  arXiv:1506.00285}}].

\end{thebibliography}\endgroup
\bibliographystyle{JHEP}


\end{document}